\documentclass{aa}  

\usepackage{graphicx}
\usepackage{tabularx}
\usepackage{txfonts}
\graphicspath{{img/}}
\usepackage{color}
\usepackage{algorithm}
\usepackage{algorithmic}
\usepackage{hyperref}
\hypersetup{colorlinks=True,citecolor=blue}

\definecolor{owngreen}{rgb}{0.0, 0.5, 0.0}

\newcommand{\changes}[1]{\textcolor{black}{#1}}
\newcommand{\newchanges}[1]{\textcolor{black}{#1}}

\begin{document}

    \title{Modeling Doppler Shifts in Radial-Velocity Data with Deep Learning toward Earth-mass Exoplanet Detection}

   
    \author{Isidro G{\'o}mez-Vargas\inst{1, 2, 3}
    \and Xavier Dumusque\inst{1}
    \and Yinan Zhao\inst{4}
    \and Khaled Al Moulla\inst{5}
    \and Michael Cretignier\inst{6}
    }
   
    \institute{
    Department of Astronomy, University of Geneva 51 chemin de Pegasi, 1290 Versoix, Switzerland.\\
    \email{isidro.gomezvargas@unige.ch}
    \and Instituto de Astrofı\'isica de Andaluc\'ia (CSIC), Glorieta de la Astronom\'ia s/n, E-18008 Granada, Spain.
    \and Institute of Space Sciences (CSIC), Carrer de Can Magrans s/n, E-08193 Barcelona, Spain.
    \and Department of Astronomy, University of Texas at Austin, 2515 Speedway, Austin, TX 78712, USA.
    \and Instituto de Astrofísica e Ciências do Espaço, Universidade do Porto, CAUP, Rua das Estrelas, 4150-762 Porto, Portugal.
    \and Department of Physics, University of Oxford, OX13RH Oxford, UK. \\
    }

   \date{Received February, 2026; }

\titlerunning{Modeling Doppler Shifts in RV Data with Deep Learning}
\authorrunning{Isidro G{\'o}mez-Vargas, et al.}

 
\abstract
{Detecting the tiny Doppler shifts induced by Earth-mass planets in stellar radial-velocity measurements remains extremely challenging due to stellar activity. Despite substantial progress in statistical and machine-learning techniques, many deep-learning methods performing well on simulated data remain difficult to apply reliably on real stellar spectra.}
{The aim of this work is to develop a deep-learning framework that generalizes to real, unseen spectra and improves the detectability of Earth-mass planets in radial-velocity data.}
{We train artificial neural networks on HARPS-N solar spectra with injected planetary signals, using physics-motivated spectral representations based on flux and line-formation temperature, together with their velocity gradients. Two training strategies are explored: hold-out testing, which provides a direct assessment of generalization to unseen spectra, and cross-validation, which evaluates performance across multiple folds of the dataset. Model robustness is enhanced through genetic-algorithm-based hyperparameter optimization, and predictive uncertainty is quantified using Monte Carlo dropout.}
{\newchanges{Our most precise neural network model reliably retrieves, under the CV strategy, the amplitudes, phases, and orbital periods of planetary signals with amplitudes greater than or equal to 25 cm/s and periods between 10 and 550 days. In addition, in all cases tested here, the successfully recovered signals correspond to the most significant peaks in the periodograms of the Doppler-shift predictions.}  Temperature-based spectral-shell representations consistently outperform flux-based shells, particularly in terms of predictive uncertainty and generalization to unseen data. As a byproduct, we release \texttt{doppleriann}, a Python package implementing the proposed spectral representations and deep-learning framework for Doppler-shift modeling.}
{Our results demonstrate that combining physically motivated spectral representations with deep learning provides a promising pathway toward the detection of Earth-mass planets in radial-velocity data from real observations, supported by a modeling framework that is both physically grounded and statistically rigorous, incorporating uncertainty quantification, and optimized training strategies.}


   \maketitle
%

\section{Introduction}
\label{sec:intro}

The Doppler spectroscopy technique, also known as the radial velocity (RV) method, is one of the most relevant ways to detect exoplanets. It was crucial in the discovery of the first exoplanet orbiting a solar-type star in 1995, the 51 Pegasi b hot-Jupiter \citep{mayor1995jupiter}, and, to date, has identified over 1000 exoplanets\footnote{See \url{https://exoplanet.eu/catalog/} for verification.}. The RV signal of a planet is directly proportional to the mass of the planet and inversely proportional to the cube root of its period. Therefore, detecting Earth-mass planets orbiting solar-type stars in their habitable zone (HZ), at orbital periods of a few hundred days, is extremely challenging, as it requires radial-velocity measurements with a precision of $\sim$0.1 m/s. 

Besides the technological challenge to reach a 0.1 m/s RV precision on timescales ranging from a few minutes to several years, the RV method is sensitive to stellar variability \citep[e.g.][]{saar1997activity, Lindegren-2003, fischer2016state, Dumusque:2017aa, meunier2019impact, luhn2023impact, Burt:2025aa_RV_review}, which further complicates the detection of Earth-mass planets. Stellar variability arises from different physical processes, including the propagation of pressure waves within the stellar photosphere, manifested as stellar oscillations \citep[e.g.][]{Dumusque-2011a, Luhn:2025aa_stellar_oscillations}), convection-driven variations in surface brightness associated with granulation \citep[e.g.][]{Dumusque-2011a, AlMoulla:2023aa, lakeland2024magnetically, OSullivan(2025)}, and the emergence and evolution of surface magnetic fields responsible for magnetic activity \citep[e.g.][]{saar1997activity, Meunier-2010a, Dumusque-2014b, korhonen2015stellar, Zhao:2025aa, zhao2023soap}. 

Even with the precision achieved by state-of-the-art high-resolution spectrographs such as ESPRESSO \citep[][]{pepe2021espresso}, detecting Earth-mass planets in the HZ of solar-type stars remains beyond reach \citep{Figueira:2025aa}. In theory, if measurement noise were independent, a few hundred observations would suffice to detect an Earth-mass planet \citep{hara2023statistical}. In practice, however, stellar variability dominates the signal. For instance, solar radial velocity measurements obtained with HARPS-N exhibit a standard deviation of 2.95 m/s over 10 years \citep{Dumusque:2025aa}. Consequently, data analysis, rather than instrumental sensitivity, is likely the primary limiting factor.

Stellar variability mitigation is essential for improving exoplanet detectability when using the RV method. Several studies have addressed this issue using techniques such as Principal Component Analysis \citep[e.g.][]{davis2017insights, Collier-Cameron:2021aa, Cretignier:2022aa}, Gaussian processes \citep[e.g.][]{rajpaul2015gaussian,Barragan:2022aa,Delisle:2022aa,Hara:2025_FENRIR}, or line-by-line RV analysis \citep[e.g.][]{dumusque2018measuring, Artigau:2022aa,Cretignier:2023ab,salzer2025searching}, all aimed at disentangling planetary signals from the effects of stellar variability.

In recent years, interest in machine-learning techniques for RV exoplanet detection has grown substantially. In particular, artificial neural networks (ANNs), including deep learning (DL) methods, have demonstrated high precision when applied to both simulated and stellar spectral data \citep{de2022identifying, perger2023machine, nieto2023exoplannet, liang2023aestra, kjaersgaard2023tau, zhao2024deep, gavankar2026machine, mcwilliam2026identifying}, owing to their ability to model complex nonlinear relationships \citep{hornik1990universal}. 

Convolutional neural networks (CNNs) have been shown to be effective in several contexts. For instance, \cite{de2022identifying} applies CNNs to solar data in combination with the cross-correlation function (CCF) technique, while \cite{perger2023machine} reports promising results using CNNs on simulated data for two different stars and multiple physical observables. Binary classification of simulated RV time series to determine the presence of planetary signals is explored by \cite{nieto2023exoplannet}.

Other neural-network architectures have also been investigated. The autoencoder-based approach of \cite{liang2023aestra} demonstrates the recovery of low-amplitude signals in simulated data, and \cite{kjaersgaard2023tau} employs autoencoders to improve telluric correction and enhance planet detectability. 

Beyond simulated datasets, recent studies have demonstrated promising performance on real stellar data. In particular, the CNN models developed by \cite{zhao2024deep} achieve high precision, detecting signals at the 20 cm/s level in solar data and at 50 cm/s in observations from Alpha Centauri B and Tau Ceti. Likewise, \cite{colwell2024deep} reports detections of signals as small as 20 cm/s using an ensemble of CNNs trained on real solar spectra, although the analysis is limited to periods of 50 days and requires computationally intensive models with significant prediction offsets. 

Although these approaches show strong performance within their respective settings, most rely on neural-network models that must be retrained for each star or simulation, leaving their behavior largely untested across different data distributions, such as new stellar targets, varying observing conditions, or unseen spectral regions. As a result, while these methods are in principle applicable to multiple stars, their trained models have not yet demonstrated consistent performance when applied to data with different statistical properties. This limitation is particularly relevant for RV analyses, where real spectra vary in wavelength coverage, noise properties, and levels of stellar activity. These considerations motivate the development of more transferable and interpretable machine-learning frameworks that operate directly on real stellar spectra and maintain robust performance on genuinely unseen data.

Building on these examples, our work presents an alternative approach based on real solar data combined with physically motivated data compression to train neural-network models conceptually similar to those of \cite{zhao2024deep}. As in \cite{colwell2024deep}, our method directly predicts the Doppler shift (DS) induced by planetary signals, thereby extracting information that is not directly accessible from classical CCF-based activity indicators. \changes{Compared to \cite{colwell2024deep}, which focuses on short-period signals (up to 50 days) using computationally intensive ensemble models, our framework extends the analysis to a broader range of orbital periods (up to 550 days), while employing physically motivated spectral representations that enable lightweight neural-network architectures. We incorporate advanced hyperparameter optimization and uncertainty quantification to obtain more robust models and prevent overfitting. The trained networks are evaluated on spectral-shell samples that are completely held out from training and validation, providing a rigorous test of their ability to generalize to genuinely unseen temporal and spectral data.}

We develop models using solar data from the HARPS-N telescope \citep{dumusque2015harps, Phillips:2016aa, Dumusque:2025aa}, based on both stellar flux and line-formation temperature, demonstrating improved performance with the latter. This behavior is consistent with previous findings \citep{al2022measuring, al2024measuring}, which show that stellar activity signals depend on the formation temperature of spectral lines, mainly due to the suppression of convective motions in active regions, thereby facilitating the disentanglement of stellar activity and planetary signals.

This paper is structured as follows. In Section \ref{sec:shells}, we describe the spectral-shell representation used as input to our deep-learning framework, considering flux and temperature information. Section \ref{sec:data_processing} details the spectral data processing. In Section \ref{sec:dl}, we present our deep-learning framework, outlining the core methodology as well as the training and prediction strategies, including the hold-out and cross-validation schemes. Section \ref{sec:results} reports the results of the planetary recovery tests for both hold-out and cross-validation approaches, using flux- and temperature-based representations, together with a discussion of their implications. Finally, Section \ref{sec:conclusions} summarizes the main conclusions of this work. Appendix~\ref{sec:app_methodological} presents additional validation tests of the methodology, while Appendix \ref{sec:app_uncertainty} provides an analysis of the predictive uncertainties of the neural-network models to assess their robustness.

\section{Spectral-Shell Representations}
\label{sec:shells}
\begin{figure}[htbp]
    \centering
    \includegraphics[trim=2mm 4mm 1mm 1mm, clip, width=\linewidth]{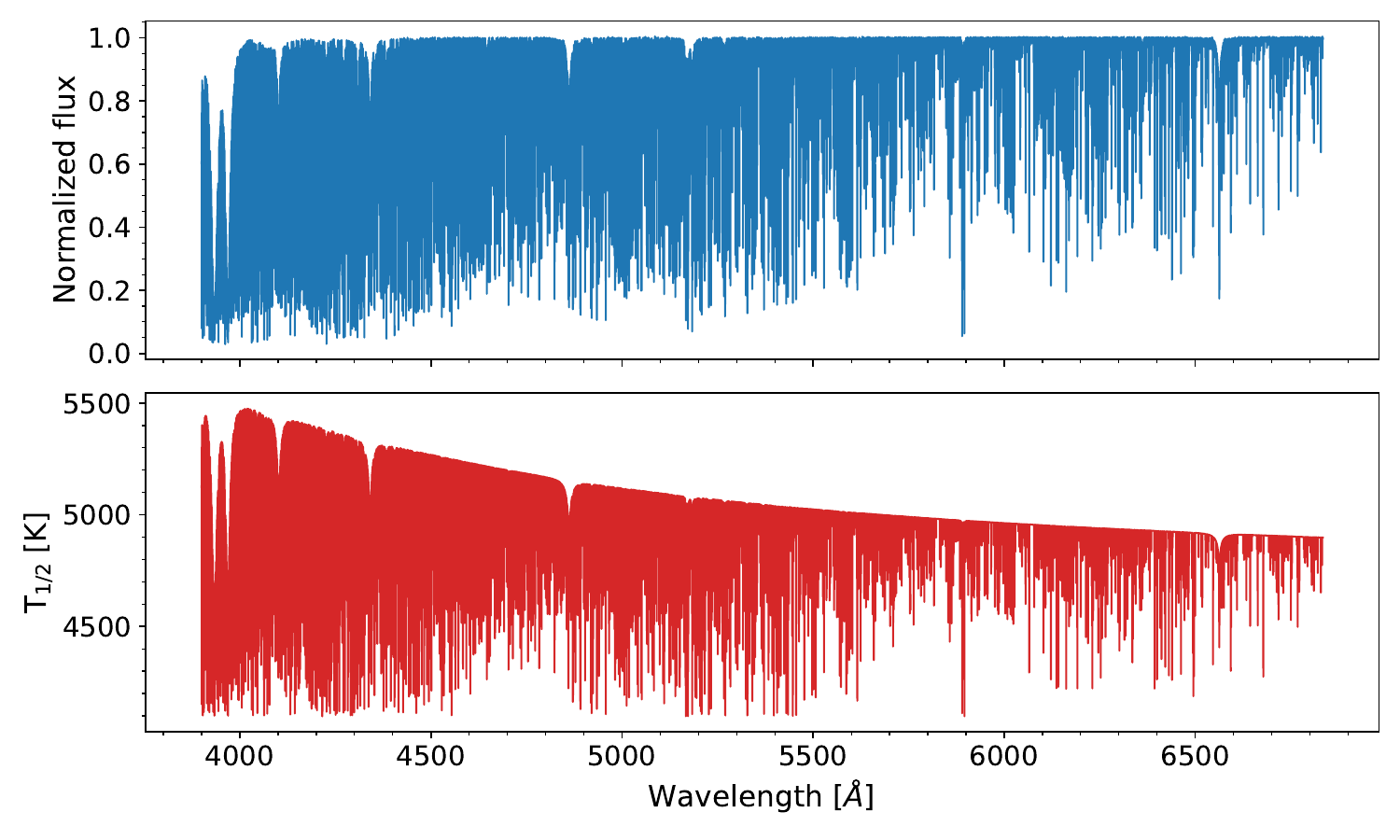}
    \caption{Example of a single HARPS-N solar spectrum. The top panel shows the flux, normalized to its continuum level, as a function of wavelength. The bottom panel shows the corresponding average formation temperature, $T_{1/2}$, corresponding to the same wavelengths.}
    \label{fig:flux_temp_spectra}
\end{figure}

\begin{figure*}[htbp]
    \centering
    \includegraphics[trim=35mm 12mm 25mm 15mm, clip, width=\linewidth]{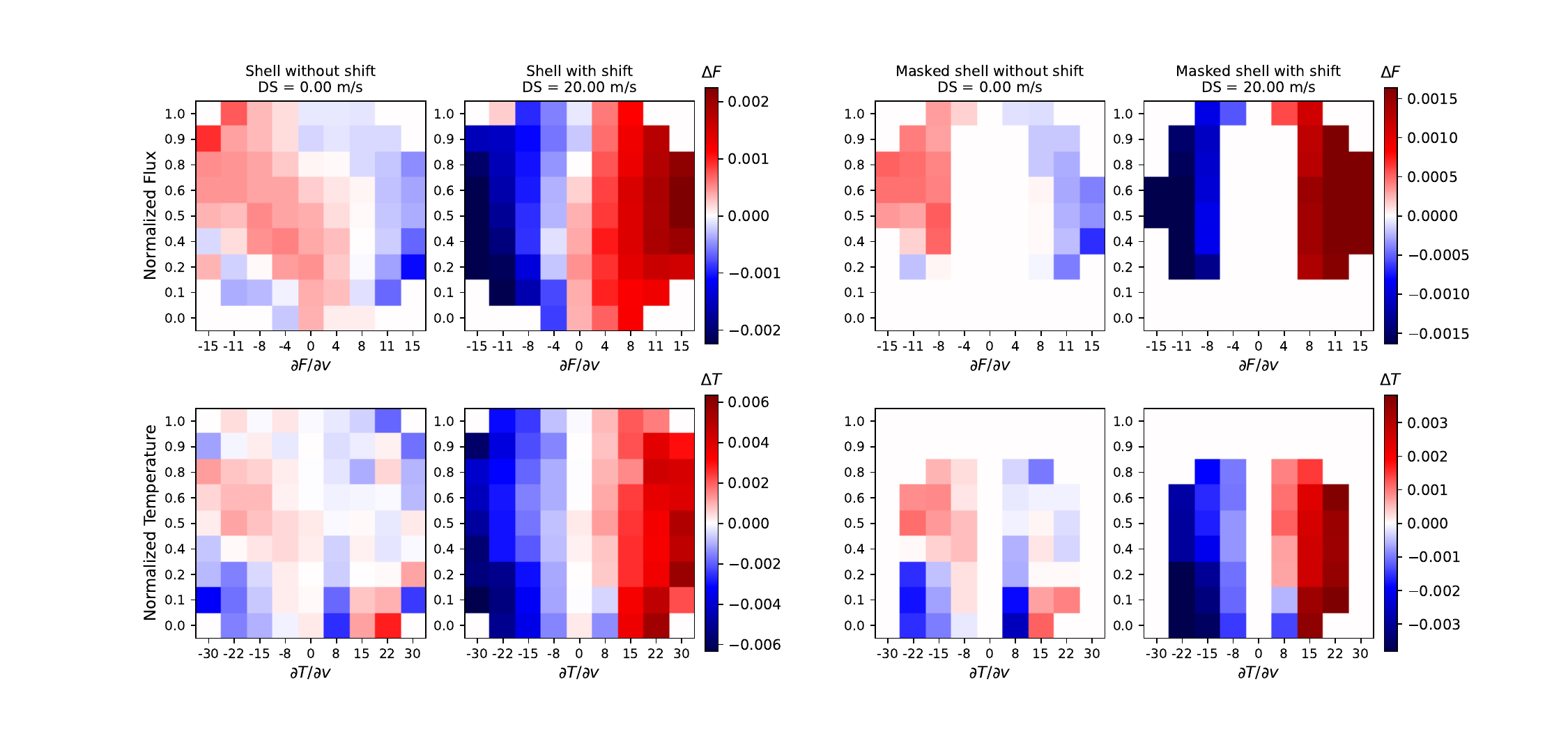}
    \caption{Shell representations for flux and temperature at the time of a maximum DS of 20 m/s. \changes{The masked shell is defined as the element-wise product between the spectral shell and the associated density-weight map. In the shell construction, a cell is assigned zero only when no spectral pixels populate the corresponding bin. Although the same selected spectral pixels are used for both the flux- and temperature-based shells, they are projected onto different shell spaces, $(F,\partial F/\partial v)$ and $(T,\partial T/\partial v)$, so the occupancy of the shell cells is not identical between the two representations.} \textit{Top row:} Flux representations, with the first two panels unmasked and the last two masked. \textit{Bottom row:} Same configuration for the temperature-based representation.}
    \label{fig:shells}
\end{figure*}

In this work, we use \textit{spectral shells} as input to our deep-learning framework, rather than stellar spectra or CCF-based representations. Spectral shells, initially introduced in \cite{cretignier2022stellar}, are a reduced-dimensionality projection of high-resolution spectra designed to preserve sensitivity to Doppler shifts and stellar activity effects. They have a lower dimensionality than stellar spectra, significantly improving neural-network training. Unlike CCFs, which are also low-dimensionality, they have been shown to better probe stellar variability signals \citep{cretignier2022stellar}. Moreover, this representation compresses the spectral data into a two-dimensional grid defined by normalized flux as a function of the flux gradient. The flux gradient is the same for all spectra and is derived from a high signal-to-noise master spectrum constructed as the average of all available stellar spectra.

A shell grid is then formed by binning the space defined by the flux gradient and normalized flux. Within each bin, the mean flux variation relative to the master spectrum is recorded. This transformation retains localized spectral variations with respect to the master spectrum that are sensitive to line shape variation, while mitigating high-frequency noise. 

In \cite{zhao2024deep}, rather than using the originally defined gradient of flux with respect to wavelength, the authors show that adopting the gradient of flux with respect to velocity is more appropriate and enables improved modeling of stellar variability. This is because spectral lines from different wavelengths overlap in the spectral-shell representation when expressed in velocity space, thereby correcting for dispersion.

Our methodology builds upon the use of spectral shells with a gradient defined with respect to velocity and extends this approach by replacing normalized flux with formation temperature. This modification is motivated by the work of \cite{al2022measuring} \citep[see also][]{Siegel:2022aa, Siegel:2024aa, al2024measuring}, which demonstrates that stellar variability depends on formation height in the stellar photosphere (or, equivalently, on formation temperature). Each sampled spectral point is associated with its average formation temperature, $T_{1/2}$, determined through radiative-transfer modeling and defined as the photospheric temperature at which the cumulative flux contribution function reaches 50\% of its maximum. This enables a mapping of the spectral information with respect to photospheric depth. Figure~\ref{fig:flux_temp_spectra} shows a comparison of the same solar spectrum using either normalized flux or formation temperature. This temperature-dependent refinement enhances the capability of the shell representation to probe activity-induced line profile variations.

In the following subsections, we describe the formalism for constructing shell representations based on both flux and temperature.

\subsection{Flux-Based Shell Representation}

Let \( F(\lambda) \) and \( F_0(\lambda) \) be two spectra defined on the same wavelength grid, with \( F_0(\lambda) \) denoting a master or reference spectrum. Under a linear approximation, when \( F(\lambda) \) is Doppler-shifted by a small velocity \( \Delta V \), the observed flux variation at pixel \( i \) is given by \citep[see][]{Bouchy-2001b}:
\begin{equation}
F(\lambda_i) - F_0(\lambda_i) = \Delta F_i = \frac{\partial F_0(\lambda_i)}{\partial \lambda_i} \cdot \Delta \lambda_i = \frac{\partial F_0(\lambda_i)}{\partial \lambda_i} \cdot \frac{\Delta V_i}{c} \lambda_i,
\end{equation}
which leads to:
\begin{equation}
\Delta V_i = \frac{c}{\lambda_i} \cdot \Delta F_i \cdot \left( \frac{\partial F_0(\lambda_i)}{\partial \lambda_i} \right)^{-1}.
\end{equation}

To eliminate chromatic effects from dispersion, the shell representation is constructed using the flux derivative with respect to velocity rather than wavelength:
\begin{equation}
\Delta V_i = \Delta F_i \cdot \left( \frac{\partial F_0(v_i)}{\partial v_i} \right)^{-1}.
\end{equation}

In this formulation, each shell is defined as the set of tuples \changes{\( (F_0, d F_0 / d v) \)}, and bins are formed in this space. The weight associated with each velocity bin is defined as:
\begin{equation}
W_{v,i} = \left[ \frac{\partial F_0(v_i)}{\partial v_i} \cdot \frac{1}{\sigma_{F_i}} \right]^2,
\end{equation}
where \( \sigma_{F_i} \) denotes the noise associated with the observed flux \( F(v_i) \). This quantity, equal to \( \sqrt{F(v_i) + \sigma_{det,i}^2} \), where \( \sigma_{det,i} \) is detector noise at the pixel $i$, is generally given by the data extraction pipeline. The total weight in each cell of the shell grid is then calculated as:
\begin{equation}
W_{v,\text{cell}} = \sum_{i \in \text{cell}} W_{v,i}.
\end{equation}
These weights $W_{v,\text{cell}}$ \changes{are used to construct the masked shell representation (illustrated in Figure \ref{fig:shells}), obtained by the element-wise product between the original spectral shell and the associated weight map. In our experiments, this weighting leads to more stable representations and neural-network training by reducing the contribution of low-information or noisy regions.}

\subsection{Temperature-Based Shell Representation} \label{temp_shells}

To account for the fact that stellar activity has a photospheric depth dependence \citep{al2022measuring}, we extended the shell framework using a formation temperature-binned approach. Each spectral pixel is associated with its average formation temperature \( T_{1/2} \).  In analogy with the flux-based formalism, we define an equivalent velocity change in temperature space as:

\begin{equation}
\Delta V_T = \frac{\Delta T}{g_T}, \qquad \text{with} \quad g_T = \frac{\partial T}{\partial v},
\label{eq:deltaV}
\end{equation}
where $g_T$  denotes the local numerical slope,  $\frac{\partial T}{\partial v}$, of the temperature spectrum computed along the Doppler velocity grid. This term describes how the temperature values derived from the flux-temperature mapping (described in the following section) vary with velocity, and it should not be interpreted as a physical temperature gradient. 

The corresponding uncertainty in velocity space becomes:
\begin{equation}
\sigma_v = \frac{\sigma_T}{g_T}.
\label{eq:sigmav}
\end{equation}

Here, \( \sigma_T \) is not a directly observed quantity; it is derived by propagating the photon noise in flux, \( \sigma_{F_i} \), through the flux-to-temperature mapping:
\begin{equation}
\sigma_T = f_{F \rightarrow T}(\sigma_{F_i}),
\label{eq:sigmaT}
\end{equation}
where \( f_{F \rightarrow T} \) denotes the calibration function (either empirical or model-based) relating flux uncertainties to effective temperature uncertainties at each wavelength.

The weight associated with each velocity bin in the temperature shell is thus:
\begin{equation}
W_{v,i}^{(T)} = \left[ \frac{g_T}{f_{F \rightarrow T}(\sigma_{F_i})} \right]^2,
\label{eq:weight_pix}
\end{equation}
and the total weight in a shell cell becomes:
\begin{equation}
W_{v,\text{cell}}^{(T)} = \sum_{i \in \text{cell}} W_{v,i}^{(T)}.
\label{eq:weight_cell}
\end{equation}

This formulation enables the shell representation to retain sensitivity to stellar activity shifts at different photospheric depths, as traced by spectral line formation temperatures.


\section{Data Processing}
\label{sec:data_processing}
We analyze a 10-year dataset of HARPS-N solar observations acquired between 2015 and 2024, comprising 2036 daily binned high-resolution spectra \citep{dumusque2021three, Dumusque:2025aa}. Each spectrum consists of 293,401 flux measurements spanning the wavelength range from 3900 to 6900 \text{\AA}. The raw spectra were preprocessed using the YARARA pipeline \citep{cretignier2021yarara}, which corrects for instrumental systematics, telluric absorption, and stellar activity. These corrections are important to mitigate systematic features unrelated to planetary Doppler shifts. We note that the YARARA activity correction was re-injected into the spectra we analyzed, forming the so-called "YVA" dataset \citep[ e.g.][]{Dalal(2024), OSullivan(2025)}, as our goal is to develop a machine learning framework that can detect small-mass planetary signals in the presence of stellar signals.

Keeping the same timestamps of the original spectra, we inject synthetic Keplerian planetary signals into them to augment the training dataset and introduce controlled DS. We used the formalism described in \citet{Bouchy-2001b} to inject a planetary signal into the spectra. For two spectra Doppler-shifted by a certain amount and sampled on the same wavelength grid, the wavelength difference at each point is, to first order, equal to the flux difference divided by the local gradient. This wavelength difference can then be transformed into velocity by dividing by the wavelength and multiplying by the speed of light. Given the spectral gradient derived from the master spectrum, we modified the flux of each spectrum at each wavelength bin according to the velocity of the injected planetary signal. The simulated signals have DS ranging from 0.1 to 5 m/s, uniformly distributed orbital periods between 10 and 100 days, and random phases. For each modified spectrum, we compute the CCF using the ESPRESSO G2 weighted line mask \citep{pepe2002coralie, pepe2021espresso} to derive the total observed RV, which includes both the stellar and the injected planetary components, in addition to residual instrumental and telluric systematics not fully corrected for by YARARA.

Subsequently, we transform the normalized flux of the spectra into formation temperature values. Following \cite{al2022measuring}, we use the precomputed solar formation temperatures presented in that work, \changes{which can be extracted from the \texttt{ARVE} code \citep[][]{AlMoulla:2025_ARVE}, are first associated with the master spectrum and subsequently propagated to each individual observation by applying the same wavelength grid, ensuring that each spectral pixel in the time series is assigned a consistent temperature value (see Eq.~\ref{eq:deltaV}).} This approach avoids full spectral synthesis, remaining computationally efficient while preserving consistency with the solar photospheric structure. The resulting temperatures are then used to build the temperature-based shell representations (see Section~\ref{temp_shells}). For stellar sources other than the Sun, the \texttt{ARVE} code can be used to extract precomputed information for a limited set of spectral types; otherwise, equivalent formation temperatures would need to be derived from dedicated spectral synthesis \citep[as done in][]{al2024measuring}.

Following the procedure described in Section \ref{sec:shells}, we compute the shell representation and associated weights for each spectrum, \changes{by only using spectral lines that can be properly model by spectral synthesis and that are therefore void of systematics induced by tellurics or detector systematics (see Sect.~\ref{sec:DL_training_data})}. For both flux- and temperature-gradient representations, we discretize the shell space at a resolution of ten points per axis, yielding a \(9 \times 9\) grid of bins. As a result, we generate two different datasets: one comprising flux-based shells and another built from temperature-based shells, each paired with its corresponding CCF-derived RV and injected DS for supervised learning.

\changes{It is worth mentioning that the injected Doppler shifts are generated using the gradient of the master spectrum and not an analytical model. The latter would imply modifying the wavelengths associated with each spectral bin, differently for each spectrum, which would then significantly increase the complexity in deriving temperature-based spectra and in building a spectral shell. A simple solution could be to resample all the spectra to a common wavelength grid, but this would add some unwanted systematics for each spectrum due to interpolation. We therefore preferred to use the linear approximation described in \citet{Bouchy-2001b} as in this case, only the systematics involved in the computation of the master spectrum gradient contaminated the data. As this master spectrum has a much higher S/N than individual spectra, the induced systematics are smaller than those obtained by performing a resampling of each spectrum, \newchanges{providing a stable reference for constructing the shell representation. The use of a fixed master spectrum remains an approximation, since stellar activity induces time-dependent spectral variations that are not explicitly incorporated into the shell construction. We note that all precise RV extraction methods are inherently based on relative measurements. Methods such as CCF, template matching, and line-by-line analyses rely on a synthetic or observed reference template, which in our case corresponds to the master spectrum.}}


\begin{figure}[ht]
    \centering
    \includegraphics[trim=90mm 45mm 95mm 60mm, clip, width=0.5\textwidth]{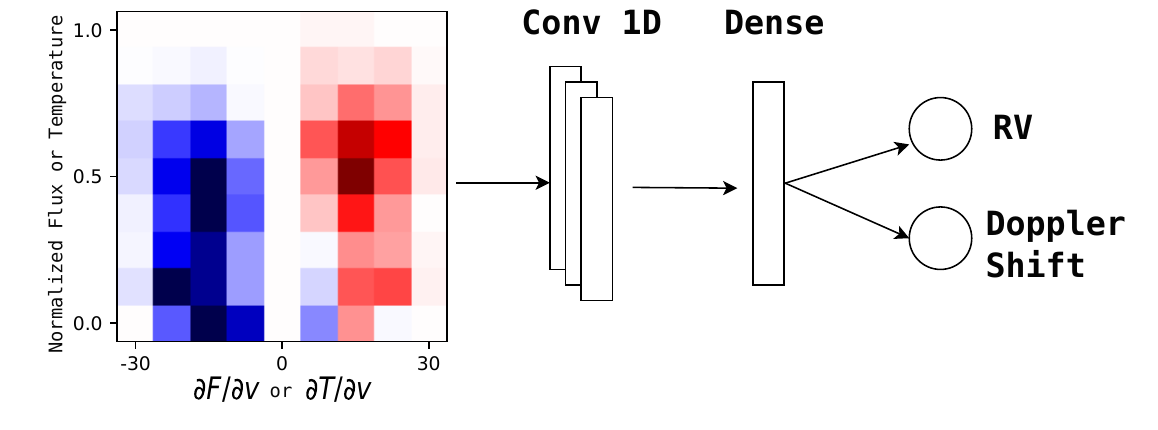}
    \caption{Schematic of the 1D CNN architecture used to predict RV and DS from shell inputs. }
    \label{fig:cnn_architecture}
\end{figure}

\section{Deep-Learning Framework}
\label{sec:dl}

\begin{table}[t]
\centering
\caption{Notation used for training, validation, and testing datasets.}
\label{tab:notation}
\begin{tabular}{ll}
\hline\hline
Symbol & Description \\
\hline
$\mathcal{S}$ & Full set of HARPS-N solar spectral shells (2036) \\
$\mathcal{V}$ & Hold-out test set (subset of $\mathcal{S}$) \\
$\mathcal{D}$ & Development set, $\mathcal{D}=\mathcal{S}\setminus\mathcal{V}$ \\
$\mathcal{T}$ & Training set (subset of $\mathcal{D},  \approx 90\%$) \\
$\mathcal{V}_{\mathrm{int}}$ & Internal validation set (subset of $\mathcal{D}, \approx 10\%$) \\
$\mathcal{J}_{\mathrm{train}}$ & Injection grid used for training: \\
 & RV semi-amplitude $\in \{0.1, 0.2, 0.3, 0.5, 1, 2, 5\}$ m/s, \\
 & $P \in \{20, 40, 60, 80, 100\}$ days, with \\
 & random phases \\
$\mathcal{J}_{\mathrm{all}}$ & Full injection set used for evaluation: \\
 & RV semi-amplitude $\in [0.1, 0.45]$ m/s in steps of $0.05$, \\
 & $P \in [10, 100]$ days in steps of $10$, plus \\
 & $P \in [150, 550]$ days in steps of $50$, with \\
 & random phases \\
$M_k$ & Neural network trained in the $k$-th CV fold \\
$M$ & Neural network trained in hold-out setting \\
DS & Doppler shift (in m/s) \\
$P$ & Orbital period (in days) \\
\hline
\end{tabular}
\end{table}

We train supervised convolutional neural networks that receive as input a physics-informed shell representation, based on flux or temperature, with dimensions of $9 \times 9$. The neural network is designed to predict two outputs: the total RV estimated from the CCF, and the DS manually injected into the spectrum. This dual-output configuration was empirically \changes{adopted to include the RV as a quantity that can be independently computed from the spectra using the CCF, enabling a direct comparison between traditional and neural-network predictions, while the DS output provides access to the injected planetary signal, which is not directly available from the CCF.} A schematic overview of the model architecture is presented in Fig. \ref{fig:cnn_architecture}; details are provided below, and Table \ref{tab:notation} summarizes the notation used hereafter.

\subsection{Training Data}
\label{sec:DL_training_data}

Our analysis is based on 2036 HARPS-N solar spectra with corresponding observation times. We restrict the analysis to the wavelength regions selected by the KITCAT line mask from \cite{al2022measuring}, yielding a total of 31,066 wavelength samples. \changes{Those regions have been selected because the observed solar spectra agree well with spectral modeling \citep[using pySME][]{Wehrhahn:2023aa_PySME}, which is required to derive precise temperature information. In addition, this ensures that the selected spectral lines are not significantly contaminated by tellurics and other instrumental systematics.} In these masked spectra, we inject synthetic Keplerian planetary signals with random phases, varying RV semi-amplitudes, and orbital periods ($P$) into the data to augment the training set. We consider two training strategies, whose pseudocode is detailed in Algorithms~\ref{alg:HO} and \ref{alg:CV}.

\begin{itemize}
    \item \textbf{Algorithm \ref{alg:HO}: Hold-out testing (HO).} We randomly select 400 spectra as a fixed test set $\mathcal{V}$, ensuring that these samples are completely unseen during training. The remaining 1636 spectra form the training set $\mathcal{T}$, which is augmented using the injection grid $\mathcal{J}_{\mathrm{train}}$. A single neural network $M$ is then trained on $\mathcal{T}$. For evaluation, we preserve the strict train–test separation by keeping the 400 real test spectra unchanged during training. For each planetary configuration $j \in \mathcal{J}_{\mathrm{all}}$, we create an injected version of the test set $\mathcal{V}_j$ by adding the corresponding synthetic planetary signal. The trained model $M$ is applied to each injected test set $\mathcal{V}_j$ (without retraining), yielding a predicted Doppler-shift time series for that specific configuration. This strategy ensures strict separation between training and evaluation data while enabling an extensive exploration of the injection parameter space. 

    \item \textbf{Algorithm \ref{alg:CV}: Cross-validation training (CV).} We implement an $N_{\mathrm{folds}}$-fold cross-validation strategy, in which the dataset $\mathcal{S}$ is partitioned into $N_{\mathrm{folds}}$ complementary subsets. For each fold $k$, one subset $\mathcal{V}_k$ is used as an unseen evaluation set, while the remaining data define a development set $\mathcal{D}_k$, further split into training and internal validation subsets. Synthetic planetary signals drawn from the full injection grid $\mathcal{J}_{\mathrm{all}}$ are injected into the training data only. A separate neural-network model $M_k$ is trained for each fold and applied to the corresponding unseen subset $\mathcal{V}_k$, without retraining. This per-fold procedure mirrors the hold-out strategy in terms of data splitting and synthetic signal injection, with the difference that the role of the evaluation set is rotated across folds.
\end{itemize}

Detectability in the hold-out setting is assessed via periodogram analysis of the predicted Doppler-shift time series obtained from the fixed test set of 400 spectra. In the cross-validation setting, after all folds are completed, predictions from the disjoint evaluation subsets are concatenated to reconstruct a time series spanning the full dataset. This ensures that each spectrum is treated as unseen data exactly once, while allowing the final periodogram analysis to exploit all 2036 samples, in contrast to the hold-out strategy, which relies on a single fixed test set.

In the main analysis, we adopt $N_{\mathrm{folds}} = 5$, yielding evaluation subsets of approximately 400 spectra per fold, comparable in size to the hold-out test set. For completeness, results obtained with $N_{\mathrm{folds}} = 10$ are presented in Appendix~\ref{sec:app_10folds}. In both the HO and CV approaches, a wide range of orbital periods, semi-amplitudes, and phases can be explored thanks to the data-augmentation strategy based on synthetic planetary injections. This constitutes a subtle but important difference from \citet{zhao2024deep}, as our cross-validation approach requires only one trained model per fold to evaluate multiple injected planetary configurations.

Under both training strategies, for each spectrum, we generate both flux-based and temperature-based shell representations as described in Section~\ref{sec:shells}. RVs are computed from the CCF for each spectrum. The CNN input consists of the shell (flux or temperature) multiplied element-wise by its corresponding weight matrix (see Figure~\ref{fig:shells}), forming a $9 \times 9$ feature map. This is done for all augmented spectra, for example, $\approx$ 57,200 elements in a hold-out test scenario, or a single fold of a 5-fold CV approach, using $\approx$400 spectral samples for testing. The two target outputs are the CCF-derived RV and the true DS induced by the injected planetary signal, which varies in absolute magnitude between 0 and the injected RV semi-amplitude, depending on the observed time of each spectrum.  

Although time is used to inject Keplerian signals into the spectra across the original HARPS-N dataset for data augmentation, temporal information is not preserved during training. The spectral shells are randomly selected, and the network is therefore trained on samples corresponding to different injected Doppler shifts without explicit temporal ordering. Time information is only reintroduced during testing (see Section~\ref{sec:DL_model_predictions}), when the predicted DS are associated with the corresponding observation times of the spectral-shell time series and analyzed via periodograms to assess whether the injected planetary signals can be recovered.

The values for the injected RV semi-amplitude and orbital periods were chosen by trial and error, based on how well the neural network performed during training. We found that using a mix of small DS values (like 0.1 m/s) and moderate ones (1 - 5 m/s) helped the model learn to recognize both subtle and stronger patterns in the data. Adding very large DS values (like 20 m/s or more) caused the model to focus too much on those signals and made it harder for it to learn the smaller ones. On the other hand, using only very small DS values (0.1 and 0.2 m/s) does not give the model enough variety to learn general patterns. The same idea applies to the choice of orbital periods: we selected them such that several phases of the injected signal are sampled within the full timespan of our dataset, ensuring the presence of at least one clear peak in the RV signal. This choice ensures that each injected signal exhibits sufficient temporal variability to provide informative training samples, given the limited temporal sampling of the dataset.

\subsection{Neural-network design}
\label{sec:DL_neural_design}
We design two separate neural-network models: one for flux-based shells and another for temperature-based shells. Both models share the same general architecture, and their hyperparameters were optimized independently of each other. All models are implemented using the \texttt{TensorFlow} library \citep{abadi2016tensorflow}.

To estimate predictive uncertainty and improve model robustness, we apply the Monte Carlo dropout (MC-DO) technique \citep{gal2016dropout}. This method introduces stochasticity during inference by performing multiple forward passes with dropout enabled, and then taking the mean prediction. This approach has been shown to provide a Bayesian-like estimate of uncertainty and has been successfully used in other small astrophysical datasets for improving regression performance \citep{gomez2023neuralrec, mitra2024dark}.

Each neural network begins with two 1D convolutional layers. Although the shell representation has a two-dimensional layout, its entries are not pixel-based visual features as in classical image processing. Instead, they contain continuous numerical values with physical meaning, normalized flux or line-formation temperature gradients, showing correlations primarily along the velocity-gradient axis. Given their small size (9 $\times$ 9) and the structured nature of the data, 1D convolutions offer a more stable and efficient representation than 2D convolutions. Applying 1D convolutions along the rows allows the network to capture local variations across gradient bins while preserving the interpretation of each row as a slice of constant normalized flux or temperature. This configuration also reduces the number of trainable parameters and mitigates overfitting without loss of relevant information, since neighboring gradient values remain correlated due to the smooth variation of normalized flux or temperature with respect to velocity.

To optimize the model architecture and training configuration, we used the Optuna library \citep{akiba2019optuna} to employ the Nondominated Sorting Genetic Algorithm II (NSGA-II) \citep{deb2000fast} for exploring the hyperparameter space (Table~\ref{tab:hyperparams}) and minimizing the training loss. This algorithm is well-suited for exploring multi-objective optimization problems. It has been mathematically demonstrated that genetic algorithms with elitism, such as NSGA-II, are guaranteed to converge toward the best solution \cite{rudolph1994convergence}. Moreover, they have been successfully applied to astrophysical regression tasks using neural networks \citep{hetem2007use, gomez2023neural, mitra2024dark, gomez2024deep}.

The hyperparameter search targeted key elements of the architecture: batch size, learning rate, convolutional layer configuration, and dense layer depth. We fixed the number of training epochs at 1000 and employed early stopping with a patience of 40 epochs based on validation loss, to prevent overfitting. Learning rates were sampled logarithmically in the range $10^{-4}$ to $10^{-2}$; convolutional layers were explored through combinations of kernel sizes and filter depths; and dense layers followed several compression structures. Dropout was fixed at 0.2, \changes{and the SELU activation function (Eq.~\ref{eq:selu}) was used throughout for its stable behavior during training.}

\begin{equation}
\label{eq:selu}
    \mathrm{SELU}(x) = \lambda 
\begin{cases}
x & \text{if } x > 0 \\
\alpha \left(e^{x} - 1\right) & \text{if } x \leq 0
\end{cases}
\end{equation}

All models were trained using the Adam optimizer and the mean squared error (MSE) loss function, defined in Eq.~\ref{eq:loss}:

\begin{equation}
\label{eq:loss}
\mathrm{MSE} =  \frac{1}{N} \sum_i \left[(RV_{\mathrm{CCF}, i} - RV_{\mathrm{pred}, i})^2 
+ (DS_{\mathrm{inj}, i} - DS_{\mathrm{pred}, i})^2 \right], 
\end{equation}
where $RV_{\mathrm{CCF}}$ denotes the radial velocity measured from the CCF, $RV_{\mathrm{pred}}$ is the neural-network prediction, $DS_{\mathrm{inj}}$ is the injected Doppler shift associated with the synthetic planetary signal, and $DS_{\mathrm{pred}}$ is the predicted Doppler shift.

\changes{Overall, the hyperparameter-optimization procedure involved evaluating a large number of candidate models, of order tens to hundreds of configurations when accounting for the combinations in Table~\ref{tab:hyperparams} and the learning-rate sampling, providing a broad exploration of the hyperparameter space. The best configurations were selected based on validation performance using this loss function. The robustness of the selected models is further supported by the uncertainty and residual analyses presented in Appendix~\ref{sec:app_uncertainty}, which show no evidence of overfitting or unstable behavior.}

For both the flux and temperature-based models, the best-performing configurations shared a single dense layer with 512 units. The convolutional layers found for the flux shells were $[(128, 3), (256, 3)]$, and  $[(256, 5), (512, 5)]$ for the temperature-based. The optimal batch sizes determined were 256 for the flux model and 128 for the temperature model, with corresponding learning rates of 0.002 and 0.0002, respectively. These configurations consistently yielded the best validation performance and stable convergence across multiple runs.
\newchanges{As discussed in Section~\ref{sec:DL_training_data}, although the dataset comprises 2036 independent solar spectra, the training procedure uses 35 synthetic planetary injections, yielding approximately 57,000 augmented training shell samples in each training run of both the hold-out and cross-validation approaches. Each training sample is represented by a 9 $\times$ 9 shell representation matrix. While these realizations are not independent observations, they provide a broad range of planetary amplitudes and phases, substantially increasing the diversity of the training examples. The final architectures selected through the hyperparameter optimization consist of two convolutional layers and one dense layer, and contain 759,042 trainable parameters for the flux representation and 931,330 trainable parameters for the temperature representation. Importantly, in neural networks, the raw number of trainable parameters is not, by itself, a sufficient diagnostic of overfitting; generalization may depend on effective capacity measures, regularization, and the training procedure rather than solely on the nominal parameter count \citep[e.g.,][]{bartlett1996valid,ingrassia2005neural,belkin2019reconciling,nakkiran2021deep}. In our case, dropout and early stopping constrain the effective complexity of the trained models, while the additional tests presented in Appendix~\ref{sec:app_methodological} and Appendix~\ref{sec:app_uncertainty} assess their robustness through zero-injection, cross-validation, dominant-peak, chronological hold-out, injection-recovery, and uncertainty-quantification analyses.}

\begin{table}[htbp]
\centering
\footnotesize
\caption{Hyperparameter combinations and best values found for batch size (BS), learning rate (LR), Convolutional layers (CL), and Dense layers (DL).}
\label{tab:hyperparams}
\begin{tabular}{|c|c|c|c|}
\hline
\textbf{} & \textbf{Explored Values} & \textbf{Flux} & \textbf{Temp} \\
\hline
BS & \{128, 256\} & 256 & 128 \\
\hline
LR & $\log\;\mathcal{U}(10^{-4}, 10^{-2})$ & 0.002 & 0.0002 \\
\hline
CL & 
\{[(128, 3), (256, 3)], & [(128, 3), & [(256, 5), \\
& [(256, 3), (512, 3)], & (256, 3)] & (512, 5)] \\
& [(128, 5), (256, 5)], & & \\
& [(256, 5), (512, 5)] & & \\
\hline
DL & 
\{[512], [512, 256], [512, 256, 128], & [512] & [512] \\
& [256], [256, 128], [256, 128, 64]\} & & \\
\hline
\end{tabular}
\end{table}

\subsection{Model Predictions and Time-Series Analysis}
\label{sec:DL_model_predictions}

Our neural network is trained to detect subtle variations in the shell representation, whether based on flux or temperature, that correspond to a DS induced by planetary signals. This is done independently of time, as no temporal information is encoded in spectral shells and no time is given during training. Time is only needed to evaluate the performance of the neural network. Indeed, when applied to a time series of spectral shells, our neural network predicts an RV (planets plus stellar activity plus instrumental residuals) and a DS (only planets) time series. Applying a standard periodogram analysis to the predicted RV and DS time series, we can evaluate the performance of the model and its ability to detect planetary signals by performing injection-recovery tests. The predicted RV time series are used to verify consistency with CCF-derived RVs and, therefore, assess the regression quality of our model. By injecting planetary signals at the level of spectral-shell time series and analyzing the predicted DS time series using a standard periodogram approach, we can assess the ability to recover planetary signals that the traditional CCF-based RVs may not capture due to stellar and instrumental signals.

\newchanges{The shell representations are constructed using a fixed master spectrum. As a result, time-dependent variations in the stellar spectrum induced by activity are not explicitly incorporated into the shell-construction stage. Instead, temporal information enters the framework only through the subsequent periodogram analysis of the predicted DS and RV time series.} In particular, we employ the Lomb-Scargle periodogram implementation from the \texttt{PyAstronomy} library \citep{czesla2019pya} to search for planetary signals in the DS time series, and require a detected signal with a False Alarm Probability (FAP) smaller than 0.1\%, as in \cite{zhao2024deep}. To evaluate the performance of our model, we adopted two strategies: \textit{hold-out testing} and \textit{cross-validation}. In the \textit{hold-out} approach (Algorithm~\ref{alg:HO}), a fixed random set of 400 spectra is reserved as an unseen test set, while a single model is trained on the remaining data and evaluated on this held-out subset.  In the \textit{cross-validation} approach  (Algorithm~\ref{alg:CV}), the model is trained on the full set of 2036 solar spectra using an $N_\mathrm{folds}$-fold scheme, where in each fold a subset of the data is treated as unseen evaluation data and the remaining spectra are used for training with synthetic planetary injections. Considering Algorithms \ref{alg:HO} and \ref{alg:CV}, note that when $N_\mathrm{folds} = 5$, the size of the evaluation subset in each fold is approximately 400 spectra; therefore, a single fold of the cross-validation procedure is similar to the hold-out training.

\begin{figure*}[t]

    \includegraphics[trim=0mm 70mm 0mm 52mm, clip, width=\textwidth]{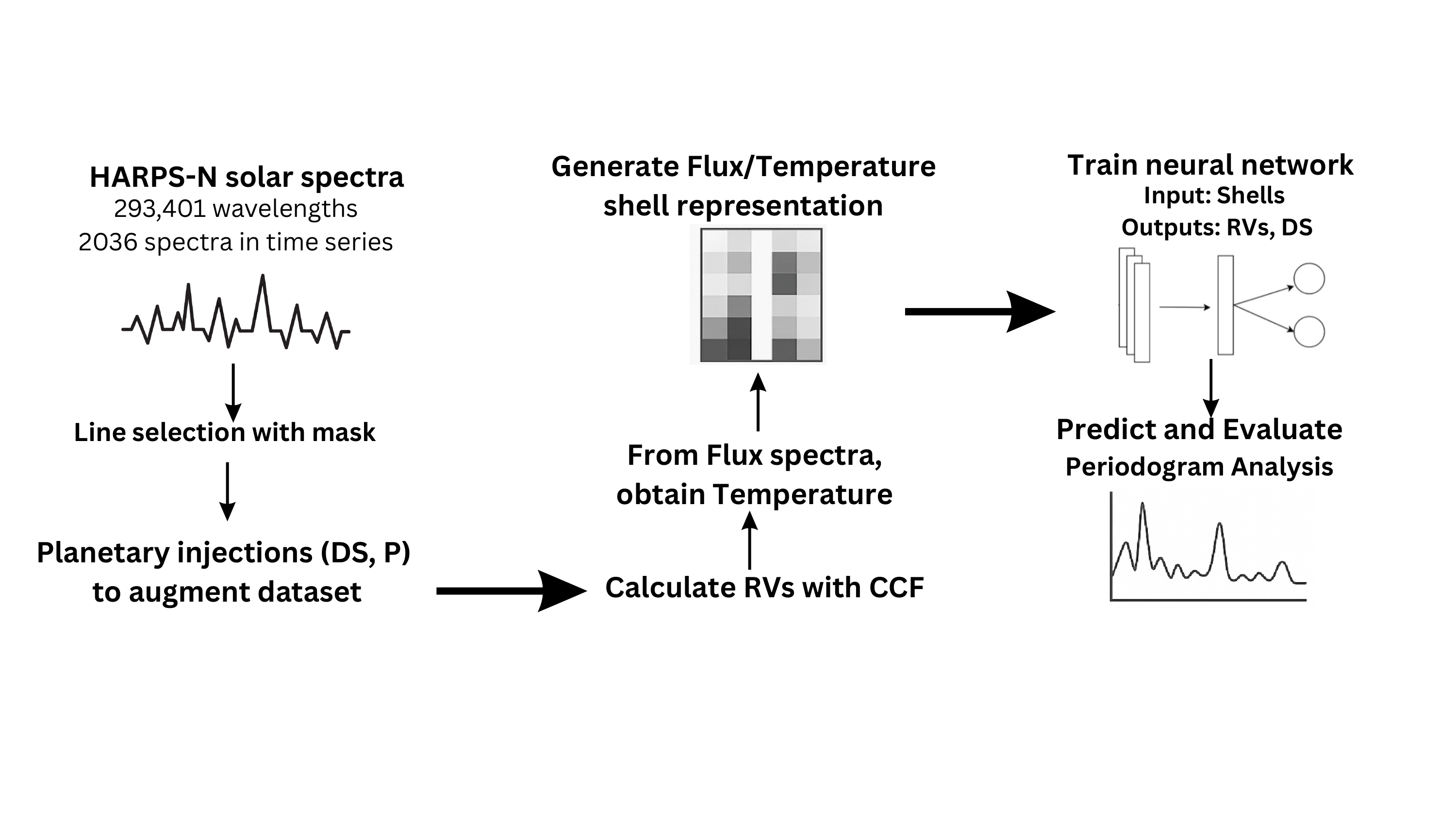}
    \caption{Schematic workflow of the methodology. HARPS-N solar spectra 
($293{,}401$ wavelengths, $2036$ spectra in a time series) are processed through a line-selection mask to select spectral regions of interest (strong lines with minimum blending). Planetary signals with specified DS and periods ($P$) are injected to augment the dataset, and their radial velocities (RVs) are calculated using the CCF method. Both flux and temperature information are used to generate shell representations, which serve as inputs to a convolutional neural network. The trained network predicts RVs and DS, and the predictions are evaluated using periodogram analysis.}
\label{fig:diagram}
\end{figure*}

This framework extends the traditional planetary detection pipeline by incorporating the DS neural network predictions, offering an additional channel of information that may uncover signals the CCF alone cannot resolve. A general summary of the workflow of our analysis is shown in Figure~\ref{fig:diagram}.

\begin{algorithm}[t]
\caption{Hold-out (HO) testing method}
\begin{algorithmic}[1]
\label{alg:HO}
\STATE \textbf{Input:} Time series $\mathcal{S}$; training injections $\mathcal{J}_{\mathrm{train}}$; all injections $\mathcal{J}_{\mathrm{all}}$
\STATE Randomly select 400 spectra as test set $\mathcal{V}$; let $\mathcal{D} = \mathcal{S} \setminus \mathcal{V}$
\STATE Split $\mathcal{D}$ into training set $\mathcal{T}$ and internal validation set $\mathcal{V}_{\mathrm{int}}$
\STATE Augment $\mathcal{T}$ with injections from $\mathcal{J}_{\mathrm{train}}$
\STATE Train a single CNN $M$ on $\mathcal{T}$ (monitoring $\mathcal{V}_{\mathrm{int}}$)
\FOR{each injection $j \in \mathcal{J}_{\mathrm{all}}$}
  \STATE Predict on $\mathcal{V}$ with $M$ (without retraining)
\ENDFOR
\STATE Evaluate detection probability, amplitude, phase, and period errors
\end{algorithmic}
\end{algorithm}

\begin{algorithm}[t]
\caption{Cross-Validation (CV) method}
\begin{algorithmic}[1]
\label{alg:CV}
\STATE \textbf{Input:} Time series $\mathcal{S}$; injection set $\mathcal{J}_{\mathrm{all}}$; number of folds $N_\mathrm{folds}$
\FOR{$k=1$ to $N_\mathrm{folds}$}
  \STATE Define evaluation fold $\mathcal{V}_k \subset \mathcal{S}$ with $|\mathcal{V}_k| \simeq |\mathcal{S}|/N_\mathrm{folds}$
  \STATE Define development set $\mathcal{D}_k = \mathcal{S} \setminus \mathcal{V}_k$
  \STATE Split $\mathcal{D}_k$ into training set $\mathcal{T}_k$ and internal validation set $\mathcal{V}_{k,\mathrm{int}}$
  \STATE Inject signals from $\mathcal{J}_{\mathrm{all}}$ into $\mathcal{T}_k$
  \STATE Train CNN $M_k$ on $\mathcal{T}_k$ (monitoring $\mathcal{V}_{k,\mathrm{int}}$)
  \STATE Predict on $\mathcal{V}_k$ with $M_k$ (without retraining) and save outputs
\ENDFOR
\STATE Concatenate predictions from all folds to recover the full time series
\STATE Evaluate detection probability, amplitude, phase, and period errors
\end{algorithmic}
\end{algorithm}

\begin{figure*}
    \centering

     \makebox[17cm][c]{
        \includegraphics[trim=0mm 0mm 30mm 0mm, clip, width=5.0cm, height=4.8cm]{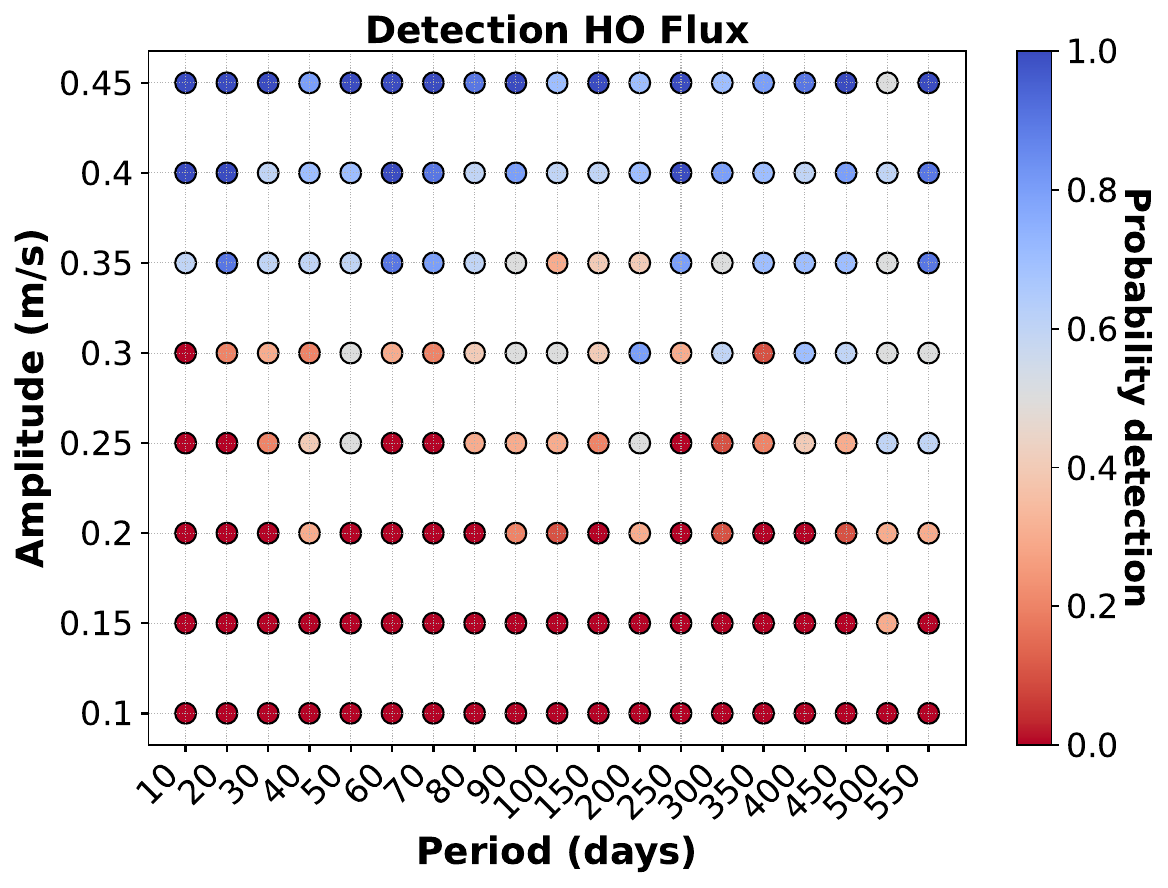}
        \includegraphics[trim=23mm 0mm 0mm 0mm, clip, width=5.2cm, height=4.8cm]{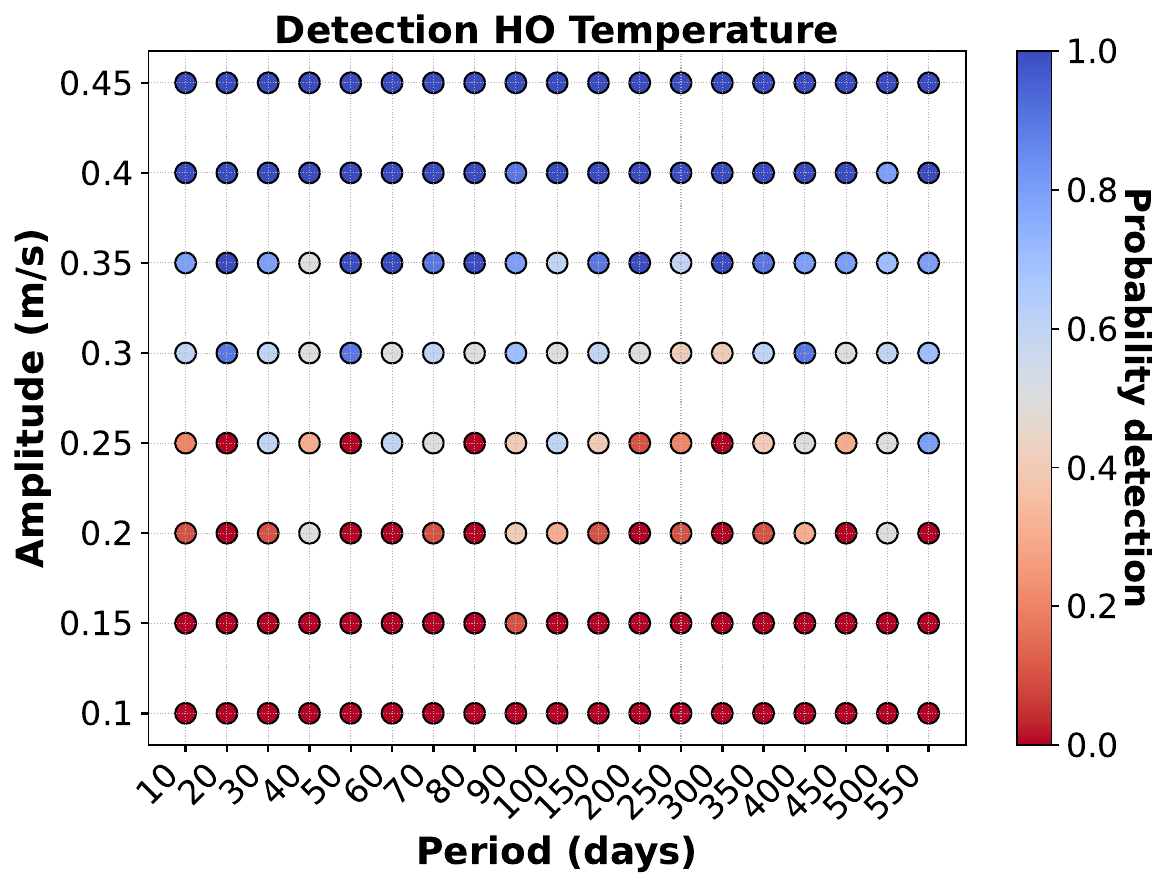}
        \includegraphics[trim=0mm 0mm 30mm 0mm, clip, width=5.0cm, height=4.8cm]{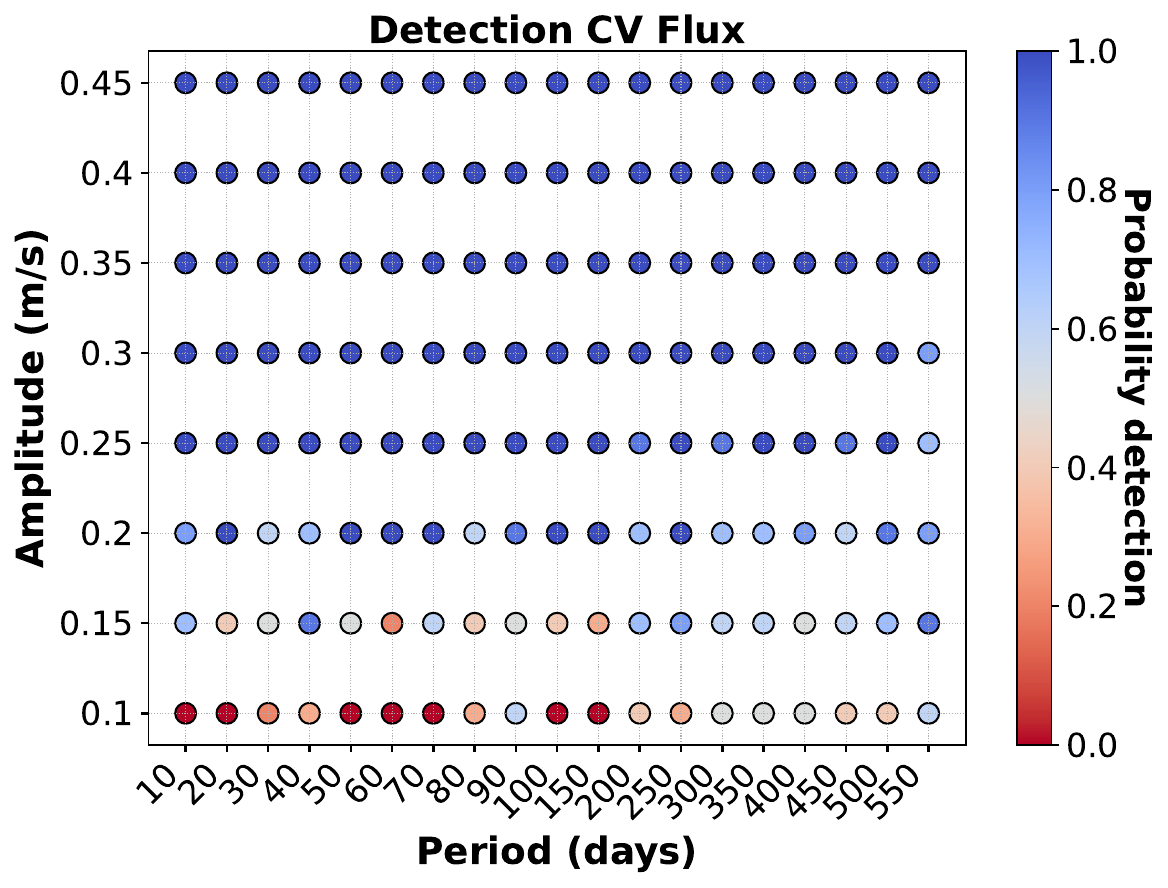}
        \includegraphics[trim=23mm 0mm 0mm 0mm, clip, width=5.2cm, height=4.8cm]{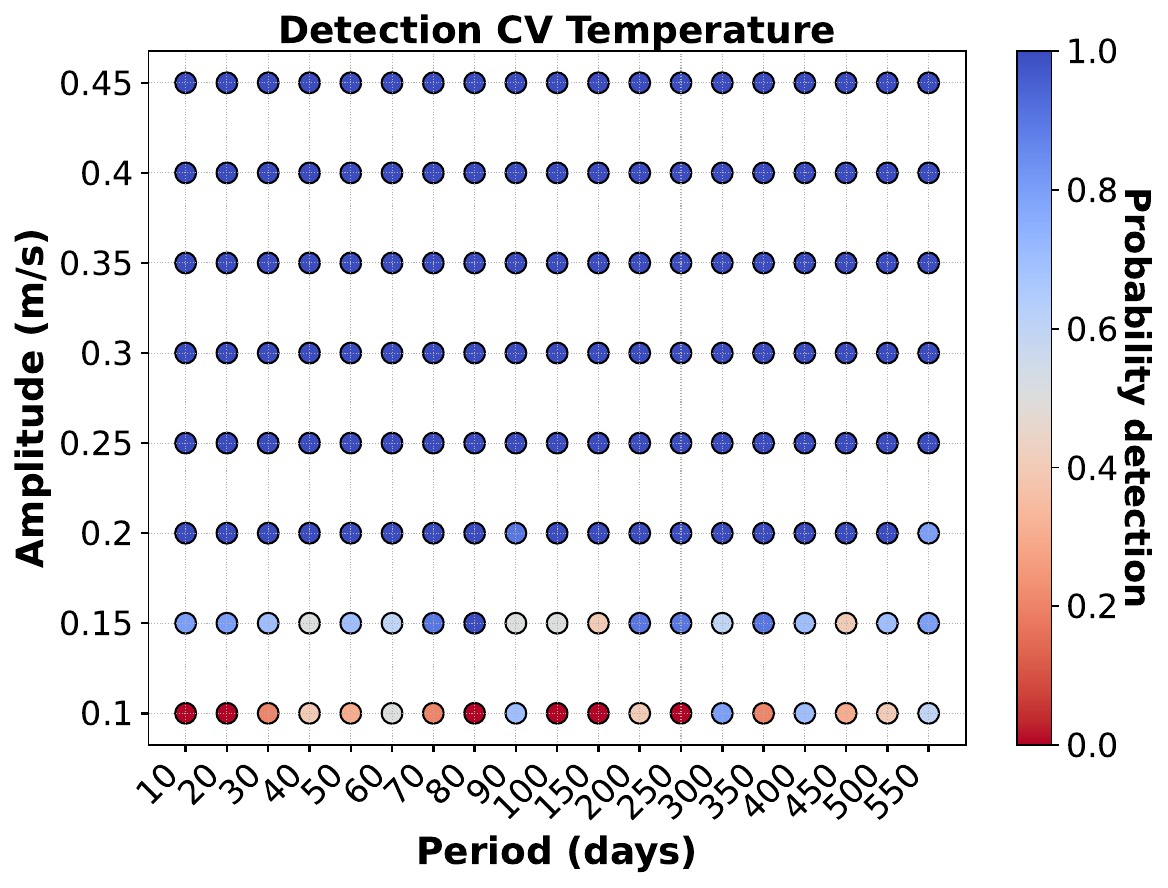}
    }

    \vspace{1em}

    \makebox[17cm][c]{
        \includegraphics[trim=0mm 0mm 29mm 0mm, clip, width=5.0cm, height=4.8cm]{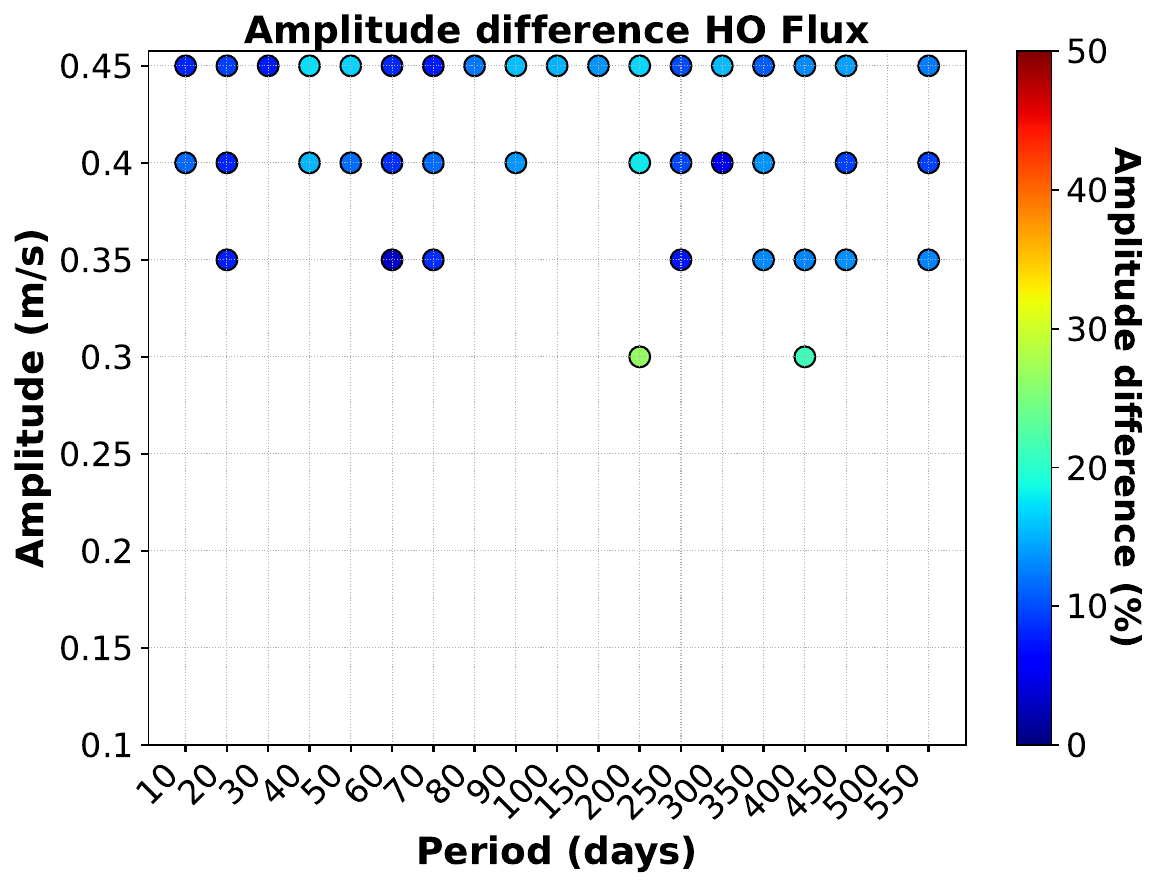}
        \includegraphics[trim=23mm 0mm 0mm 0mm, clip, width=5.2cm, height=4.8cm]{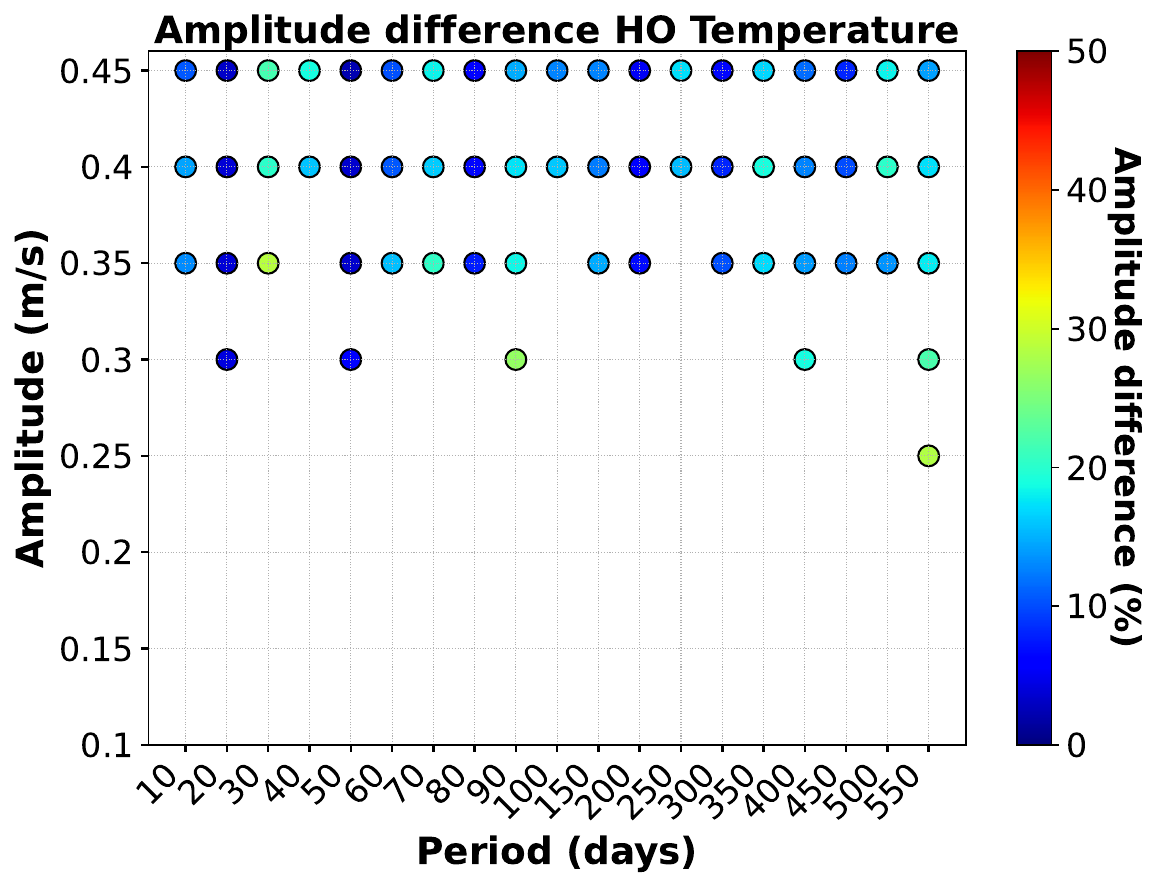}
        \includegraphics[trim=0mm 0mm 29mm 0mm, clip, width=5.0cm, height=4.8cm]{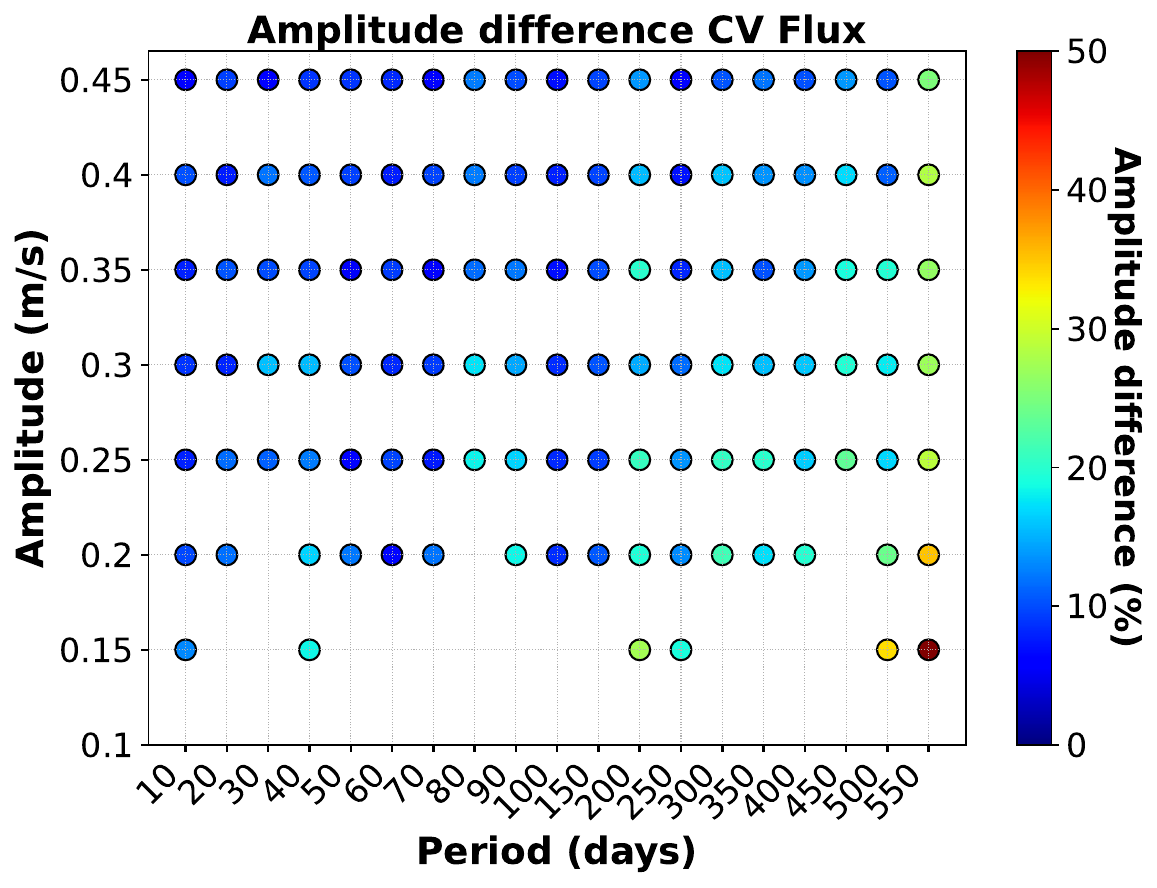}
        \includegraphics[trim=23mm 0mm 0mm 0mm, clip, width=5.2cm, height=4.8cm]{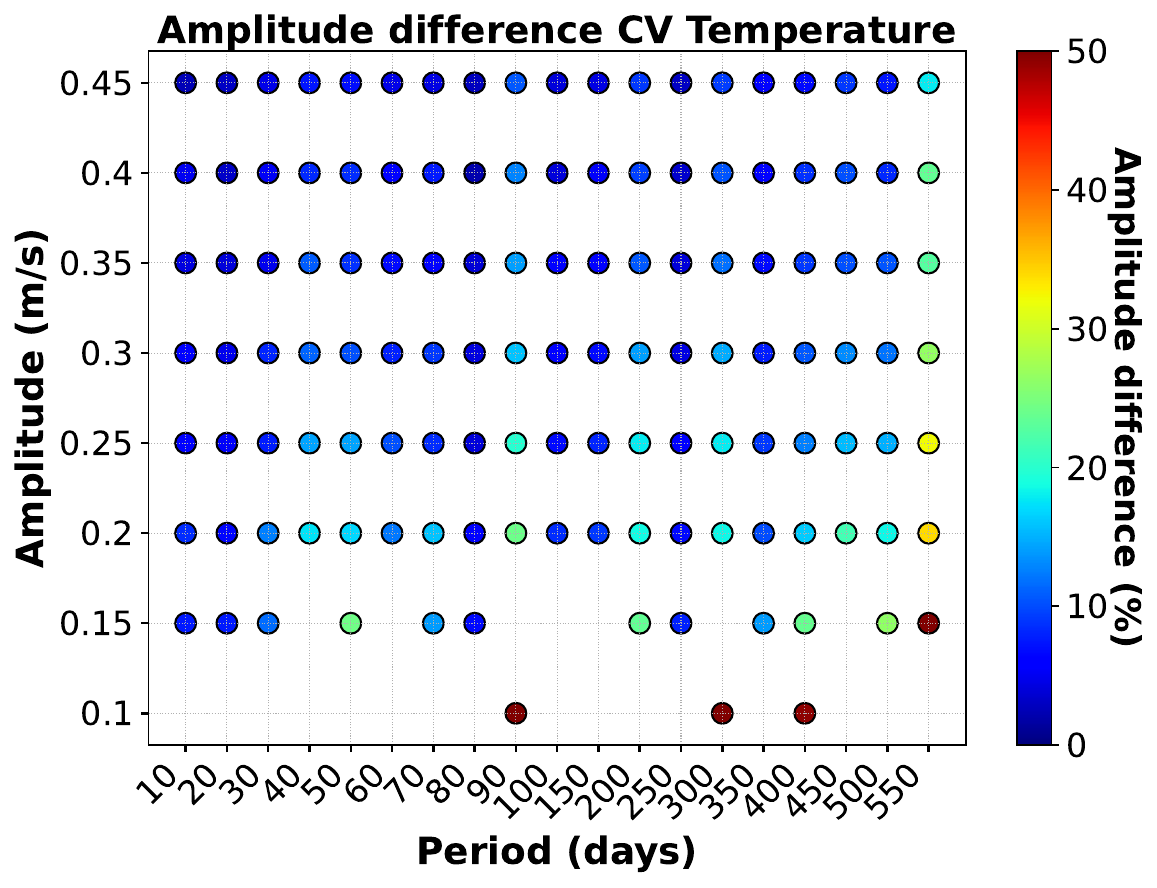}

    }

    \vspace{1em}

    \makebox[17cm][c]{
        \includegraphics[trim=0mm 0mm 30mm 0mm, clip, width=5.0cm, height=4.8cm]{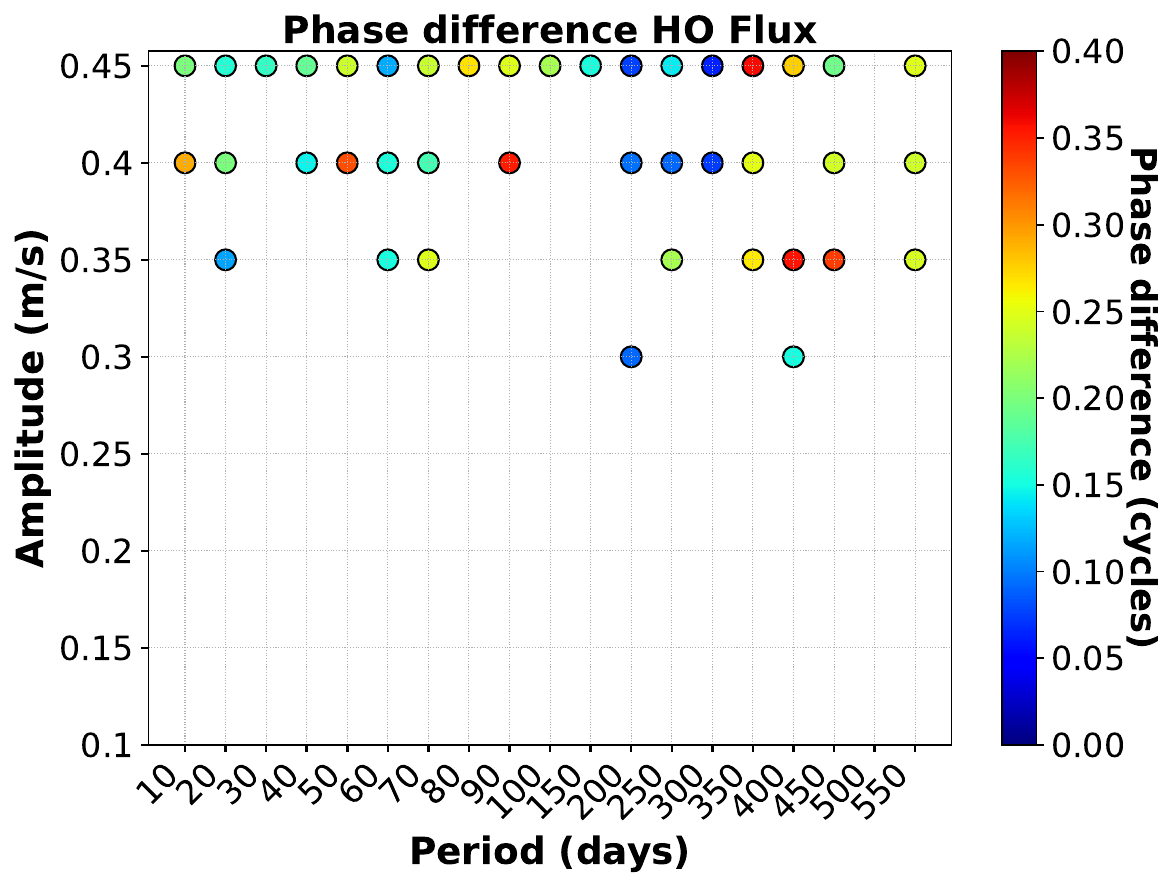}
        \includegraphics[trim=23mm 0mm 0mm 0mm, clip, width=5.2cm, height=4.8cm]{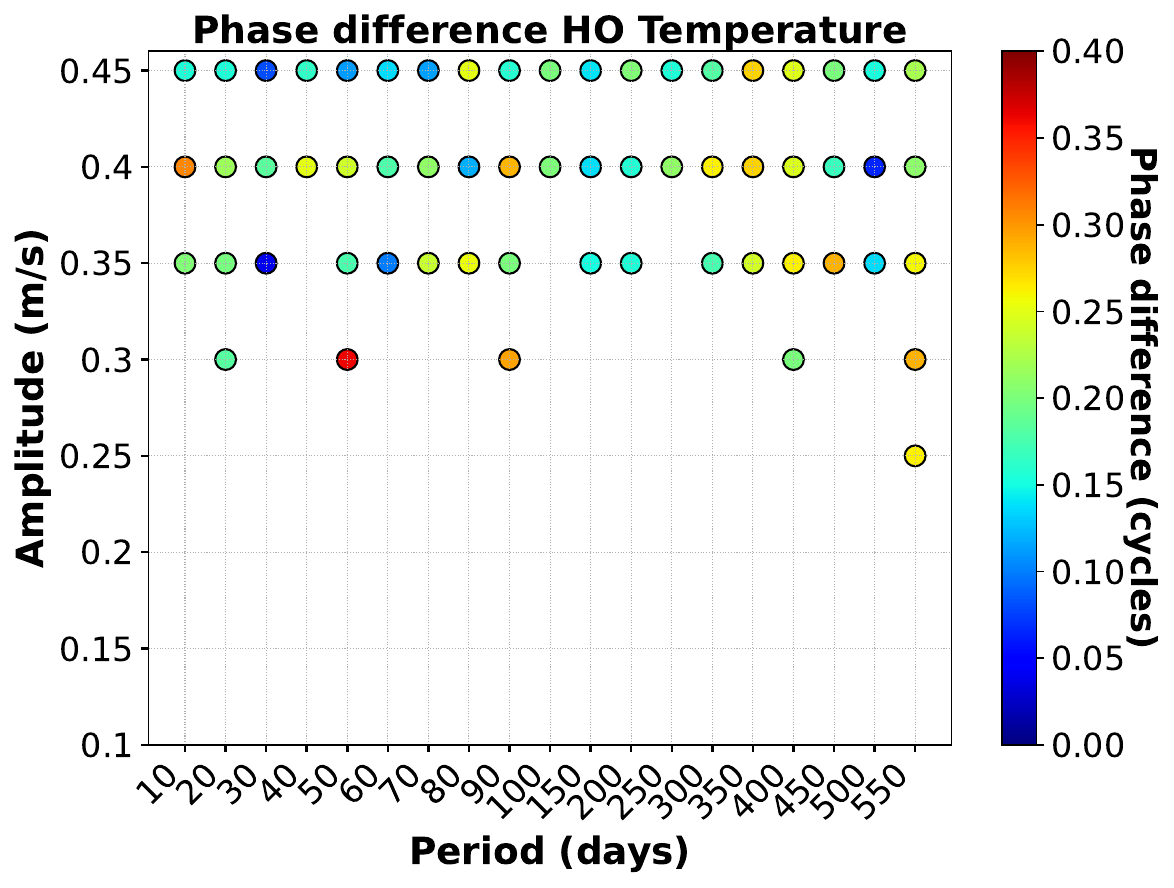}
        \includegraphics[trim=0mm 0mm 30mm 0mm, clip, width=5.0cm, height=4.8cm]{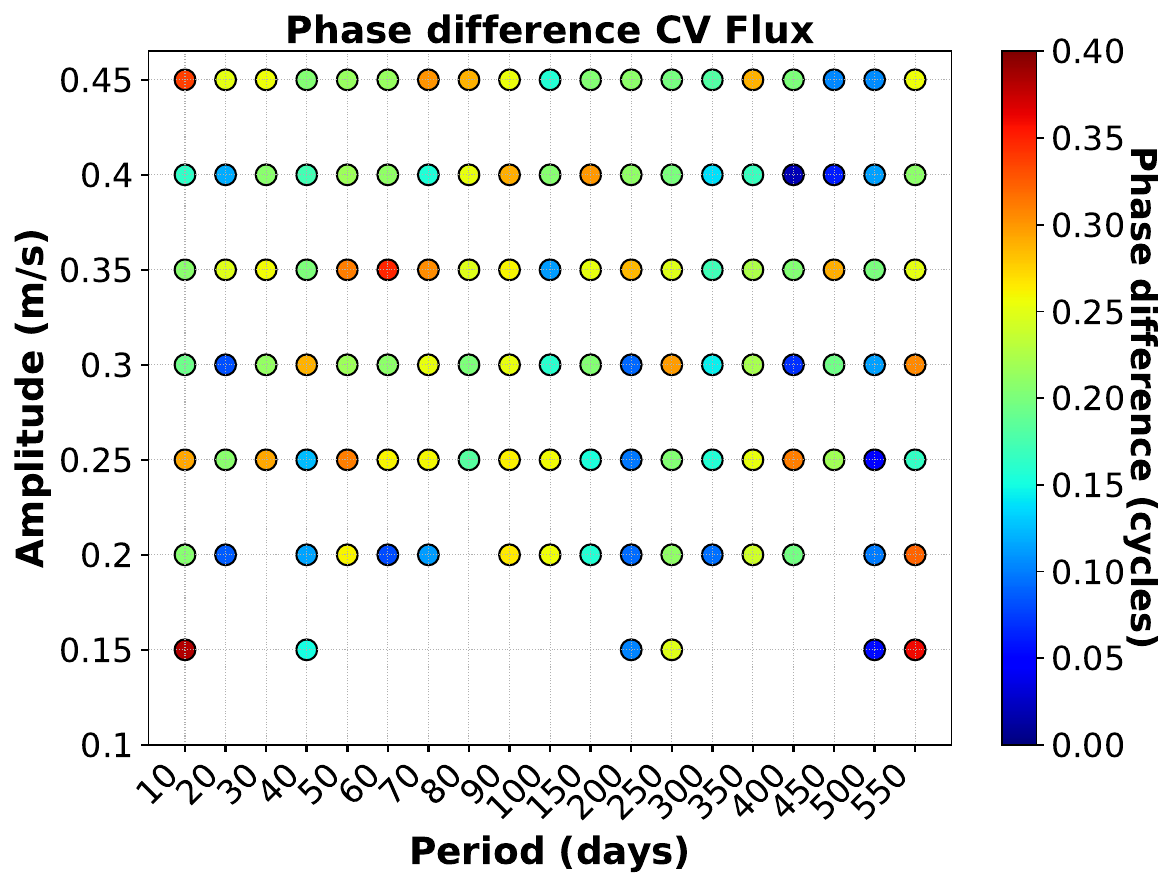}
        \includegraphics[trim=23mm 0mm 0mm 0mm, clip, width=5.2cm, height=4.8cm]{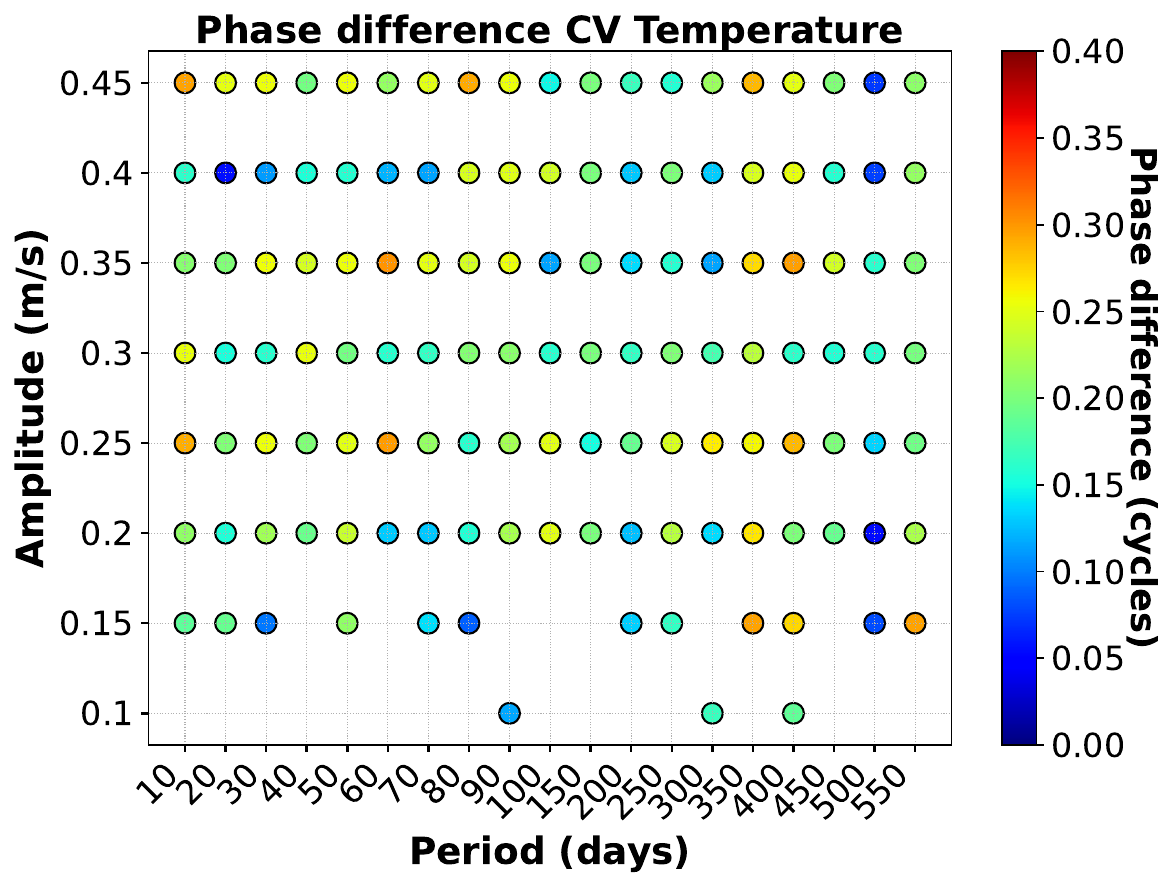}
    }

    \vspace{1em}

    \makebox[17cm][c]{
        \includegraphics[trim=0mm 0mm 27mm 0mm, clip, width=5.0cm, height=4.8cm]{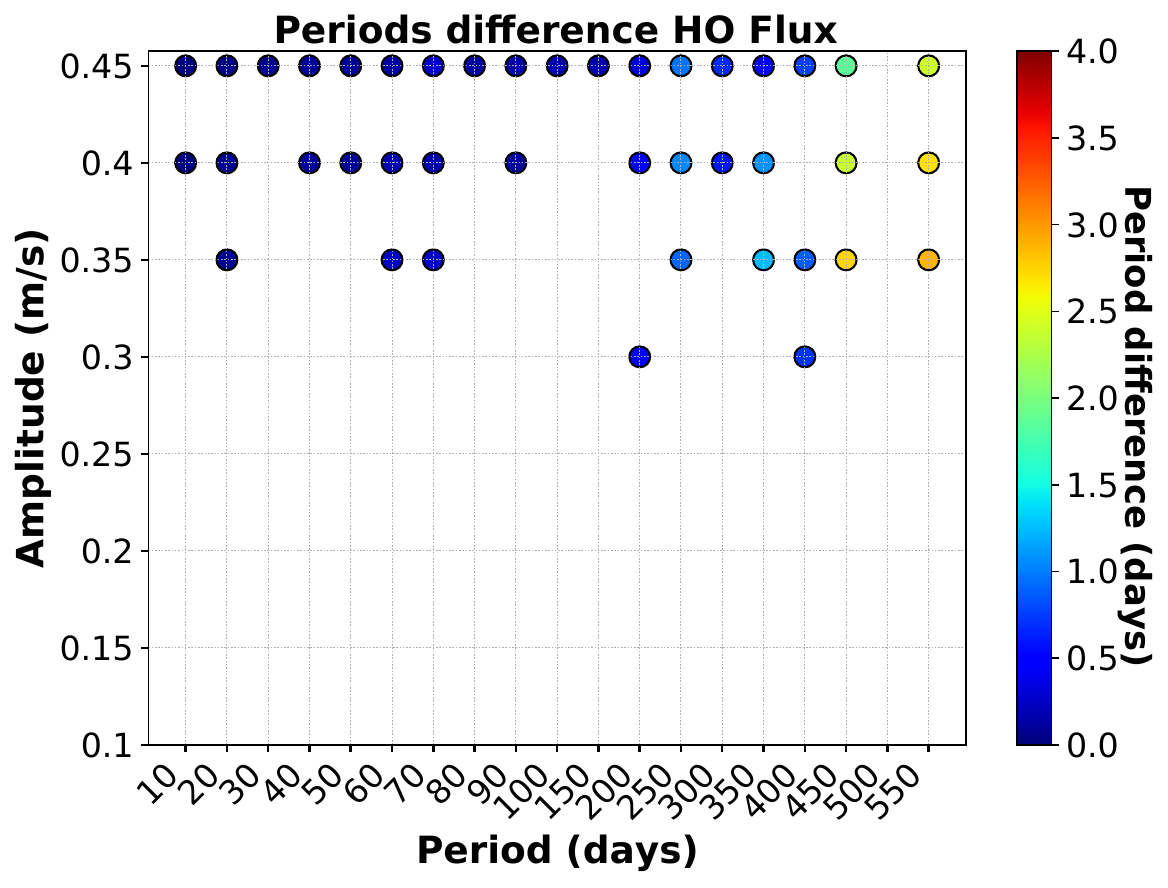}
        \includegraphics[trim=23mm 0mm 0mm 0mm, clip, width=5.2cm, height=4.8cm]{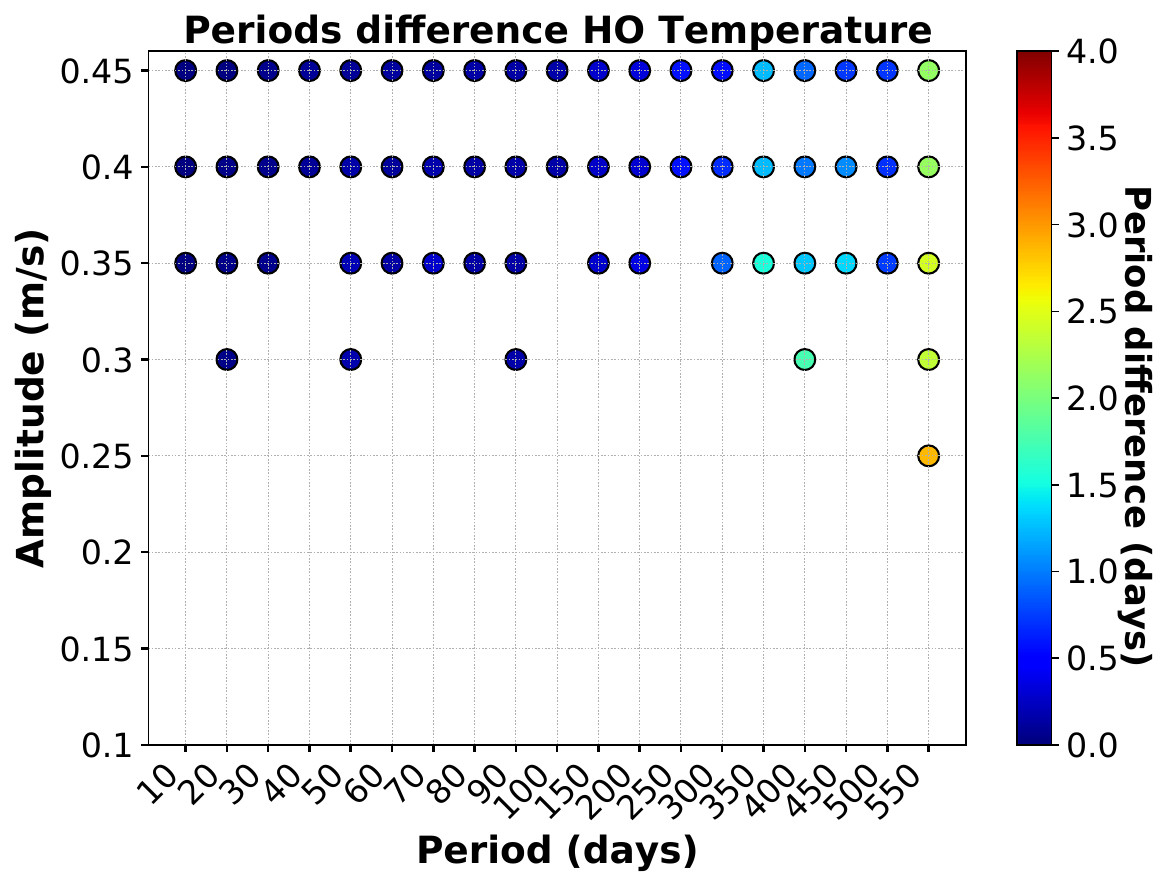}
        \includegraphics[trim=0mm 0mm 27mm 0mm, clip, width=5.0cm, height=4.8cm]{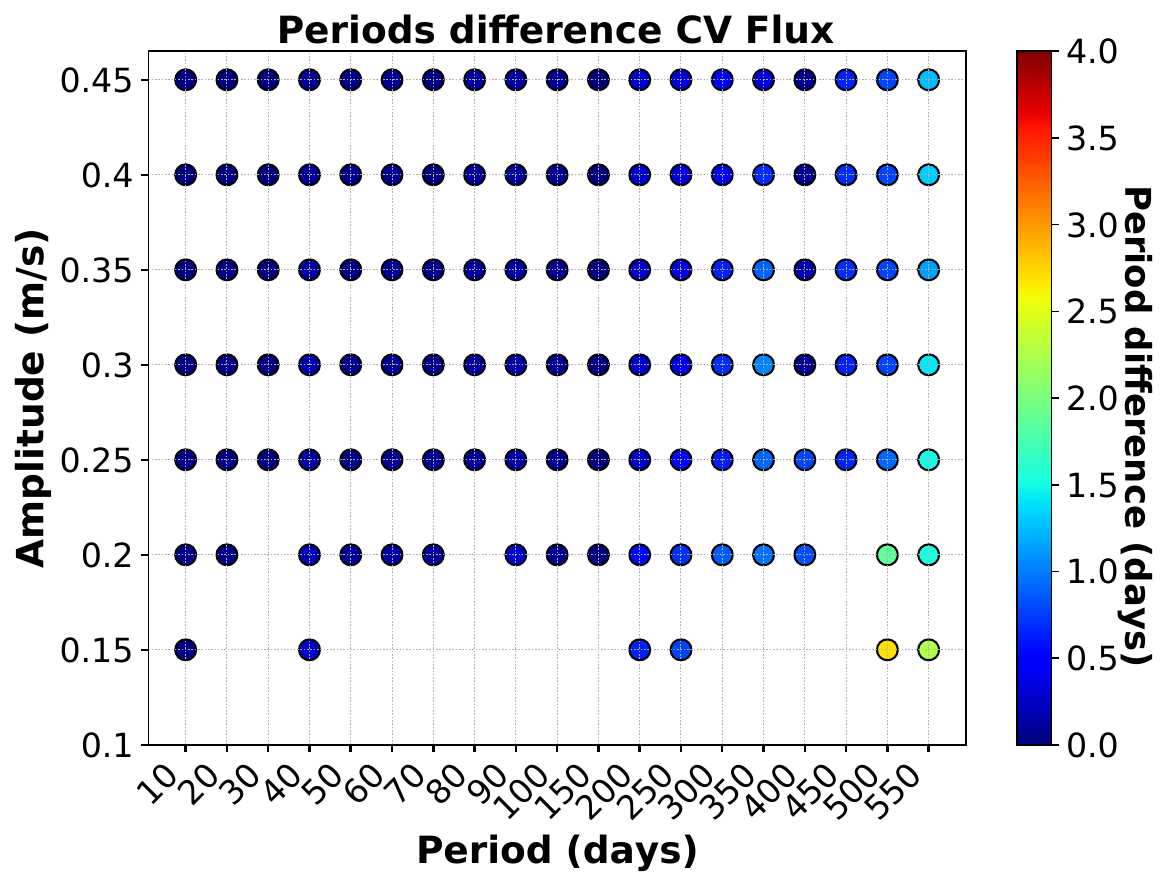}
        \includegraphics[trim=23mm 0mm 0mm 0mm, clip, width=5.2cm, height=4.8cm]{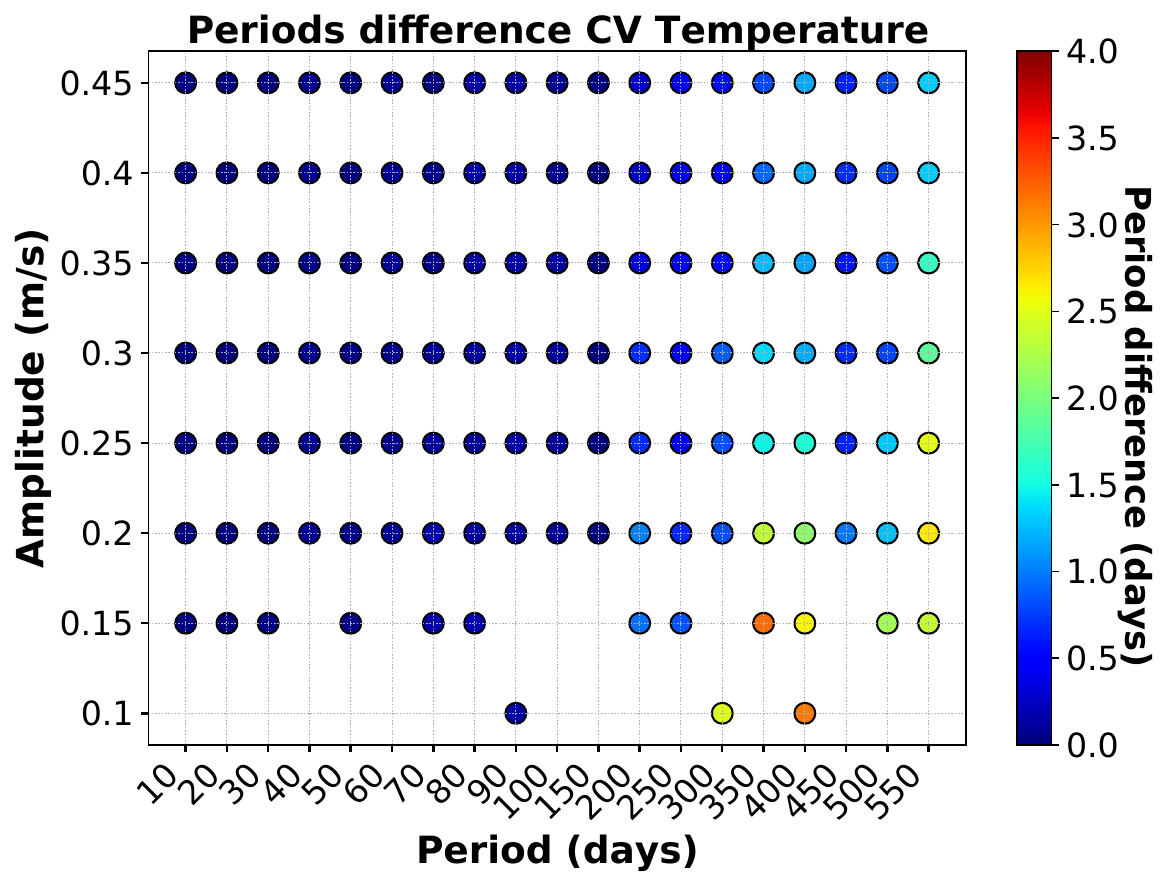}
    }
    \caption{Detection maps grouped by metric (rows) and model configuration (columns). From top to bottom: detection probability, relative amplitude difference, phase difference, and \changes{absolute period difference}. From left to right: Flux+HO, Temp+HO, Flux+CV, and Temp+CV. \changes{All quantities, except detection probability, correspond to mean values over the successful realizations.}}
    \label{fig:compact_detection_maps}
\end{figure*}

\subsection{Code availability: \texttt{doppleriann}}
\label{sec:doppleriann}

As part of this work, we developed \texttt{doppleriann}\footnote{\url{https://github.com/igomezv/doppleriann}.}, a Python package that implements the spectral-shell representations, computes CCF and extracts radial velocity and CCF-based activity indicators, performs planetary signal injections, applies mask filtering, carries out detectability tests, and integrates the deep-learning framework introduced in this study using the \texttt{TensorFlow} library. The package provides tools for constructing flux-based and temperature-based shell representations, generating synthetic planetary injections, and training convolutional neural networks to predict radial velocities and DS. \newchanges{In addition, the package includes the scripts used for the hyperparameter optimization, training, and evaluation of the neural-network models presented in this work.}
 
\texttt{doppleriann} is designed to be modular and extensible, enabling users to adapt the framework to other spectrographs and stellar targets. It includes pre-processing routines for HARPS-N solar spectra, utilities for line selection and temperature assignment, and ready-to-use CNN models consistent with the architectures described in this paper. 

The library is publicly available under an open-source license. Documentation and examples are provided to ensure reproducibility and to encourage the application of this methodology to other observational datasets.

\section{Results}
\label{sec:results}

Using the hyperparameters summarized in Table \ref{tab:hyperparams}, we define neural-network architectures for both the flux-based and temperature-based shell representations. Each model is trained and tested under two strategies: hold-out (HO) testing and cross-validation (CV). This results in four distinct model configurations, which we refer to as Flux+HO, Flux+CV, Temp+HO, and Temp+CV, respectively. 

For both approaches, each configuration is evaluated over 10 independent realizations on $\mathcal{J}_{\mathrm{all}}$. A detection is considered positive if, in at least 70\% of the trials (out of 10 realizations), the Lomb--Scargle power within $\pm 5\%$ of the injected period exceeds the 0.1\% FAP threshold. The top row of Fig.~\ref{fig:compact_detection_maps} shows the corresponding detection probability. \changes{For the remaining metrics, we report the mean values over the successful realizations only, including the mean absolute fractional amplitude difference (in percent), the mean circular phase difference (in cycles), and the mean absolute period difference.} Figure~\ref{fig:compact_detection_maps} summarizes these quantities across the explored parameter space, and each metric is discussed in detail below.

As supplementary material, Appendix~\ref{sec:app_methodological} presents additional validation tests of the methodology, including a sanity check based on a zero semi-amplitude injection (Section~\ref{sec:app_individual}), a comparison between 5-fold and 10-fold cross-validation strategies (Section~\ref{sec:app_10folds}), \changes{an analysis of whether the injected signal is recovered as the dominant peak in the periodogram (Section~\ref{sec:app_highest}), and a test of the impact of temporal correlations using a chronological hold-out split (Section~\ref{sec:app_time}),} and Section~\ref{sec:app_plantary20cms} describes specific examples of planetary recovery with small Doppler shifts injected at different periods. Appendix \ref{sec:app_uncertainty} provides additional analysis of the predictive uncertainties of the neural-network models, together with their root mean squared errors and loss function, offering further support for the robustness of our results.

\subsection{Detection probability}

The detection probability panels (top row of Fig. \ref{fig:compact_detection_maps}) show that both HO and CV models successfully recover planetary signals with semi-amplitudes \changes{above 40~cm/s } 
across the full range of orbital periods, from 10 to 550 days. At smaller amplitudes, however, differences between the configurations become more evident. 

\begin{itemize}
    \item \textbf{Hold-out models}. The Flux+HO configuration begins to detect injected signals at RV semi-amplitudes of \changes{40~cm/s, whereas this limit is about 5~cm/s better for the 
    Temp+HO configuration. Overall, the temperature-based approach exhibits consistently higher detection rates.}
    \item \textbf{Cross-validation models}. \changes{Overall, CV configurations allows to detect much smaller planetary signals, with amplitude as low as 20~cm/s confidently recovered. Like in the HO case, using temperature shells as input allows to detect signals with amplitude about 5~cm/s smaller.}

\end{itemize}

Cross-validation models exhibit higher raw detection sensitivity than their hold-out counterparts. This is partly a consequence of the evaluation protocol: in the CV approach, planet recovery is assessed over the full time series of 2036 predicted Doppler shifts, whereas in the HO case, the evaluation is performed on a fixed set of 400 unseen spectra. As a result, the CV models are able to detect weaker planetary signals, extending sensitivity to lower RV semi-amplitudes. Both evaluation strategies consistently show that temperature-based spectral-shell representations outperform flux-based models in terms of detection sensitivity and recovery rates.

\subsection{Amplitude recovery}

Amplitude recovery probes the ability of the models to reconstruct the strength of the injected planetary signal and to disentangle it from stellar-induced variability. \changes{The amplitude difference is defined as the absolute fractional difference between the recovered and injected Doppler-shift semi-amplitudes, $100\,|DS_{\mathrm{rec}} - DS_{\mathrm{inj}}|/DS_{\mathrm{inj}}$, averaged over the successful realizations for each injected configuration.} The amplitude recovery maps (second row of Fig.~\ref{fig:compact_detection_maps}) highlight systematic biases between flux- and temperature-based models.

\begin{itemize}
    \item \textbf{Hold-out models.} Both flux- and temperature-based models achieve satisfactory relative amplitude differences, remaining below 30\% over most of the explored parameter space. For larger injected RV semi-amplitudes, the amplitude differences decrease further in both cases.
    
    \item \textbf{Cross-validation models.} Using the flux-based representation, relative amplitude differences remain below 30\% for most injections, except for three cases at RV semi-amplitudes of 15~cm/s and 20~cm/s occurring at long orbital periods (500 and 550 days). In contrast, the temperature-based representation achieves more accurate signal recovery, with relative amplitude differences also remaining below 30\% across most of the parameter space. The main exceptions are the 10~cm/s injections and two cases at 20~cm/s and 25~cm/s at an orbital period of 550 days. These results further confirm the improved robustness of temperature-informed spectral shells, which more effectively capture depth-dependent stellar activity signatures.

\end{itemize}

In general, the cross-validation models achieve more reliable amplitude recovery than the hold-out counterparts, with temperature-based representations providing the most stable and accurate reconstruction of the injected Doppler-shifts.

\subsection{Phase recovery}

Phase errors are quantified as the minimum circular distance between the recovered and injected phase offsets, expressed in units of orbital cycles (i.e., fractions of a full period). The phase difference maps (third row of Fig.~\ref{fig:compact_detection_maps}) indicate that all models reproduce orbital phases with generally good fidelity. \changes{The values shown correspond to the mean phase difference, averaged over the successful realizations for each configuration.}

\begin{itemize}
    \item \textbf{Hold-out models.}  
    Flux- and temperature-based representations achieve comparable performance, with typical phase differences remaining below $\sim 0.35$ cycles, indicating that the recovered phase is usually accurate to better than one third of an orbital period.
    
    \item \textbf{Cross-validation models.}  
    Phase differences remain small across most of the explored parameter space, with values typically of order $\sim 0.3$ cycles and systematically lower errors for temperature-based representations, reflecting improved phase stability under cross-validation.
\end{itemize}

Overall, these results demonstrate that the proposed models preserve the phase information of injected planetary signals with high robustness, even under more conservative cross-validation schemes, and that temperature-based inputs provide a modest but consistent advantage in phase recovery accuracy.

\subsection{Period recovery}

 Period errors are expressed \changes{as absolute differences between the recovered and injected orbital periods, reported in days.} The period recovery maps (bottom row of Fig.~\ref{fig:compact_detection_maps}) show consistently high accuracy across all models.

\begin{itemize}
    \item \textbf{Hold-out models.}  
    Both flux- and temperature-based representations achieve relative period differences below $\sim 3\%$ over the explored parameter space. The largest discrepancies occur at long orbital periods, particularly at 500 and 550 days, where period recovery becomes more challenging.

    \item \textbf{Cross-validation models.}  
    Similarly, flux- and temperature-based representations maintain relative period differences below $\sim 3\%$ in most cases. The largest deviations are observed at the longest orbital period (550 days) and for the weakest injected signals, with RV semi-amplitudes of 10--15~cm/s, reflecting the increased difficulty of recovering long-period, low-amplitude signals.
\end{itemize}

Period recovery remains strong for both training strategies, with comparable levels of accuracy in flux and temperature representations and slightly worse performance at longer periods and low signal amplitudes.

\subsection{Discussion}

\changes{Our best neural-network model, based on cross-validation and temperature-based shell, is capable of reliably recover amplitudes, phases, and orbital periods for planetary signals with amplitude as small as 0.20~m/s and periods ranging from 10 to 550 days. Those result are based on the analysis of 2036 HARPS-N solar spectra and controlled planetary injections. Those detection limits are obtained by analyzing the periodogram of the DS output of our neural network and detecting signals at $\pm$ 5\% of the injected periods with a FAP smaller than 0.1\%. As this criterion does not ensure that the injected planetary signal corresponds to the dominant peak in the DS periodogram, we tested in Appendix~\ref{sec:app_highest} this stricter condition. In such a case, the detection limit for the cross-validation and temperature-based shell model only rises by 0.05~m/s to reach 0.25~m/s.}

The hold-out strategy evaluates performance on a fixed set of 400 unseen spectra and therefore provides a conservative assessment of generalization. Despite the reduced evaluation sample, hold-out models robustly recover signals above \changes{$\sim 0.35$~m/s} with accurate amplitude, phase, and period estimates. In contrast, cross-validation models exhibit higher raw detection sensitivity, primarily due to the evaluation protocol, which leverages the full time series of 2036 predicted Doppler shifts. \changes{More measurements, coupled with an extended temporal baseline, }enable the detection of weaker planetary signals at lower RV semi-amplitudes. At the same time, cross-validation models achieve improved recovery of amplitude, phase, and period across most of the explored parameter space, particularly when using temperature-based representations.
The remaining failures are largely confined to long-period and low-amplitude injections, where signal recovery is inherently more challenging.

\begin{figure*}
    \centering
    \includegraphics[trim=15mm 35mm 15mm 45mm, clip, width=\textwidth]{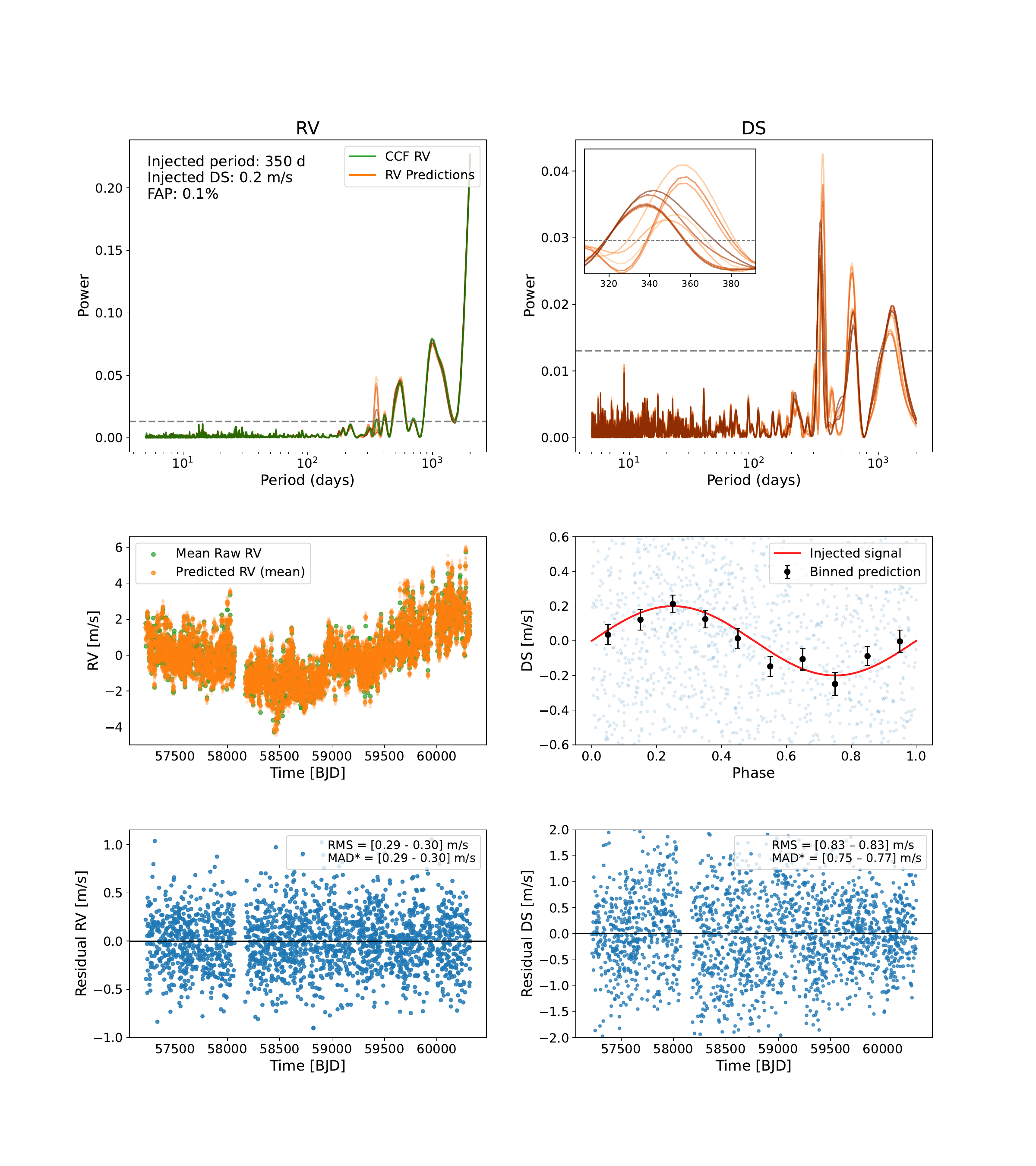}
    \caption{Example of a 20~cm/s planetary signal recovery using the cross-validation temperature-based model (CV-Temp).
The \textit{top panels} show periodograms of the mean raw CCF RVs (left) and the predicted Doppler shifts (DS; right), with the dashed line indicating the 0.1\% FAP threshold. The injected period at 350~days is clearly detected in the DS periodogram, while it remains obscured in the RVs. The \textit{middle panels} display the time series of the mean raw RVs (left) and the phase-folded DS predictions (right), \changes{where the black points represent the mean value within phase bins and the error bars correspond to the standard error of the mean within each bin. For clarity, the y-axis range is restricted in the phase-folded panel; the full dispersion of the predictions is shown in the residual panel.} The \textit{bottom panels} show the corresponding residuals, illustrating that although individual DS predictions exhibit substantial scatter, \changes{the residuals remain centered around zero without a clear systematic structure.}}
    \label{fig:20cmCV}
\end{figure*}

The superior performance of the neural network model trained with temperature-based shells is consistent with their sensitivity to depth-dependent stellar activity effects. Stellar line-profile variations are known to depend on formation depth, and encoding formation temperature therefore provides information that is not captured by flux alone. In our experiments, this results in systematically higher detection rates and more stable recovery of amplitudes, phases, and periods when using temperature-based representations. The benefits of this approach are illustrated in Fig.~\ref{fig:20cmCV}, which shows a representative recovery of a low-amplitude planetary signal using the CV-Temp model. Despite an injected Doppler-shift semi-amplitude of only 20~cm/s and a long orbital period of 350~days, the DS periodogram exhibits its \changes{strongest peak at the injected period. This peak is extremely significant with a FAP well below the 0.1\% threshold}. In contrast, the corresponding RV periodogram is dominated by long-period power and stellar variability, with no evident detection at the injected period. This highlights the advantage of temperature-informed representation in suppressing stellar-induced signals and enhancing planetary detectability. On the other hand, the phase-folded DS time series \changes{shows a coherent modulation}, with the binned predictions following the injected sinusoidal trend despite significant scatter. Although the residual DS variability remains large, the planetary signal becomes apparent when the data are phase-folded.

\changes{Our results complement recent deep-learning approaches applied to solar spectra. Relative to \cite{zhao2024deep}, we extend the physically motivated spectral representation to include line-formation temperature in addition to flux, increasing the effective training dataset while predicting Doppler shifts directly. In contrast, \cite{colwell2024deep} focuses on short-period signals, whereas our framework explores a broader range of orbital periods and emphasizes generalization to unseen data using lightweight models, achieving comparable precision over a wider parameter space.} These results highlight the potential of combining physically informed representations with deep learning to improve the detection of low-amplitude planetary signals in high-precision RV surveys.

\section{Conclusions}
\label{sec:conclusions}

In this work, we presented a deep-learning framework for modeling Doppler shifts in stellar spectra using physics-motivated spectral-shell representations. By combining data-driven neural networks with advanced hyperparameter optimization, uncertainty quantification, and regularization techniques, and by exploiting both flux and line-formation temperature information, we \changes{explored a promising approach for separating planetary Doppler signals from stellar variability} in real solar observations.

Using HARPS-N solar spectra with injected planetary signals, we showed that convolutional neural networks trained on shell representations \changes{recover injected Doppler shifts, amplitudes, phases, and orbital periods under controlled injection-recovery experiments}. Temperature-based shell representations consistently outperform flux-based shells across all evaluation metrics, highlighting the benefit of encoding depth-dependent stellar information to mitigate activity-induced variability. Under the hold-out strategy and temperature-based shell, planetary signals with Doppler-shift semi-amplitudes above \changes{0.35~m/s} are \changes{consistently recovered in our experiments}, while cross-validation models demonstrate the confident detection of signals with semi-amplitudes as low as \changes{0.20~m/s.} \changes{This detection limit is obtained by checking that in the DS output of the neural network, a peak at $\pm$ 5\% of the injected planetary period with a FAP smaller than 0.1\% exists in the corresponding periodogram. Using a stricter criterion that the planetary signal corresponds to the highest periodogram peak, the detection limit for all periods ranging from 10 to 550 days rises to 0.25~m/s. Such a detection limit thus ensures that the strongest peaks in the DS periodogram correspond to a planetary signal and not a non-related systematics.}

\newchanges{We note that the lowest detection limits are obtained with the cross-validation strategy and may still be influenced by residual temporal correlations between training and evaluation samples. A more conservative estimate is provided by the chronological hold-out experiment (Appendix~\ref{sec:app_time}), which preserves temporal separation between training and test data. The training procedure uses 35 synthetic planetary injections, yielding approximately 57,000 training realizations, all derived from the same underlying solar observations. Therefore, any remaining risk of overfitting is more likely associated with stellar or instrumental patterns than with the injected planetary signals. However, the injection-recovery experiments are limited to circular single-planet systems, and future work should investigate more complex planetary configurations.} Nevertheless, a key advantage of the proposed framework is that it operates directly on real stellar data without requiring explicit stellar-activity modeling. The improved performance achieved by temperature-informed representations indicates that physically motivated data compression can substantially enhance the sensitivity of neural-network approaches to low-amplitude planetary signals.

Future work will extend this framework to other stellar types and instruments, explore transfer-learning strategies to enable efficient adaptation across heterogeneous stellar targets and observational conditions. \changes{We note that for other stars than the Sun, telluric contamination will be more important as the excursion in barycentric correction for the Sun is minimal, thus implying smaller than one pixel shift on the detector between stellar and telluric features. Although YARARA enables to strongly mitigate telluric systematics \citep[][]{cretignier2021yarara}, it is also possible to build spectral shell from selected spectral regions, and thus reject regions strongly contaminated by tellurics.} Overall, these results \changes{highlight the potential of combining deep learning with physically informed spectral representations to improve the sensitivity of radial-velocity analyses, an important step toward detecting Earth-mass planets}.

\begin{acknowledgements}
IGV and X.D acknowledge the support from the the Swiss National Science Foundation under the grant SPECTRE (No 200021\_215200). KA acknowledges support from the Swiss National Science Foundation (SNSF) under the Postdoc Mobility grant P500PT\_230225. This work has been carried out within the framework of the NCCR PlanetS supported by the Swiss National Science Foundation under grants 51NF40\_182901 and 51NF40\_205606. IGV acknowledges current support from the Severo Ochoa Centre of Excellence grant CEX2021-001131-S (MCIN/AEI/10.13039/501100011033), project PID2023-149439NB-C42 from the 'Proyectos de Generacion de Conocimiento', and the MSCA-COFUND ALLIES-COFUND programme under Horizon Europe (Grant Agreement No. 101126626). This work made significant use of the High-Performance Computing Center of the University of Geneva.

\end{acknowledgements}

\bibliographystyle{aa}
\bibliography{bibliography}

\appendix
\section{Methodological Considerations}
\label{sec:app_methodological}

This appendix addresses three methodological aspects that deserve further discussion. First, we assess the possibility that the neural-network models could memorize or generate spurious signals in the absence of an injected Doppler-shift signal by means of a zero-signal injection test. Second, we examine whether increasing the number of folds in the cross-validation procedure from five to ten leads to a significant improvement in model performance. \changes{In addition, we analyze whether the injected planetary signal is recovered as the dominant peak in the periodogram for our best-performing model. We also investigate the potential impact of temporal correlations in the data by performing a chronological hold-out test, in which the model is trained on the first 1636 spectra and evaluated on the most recent 400 samples.} Finally, we discuss the behavior of the predictions for an injected signal of 20 cm/s in different orbital periods. These three additional analyses provide additional validation of the robustness and reliability of the proposed methodology.

\subsection{The zero signal injection}
\label{sec:app_individual}

To verify that the pipeline does not produce spurious detections in the absence of a planetary signal, we perform a zero-injection test on the CV-Temp configuration, in which the injected Doppler semi-amplitude is set to $DS = 0.0$~m/s. This test provides a baseline validation of the model behavior and ensures that detected signals arise from genuine injected variability rather than from artifacts of the neural network or the analysis procedure.
 
\begin{figure}[h!]
    \centering
    \includegraphics[width=\linewidth]{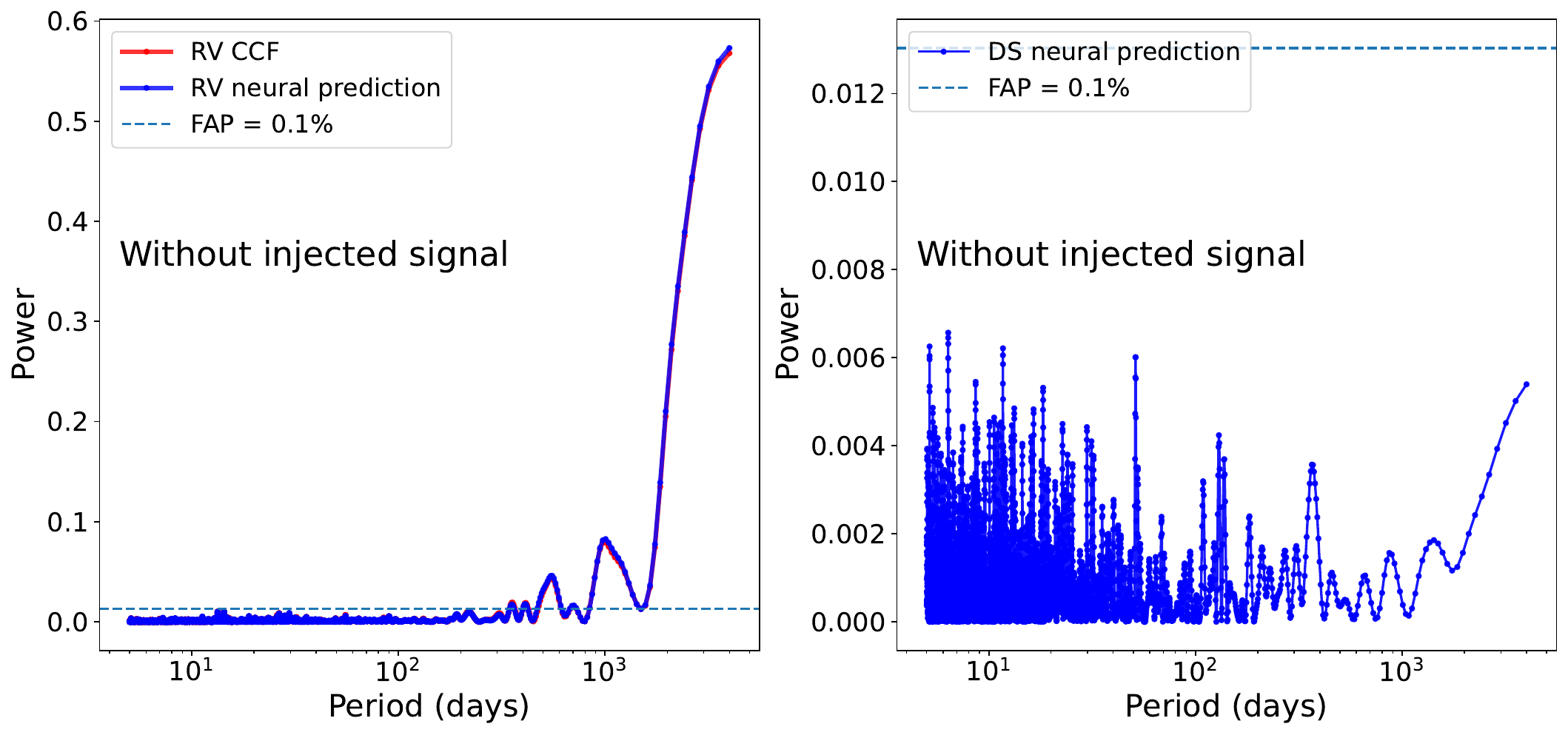}
    \caption{Case without injected planetary signal (DS = 0.0\,m/s). \textit{Left:} Periodogram of the RV from the CCF (red) and neural prediction (blue), both showing only long-term variability without significant periodic power. \textit{Right:} Periodogram of the DS prediction, confirming the absence of false planetary detections.}
    \label{fig:appendixA0}
\end{figure}

Figure~\ref{fig:appendixA0} shows the corresponding periodograms for this case. The RV periodogram computed from the CCF and the neural-network RV prediction both exhibit only long-term variability, with no statistically significant periodic power. Similarly, the DS periodogram displays no peaks exceeding the 0.1\% false-alarm probability threshold. The absence of significant peaks across all representations confirms that the pipeline does not introduce false positives and behaves consistently when no planetary signal is present.
\newchanges{We emphasize that this experiment should be interpreted as a consistency check of the pipeline rather than as a statistical estimate of the false-alarm rate, since the zero-injection case does not generate multiple independent realizations of the underlying dataset.}

\subsection{Effect of the number of folds in the cross-validation approach}
\label{sec:app_10folds}

Given that the HARPS-N solar dataset used in this work contains 2036 spectra, we consider a test sample of 400 spectra, corresponding to approximately 25\% of the dataset, to be sufficiently large for a statistically meaningful evaluation. Accordingly, in the hold-out approach, we select this number of spectra for testing, while the remaining data are used for model development. To ensure consistency between the HO and cross-validation strategies, we adopt $N_{\mathrm{folds}} = 5$ in the CV procedure, such that each evaluation fold contains approximately $2036/5 \simeq 400$ spectra, making a single CV fold directly comparable to the HO test set.

\begin{figure*}
    \centering
     \makebox[17cm][c]{
        \includegraphics[trim=0mm 0mm 30mm 0mm, clip, width=5.0cm, height=4.8cm]{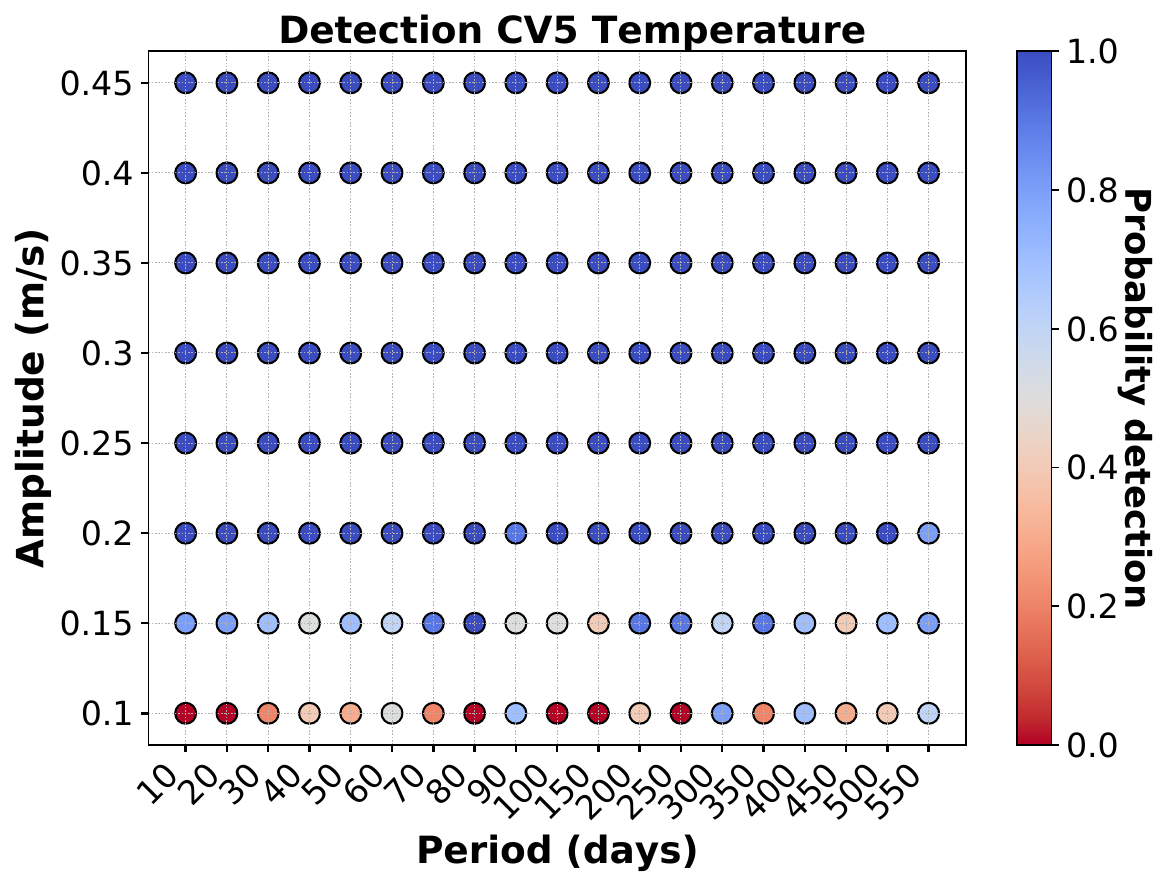}
        \includegraphics[trim=23mm 0mm 0mm 0mm, clip, width=5.2cm, height=4.8cm]{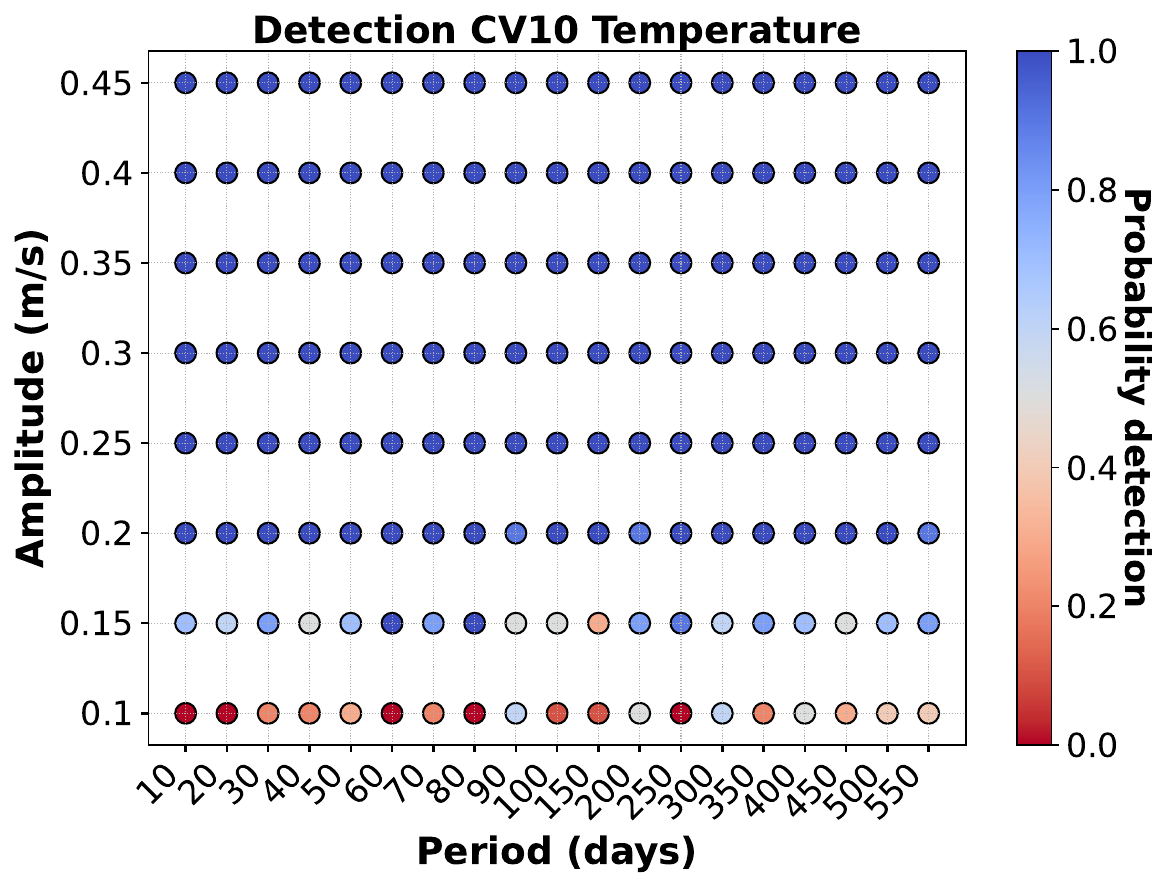}
        \includegraphics[trim=0mm 0mm 30mm 0mm, clip, width=5.0cm, height=4.8cm]{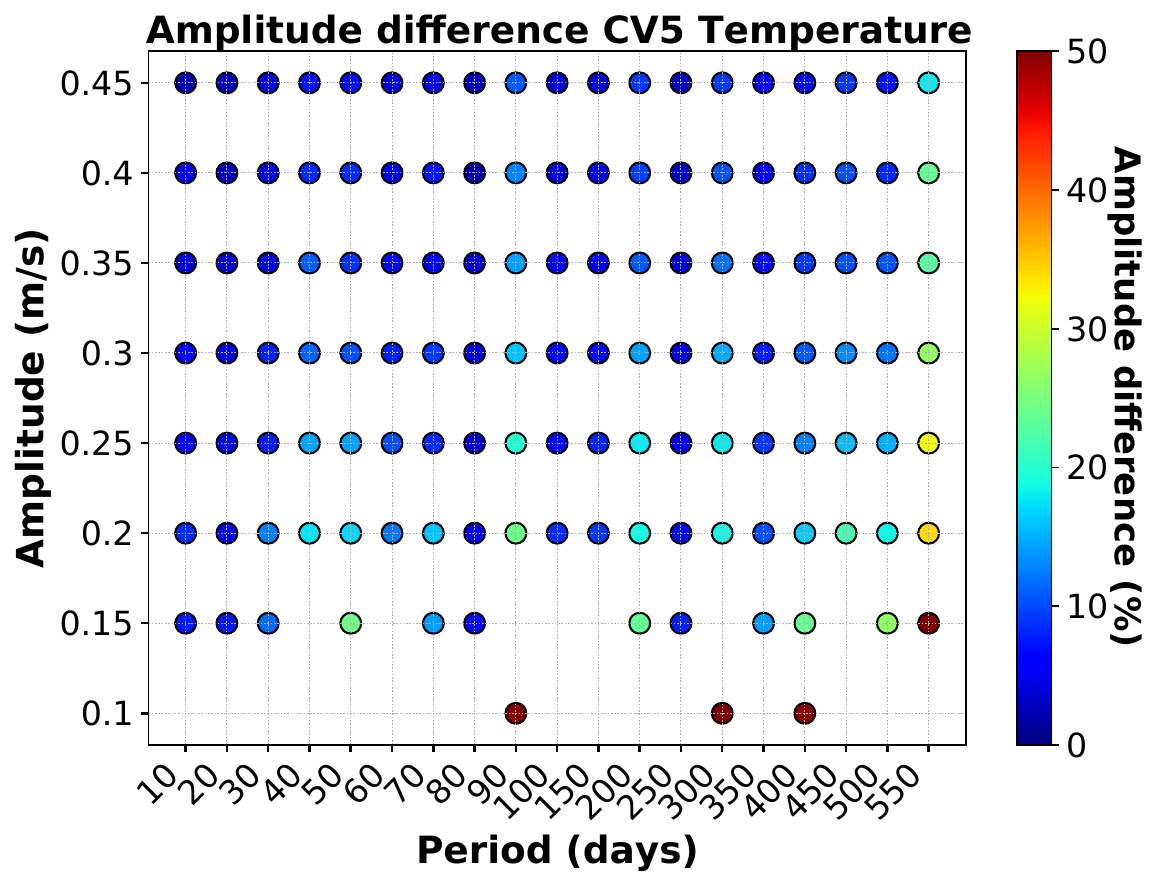}
        \includegraphics[trim=23mm 0mm 0mm 0mm, clip, width=5.2cm, height=4.8cm]{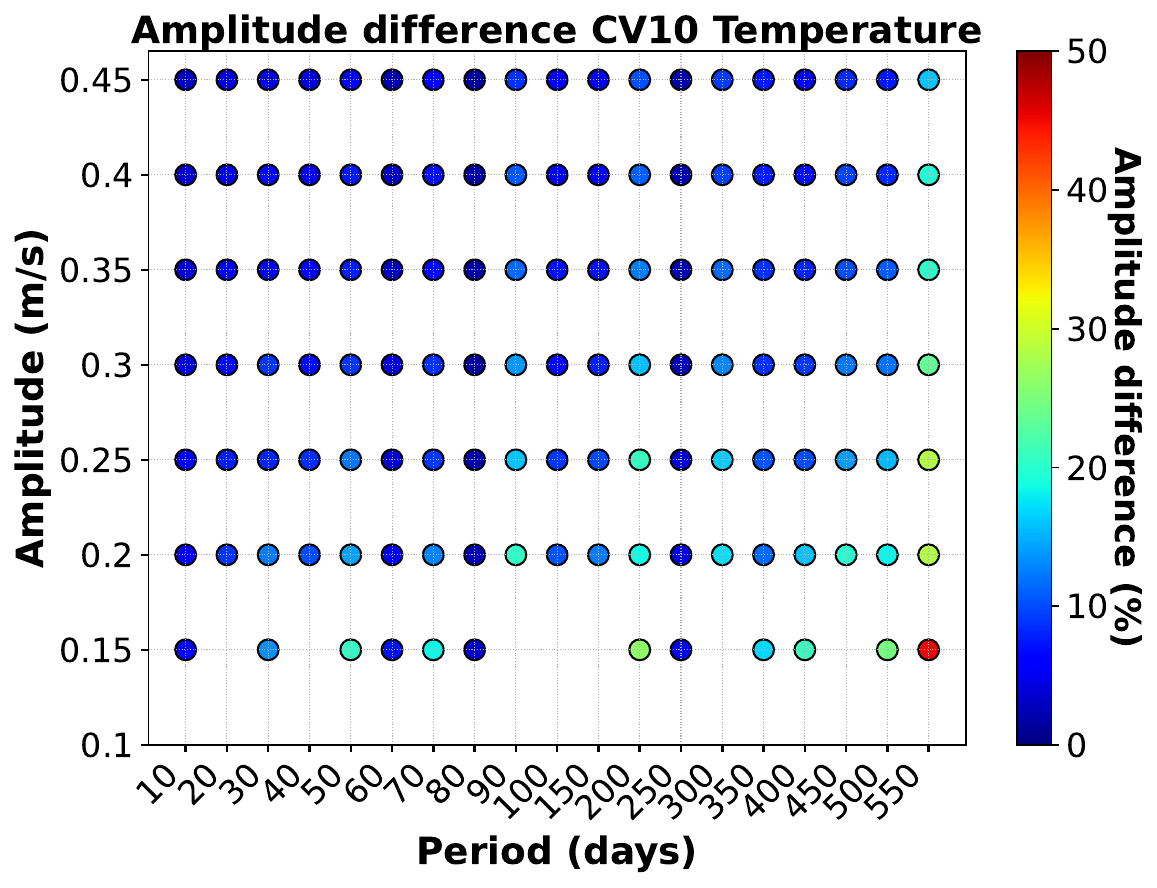}
    }

    \caption{Comparison between 5-fold and 10-fold cross-validation results for the temperature-shell CNN model. Left panels show the detection probability as a function of injected orbital period and RV semi-amplitude for $N_{\mathrm{folds}}=5$ (CV) and $N_{\mathrm{folds}}=10$ (CV10), respectively.  The right panels show the corresponding relative amplitude differences between injected and recovered signals. Overall, increasing the number of folds from 5 to 10 does not lead to a substantial improvement in detection performance or amplitude recovery, particularly above $\sim$15~cm\,s$^{-1}$, while incurring a higher computational cost.}
    \label{fig:cv10}
\end{figure*}

In principle, increasing the number of folds in the CV procedure results in smaller evaluation subsets and larger training sets, which could intuitively lead to improved predictive accuracy. However, this comes at the expense of increased computational requirements, since the number of independently trained neural-network models scales linearly with $N_{\mathrm{folds}}$. For example, adopting $N_{\mathrm{folds}} = 10$ requires training twice as many models as in the 5-fold configuration. Such an increase would be justified only if it led to a substantial improvement in performance. As shown in Figure~\ref{fig:cv10}, however, the differences between the 5-fold and 10-fold configurations are minimal. For injected amplitudes above approximately 15~cm\,s$^{-1}$, the 10-fold configuration yields slightly improved amplitude recovery in a limited fraction of cases (of order 5\%); this improvement is neither systematic nor sufficient to justify the increase in computational cost.

\subsection{Dominant peak analysis}
\label{sec:app_highest}
\changes{Planetary recovery tests are typically performed by searching for signals around the injected period. In this work, we consider a signal to be recovered if it is detected within $\pm 5\%$ of the injected period and with a FAP smaller than 0.1\%. However, in real observational scenarios, the true period is unknown, and detections must rely on identifying significant peaks in the full periodogram.}

\begin{figure}[h!]
    \centering
    \includegraphics[trim=2mm 0mm 30mm 0mm, clip, width=4.5cm, height=4.5cm]{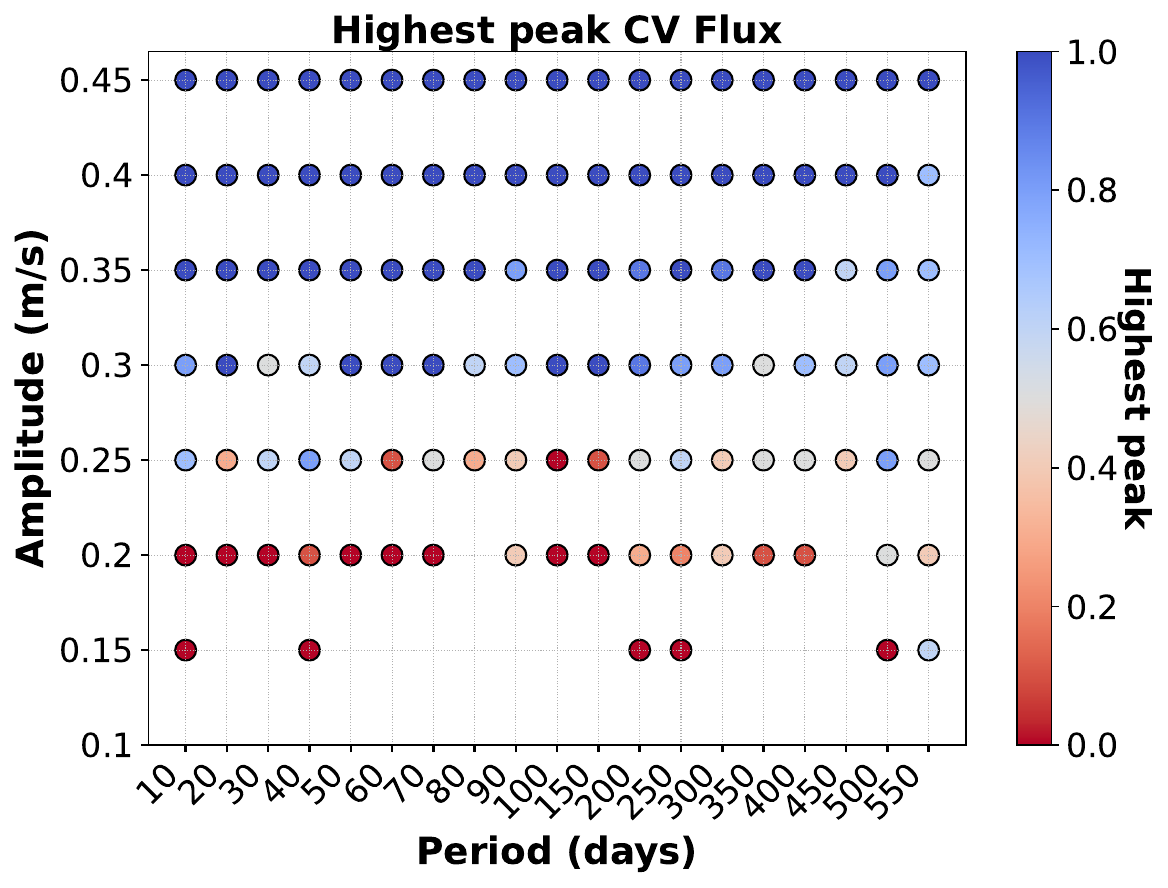}
    \includegraphics[trim=25mm 0mm 1mm 0mm, clip, width=4.2cm, height=4.5cm]{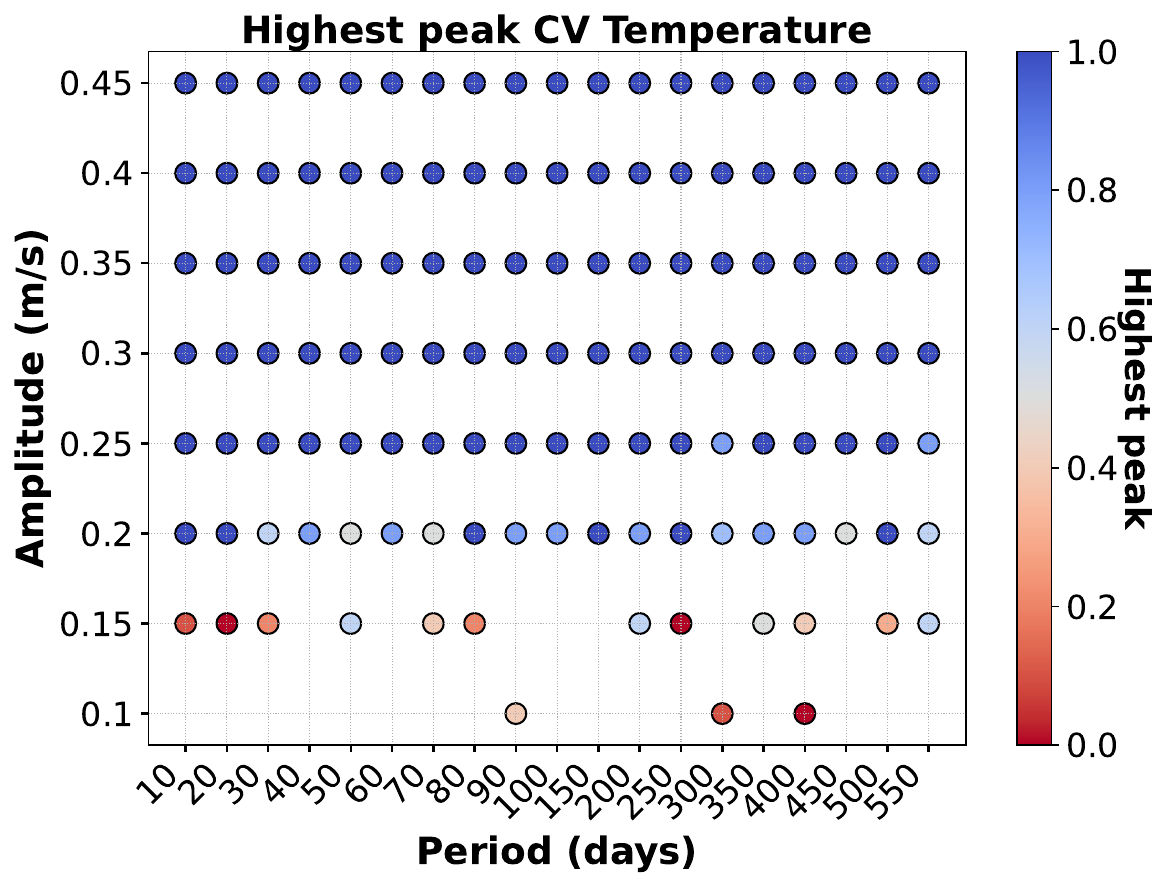}
    \caption{\changes{For the neural-network models using the cross-validation strategy, we assess whether the detected planetary signal corresponds to the dominant peak in the periodogram across 10 independent realizations with different random phase injections, for both flux- and temperature-based representations. The color bar indicates the fraction of realizations in which the injected signal is recovered as the dominant peak: a value of 1 (dark blue) corresponds to 10 of 10 successful recoveries, while 0 (dark red) indicates that the injected signal is not recovered as the dominant peak in any realization.}}
    \label{fig:highest_peak}
\end{figure}

\changes{To better reflect this situation, we analyze the cases in which our neural network models trained within the CV strategy identify the injected signal as the dominant peak in the periodogram across 10 independent realizations, following the same injection setup used in Fig.~\ref{fig:compact_detection_maps}. As shown in Fig.~\ref{fig:highest_peak},} \changes{using this stricter criterion only slightly worsen the detection limits. For the best model which is based on CV and temperature-based shells, the detection limits slightly goes up from 0.20 to 0.25~cm/s. For the CV and flux shell model, the detection limit rises slightly more, going from 0.25 to 0.35~cm/s. This test demonstrate, that when using the CV and temperature-based shell model, planetary signals with amplitude as small as 25~cm/s will correspond to the dominant peak in the analysis of the DS periodogram.}
\changes{We note that even when the injected signal does not correspond to the dominant peak, detections above the false-alarm probability threshold remain informative, as they can be further investigated as candidate signals using complementary analysis methods.}

\subsection{Chronological hold-out test}
\label{sec:app_time}

\changes{The training and validation sets adopted throughout this work are constructed using random splits, following standard machine-learning practices. However, because solar spectra samples obtained on nearby dates can exhibit temporal correlations induced by stellar activity, a random split may contain observations in the training set that are temporally close to the test samples. To evaluate whether the neural network framework discussed in this manuscript could be affected by such correlations, we performed additional chronological held-out stress tests inspired by the temporally grouped validation strategy presented in \citet{salzer2025searching}.}

\changes{In this chronological tests, the spectra samples are split chronologically rather than randomly, allowing that training and test sets are fully separated in time. Considering that the neural network generates predictions independently for each spectrum and does not explicitly use temporal information as an input feature, this test evaluate the ability of the model to generalize across different observing periods. They also reduce the possibility that temporally nearby spectra in the training set artificially improve the evaluation performance. To enable a direct comparison with the standard hold-out strategy presented in Section~\ref{sec:results}, we preserve the same number of training and evaluation samples, using 1636 spectra for training and 400 for testing.}

\begin{figure}[h!]
    \centering
    \includegraphics[trim=1mm 2mm 1mm 1mm, clip, width=0.5\textwidth]{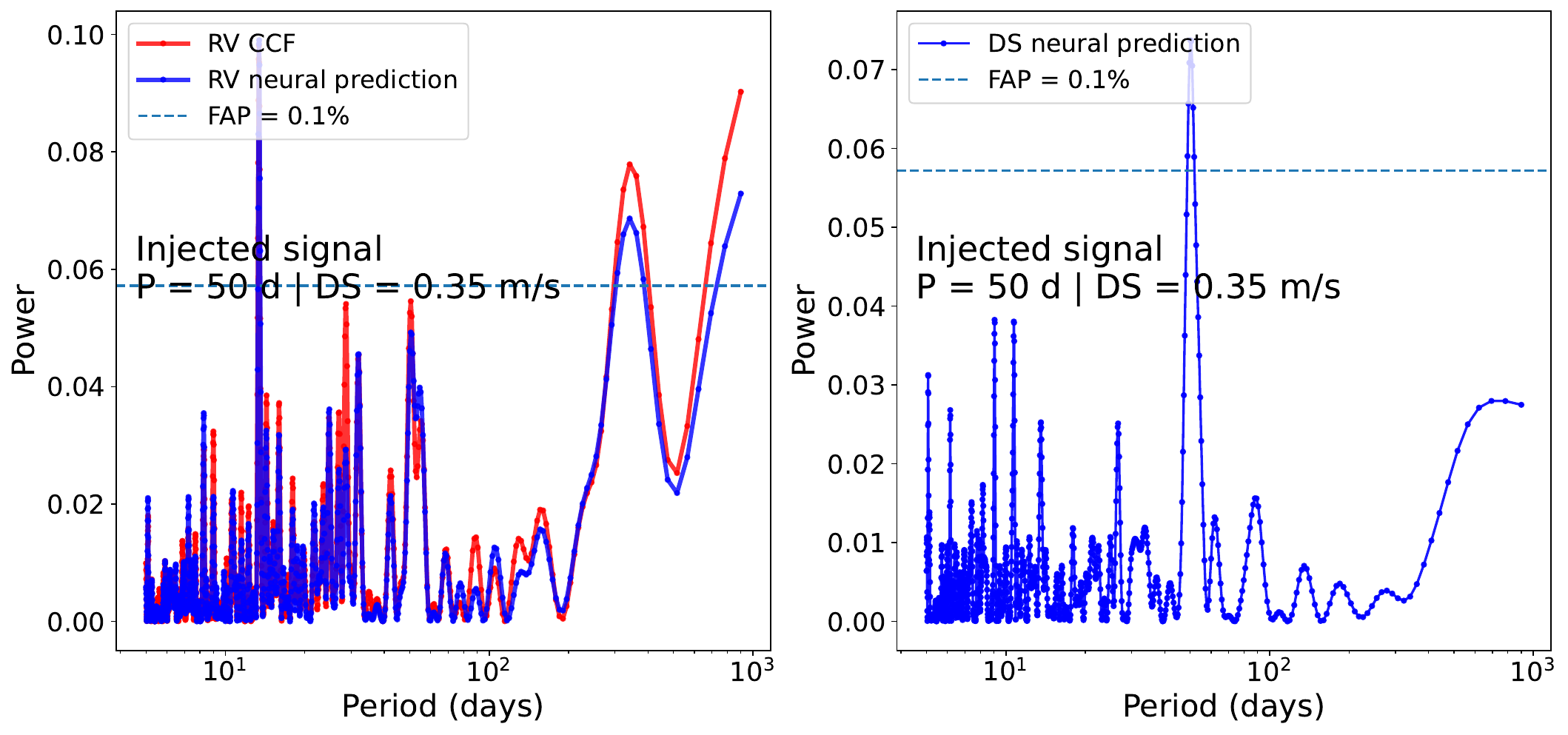}
    \caption{\changes{RV (left) and DS (right) periodograms generated using a neural network optimized under the chronological hold-out strategy, trained on the most recent 1636 spectra and tested on the oldest 400. The 0.35~m/s injected planet at 50 days is highly significant in the DS periodogram.}}
    \label{fig:first400samples}
\end{figure}

\begin{figure}[h!]
    \centering
    \includegraphics[trim=2mm 0mm 2mm 0mm, clip, width=4.8cm, height=4.5cm]{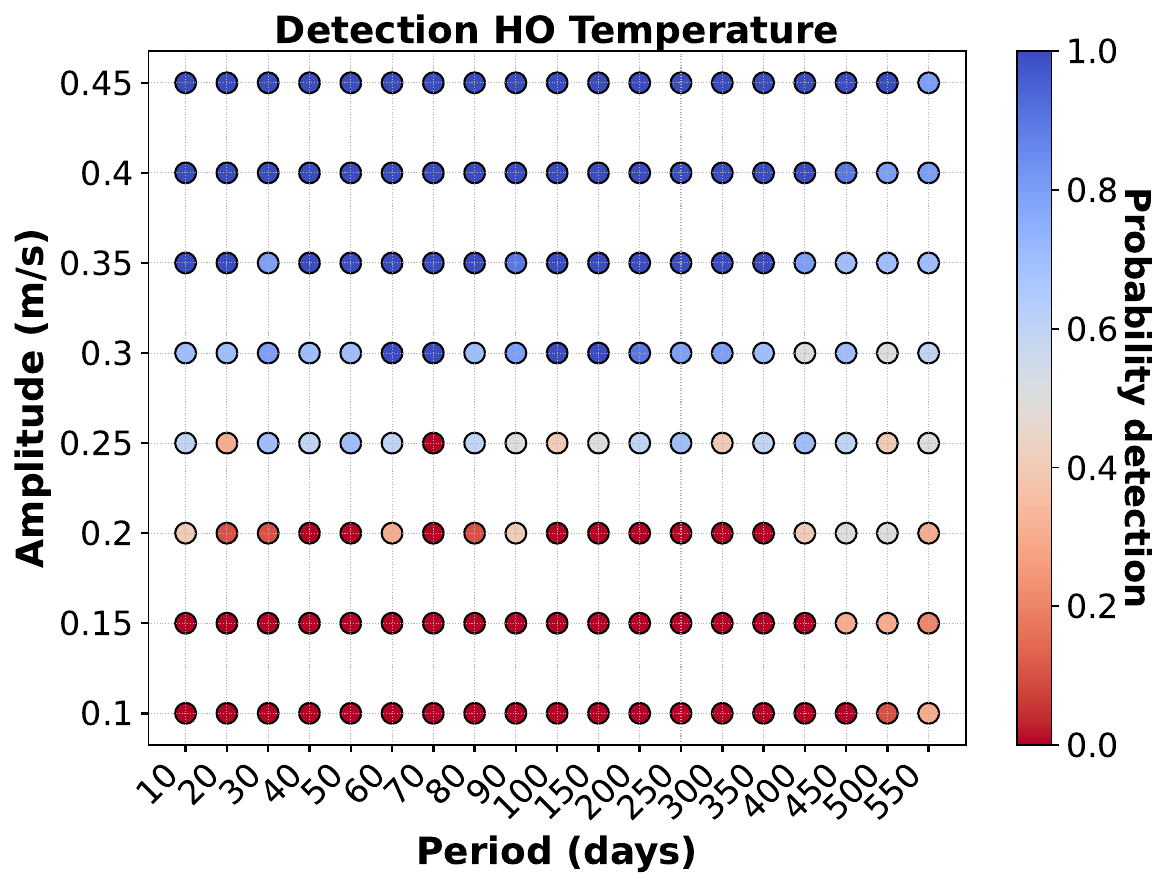}
    \includegraphics[trim=22mm 0mm 2mm 0mm, clip, width=4.0cm, height=4.5cm]{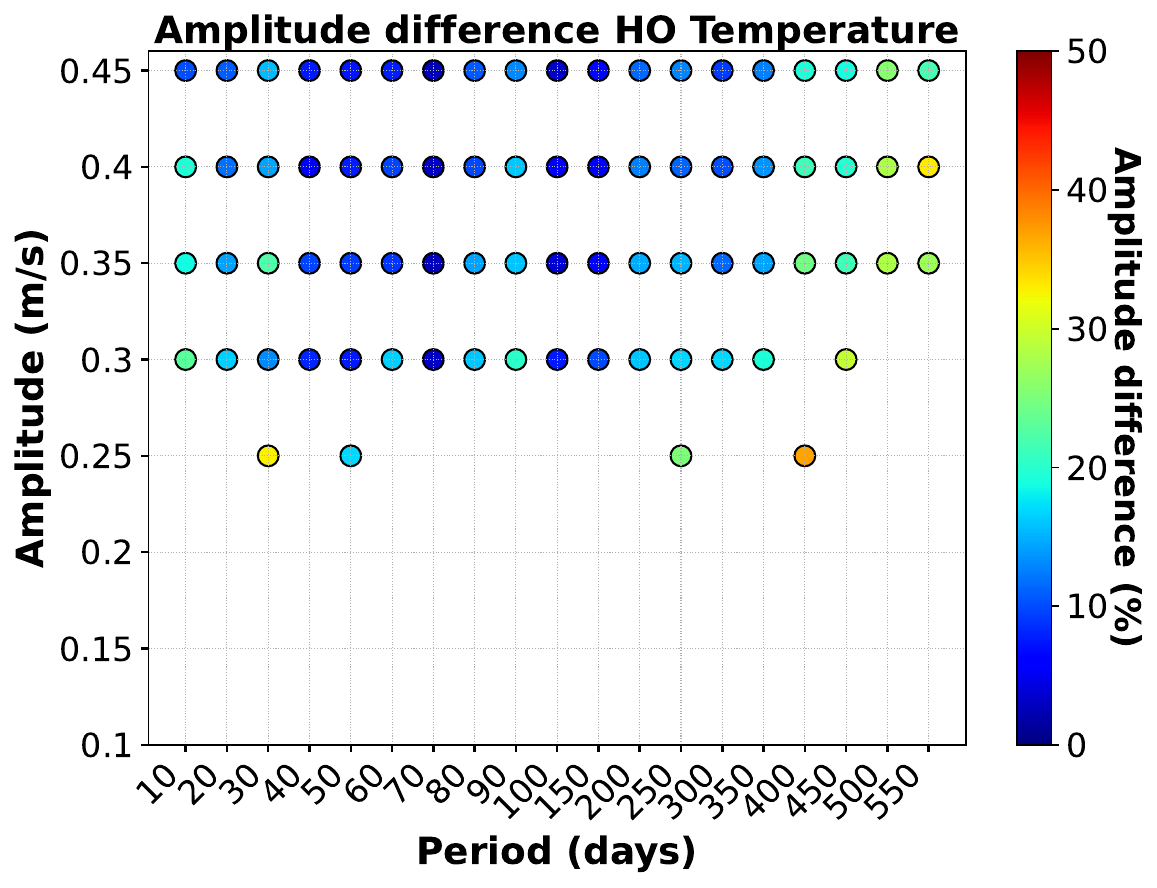}
    \caption{\changes{Detection probability for the temperature-based model under the chronological hold-out strategy, trained on the most recent 1636 spectra and tested on the oldest 400. \textit{Left}: fraction of successful recoveries over 10 realizations. \textit{Right}: relative amplitude differences for the detected signals.}}
    \label{fig:first400samplesMap}
\end{figure}

\changes{We train our neural network framework on the most recent 1636 spectra and evaluate it on the oldest 400 observations. We selected the oldest 400 spectra for testing as they correspond to a level of activity that exists in the training set; however, they are separated by half the solar cycle periodicity ($\sim$5-6 years), as the first 400 measurements correspond to the middle of the decreasing phase of solar cycle 24, while the training set includes the middle of the rising phase of solar cycle 25.} 
\changes{Figures~\ref{fig:first400samples} and~\ref{fig:first400samplesMap} show the corresponding periodograms and detection maps when testing our trained model on the oldest 400 spectra. Those results are very similar to the ones obtained when the samples are randomly shuffled, with a robust detection limit in the hold-out experiments based on the temperature shell at a level of 0.35~m/s. However, a small difference is observed for periods equal to or greater than 400 days, where we see that for planetary signals of semi-amplitude 0.35~m/s, the detection rate between the 10 different phase evaluations drops, and the relative amplitude difference between the injected and recovered signal increases. This comes from the fact that the testing set is now chronological in time, and thus the 400 spectra selected correspond to a time span of only 600 days. The signal of planets with long periods is therefore not well sampled in phase, inducing the drop in detection sensitivity.}

\changes{This test demonstrates that our neural network framework retains the ability to recover injected planetary signals even under a strict temporal split between the training and test sets, and thus temporal correlation between nearby spectra when selecting training and testing sets randomly is not inducing overfitting. The prediction of our neural network framework seems therefore, to generalize well across different observing periods.}

\subsection{Planetary Signal Recovery at Different Orbital Periods for a 20 cm/s Signal}
\label{sec:app_plantary20cms}

To illustrate the behavior of our best-performing neural-network model (CV-Temp), we analyze the recovery of low-amplitude planetary signals at different orbital periods. In analogy with the case shown in Fig.~\ref{fig:20cmCV}, which corresponds to an injected semi-amplitude of 0.2~m/s at 350~days, we perform periodogram analyses of both RV and DS predictions for injections at 10, 50, and 100 days. For each period, the analysis is repeated over 10 independent evaluation sets, in accordance with the analysis discussed in Section \ref{sec:results} and with the Figure \ref{fig:compact_detection_maps}.

\begin{figure*}
    \centering
    \includegraphics[trim=30mm 280mm 35mm 40mm, clip, width=\textwidth]{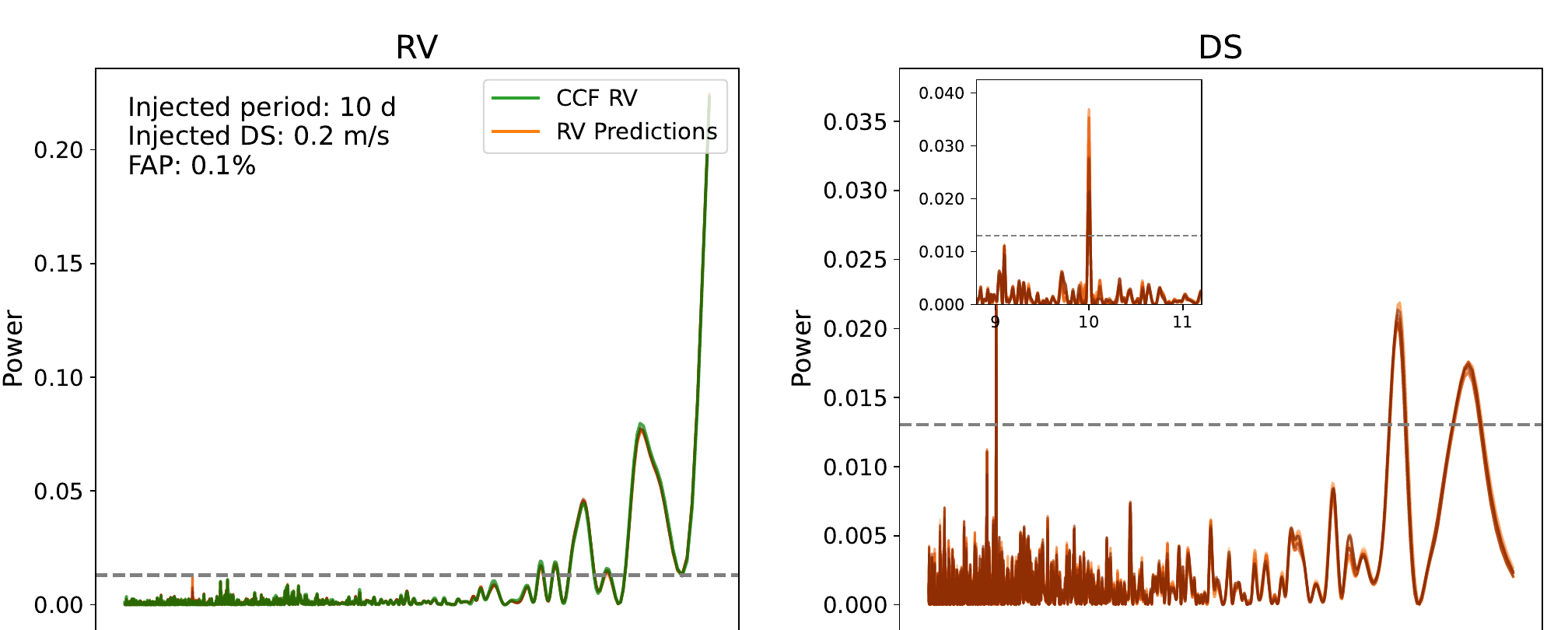}
    \includegraphics[trim=30mm 280mm 35mm 53mm, clip, clip, width=\textwidth]{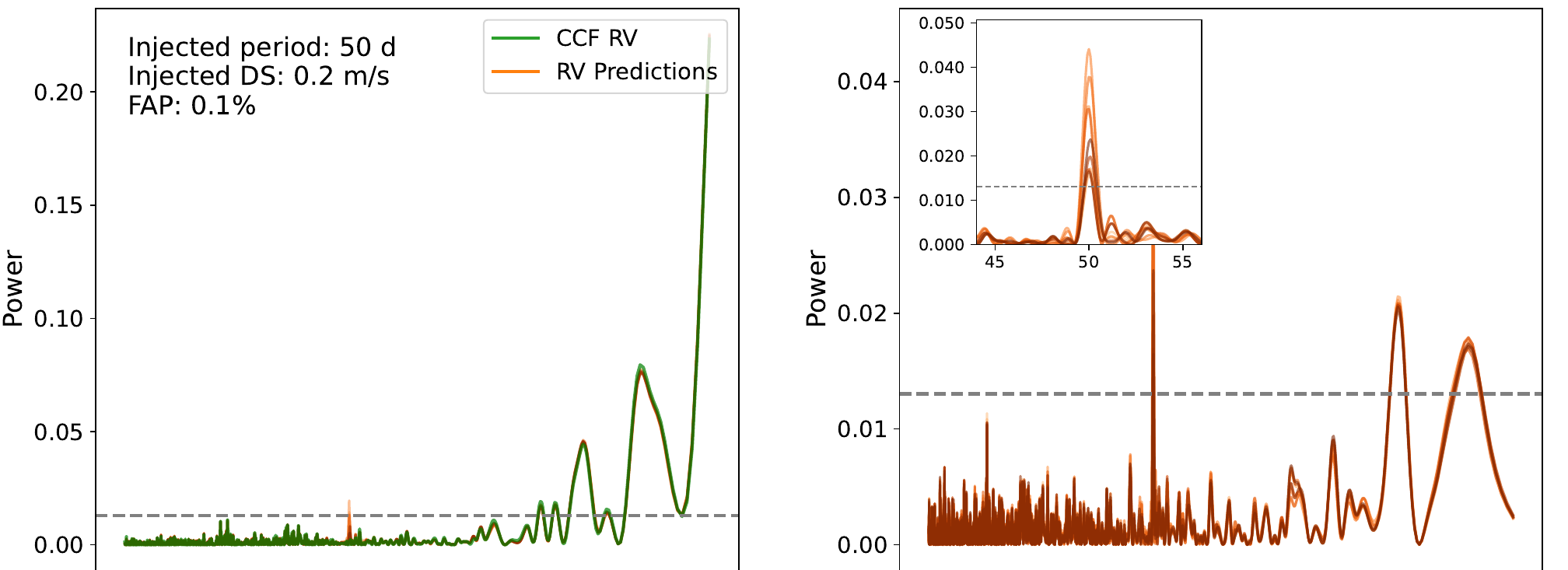}
    \includegraphics[trim=30mm 260mm 35mm 53mm, clip, clip, width=\textwidth]{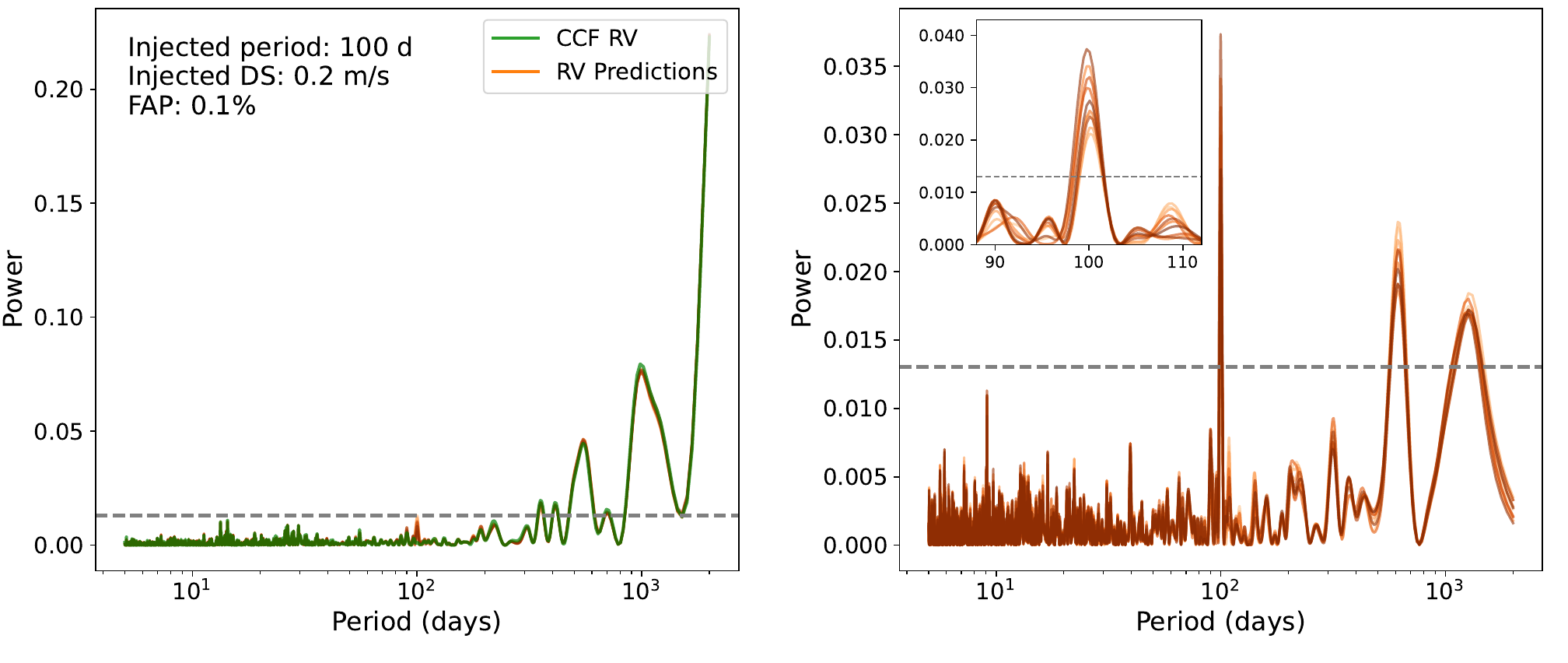}
    \caption{Recovery of a 0.2~m/s planetary signal at different orbital periods using the CV-Temp model. Each row corresponds to an injected period of 10, 50, and 100~days, respectively. Left panels show the periodograms of the traditional CCF RVs (green) and the neural-network-predicted RVs (orange), while right panels show the periodograms of the predicted Doppler shifts (DS). The dashed line indicates the 0.1\% false-alarm probability threshold, and the insets highlight the region around the injected period for the 10 independent evaluation sets. The planetary signal is consistently recovered in the DS periodograms for all tested periods, while remaining undetected in the RV-based analyses. }

    \label{fig:app_20cmCV}
\end{figure*}

Figure~\ref{fig:app_20cmCV} compares the resulting RV and DS periodograms. The second column, corresponding to the DS predictions, shows that for shorter orbital periods, the recovered signals appear as narrower and more sharply defined peaks in the periodogram. This behavior is particularly evident for the 10-day injection. As the orbital period increases, the detected power becomes more dispersed, as seen in the 100-day case, although the spread remains smaller than for the 350-day injection shown in Fig.~\ref{fig:20cmCV}. 

In all three cases, the injected signal is successfully detected in all 10 out of 10 trials using the DS predictions. In contrast, the RV periodograms shown in the first column do not exhibit significant power at the injected periods, neither for the traditional CCF-derived RVs nor for the RVs predicted by the neural network. Even when the FAP threshold is formally satisfied, the peaks associated with the injected signal are overshadowed by stronger power at longer periods, likely driven by magnetic activity. 

Overall, in these examples, we can highlight the improved sensitivity of the DS-based analysis for recovering low-amplitude planetary signals across a range of orbital periods.

\section{Model Robustness and Uncertainty Quantification}
\label{sec:app_uncertainty}

We estimate predictive uncertainties using Monte Carlo dropout, which enables stochastic inference and provides an estimate of the epistemic uncertainty of the neural-network models. \changes{These uncertainties should be interpreted as a measure of model dispersion rather than as calibrated predictive intervals, as would be obtained with methods such as conformal prediction, which we leave for future work.} All results reported in this manuscript correspond to the mean prediction over $N = 100$ stochastic forward passes, with uncertainties defined as the standard deviation across these realizations.

\begin{figure*}
    \centering

    \makebox[17cm][c]{
        \includegraphics[trim=0mm 0mm 8mm 0mm, clip, width=5.3cm, height=4.8cm]{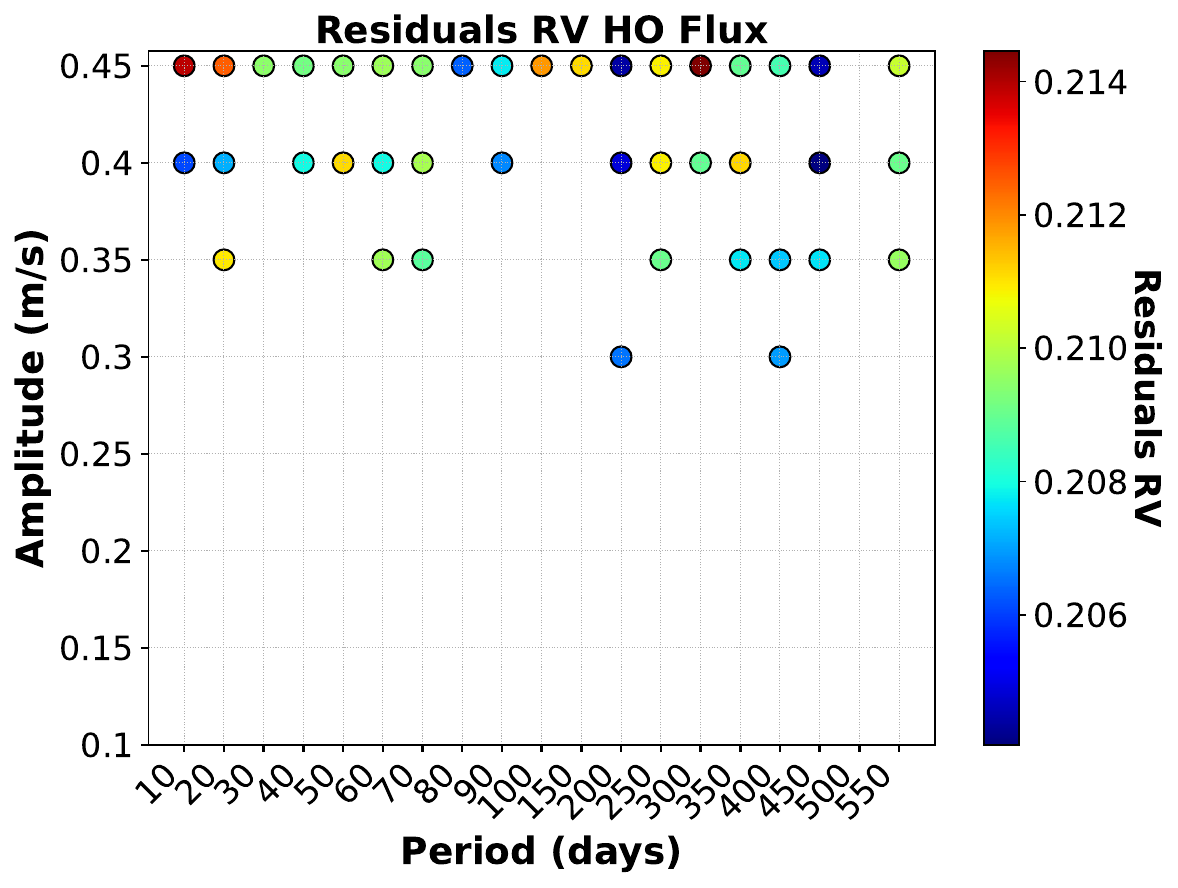}
        \includegraphics[trim=23mm 0mm 0mm 0mm, clip, width=4.9cm, height=4.8cm]{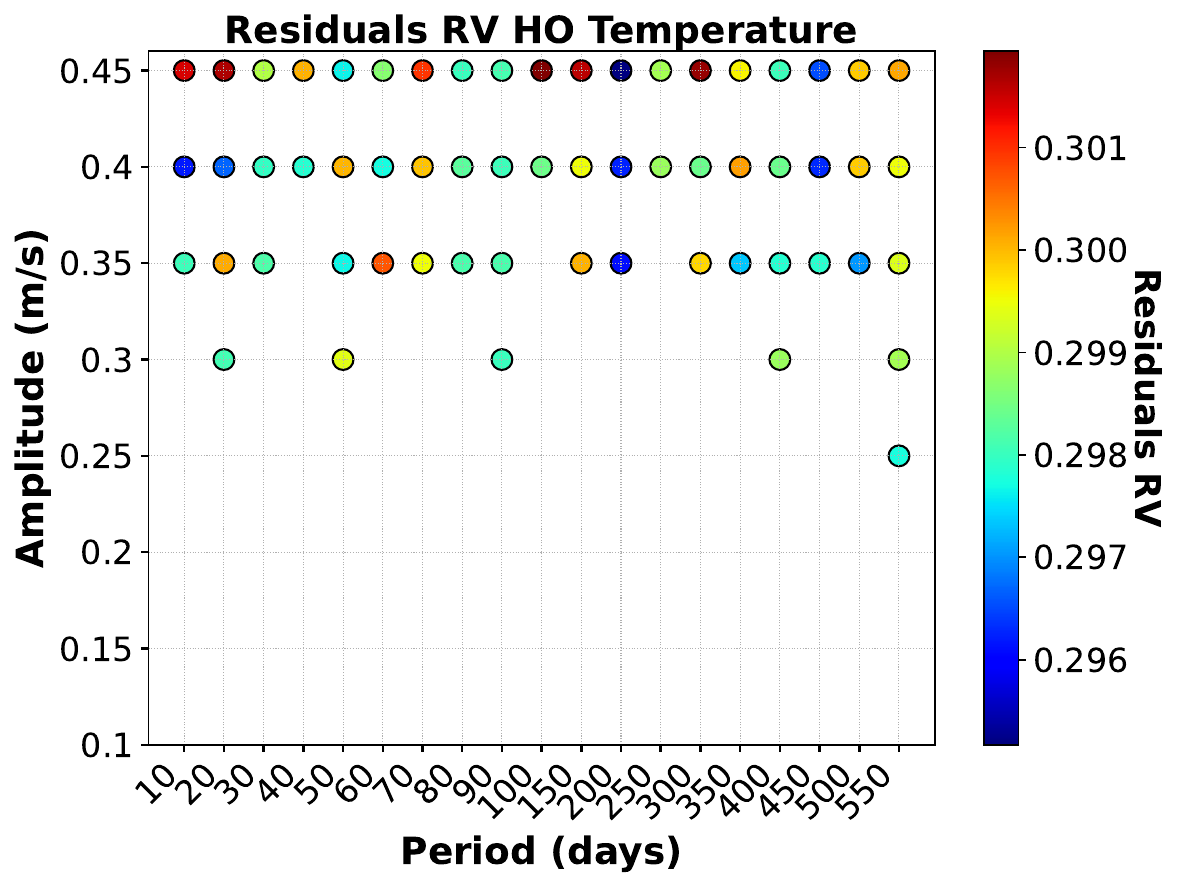}
        \includegraphics[trim=0mm 0mm 8mm 0mm, clip, width=5.3cm, height=4.8cm]{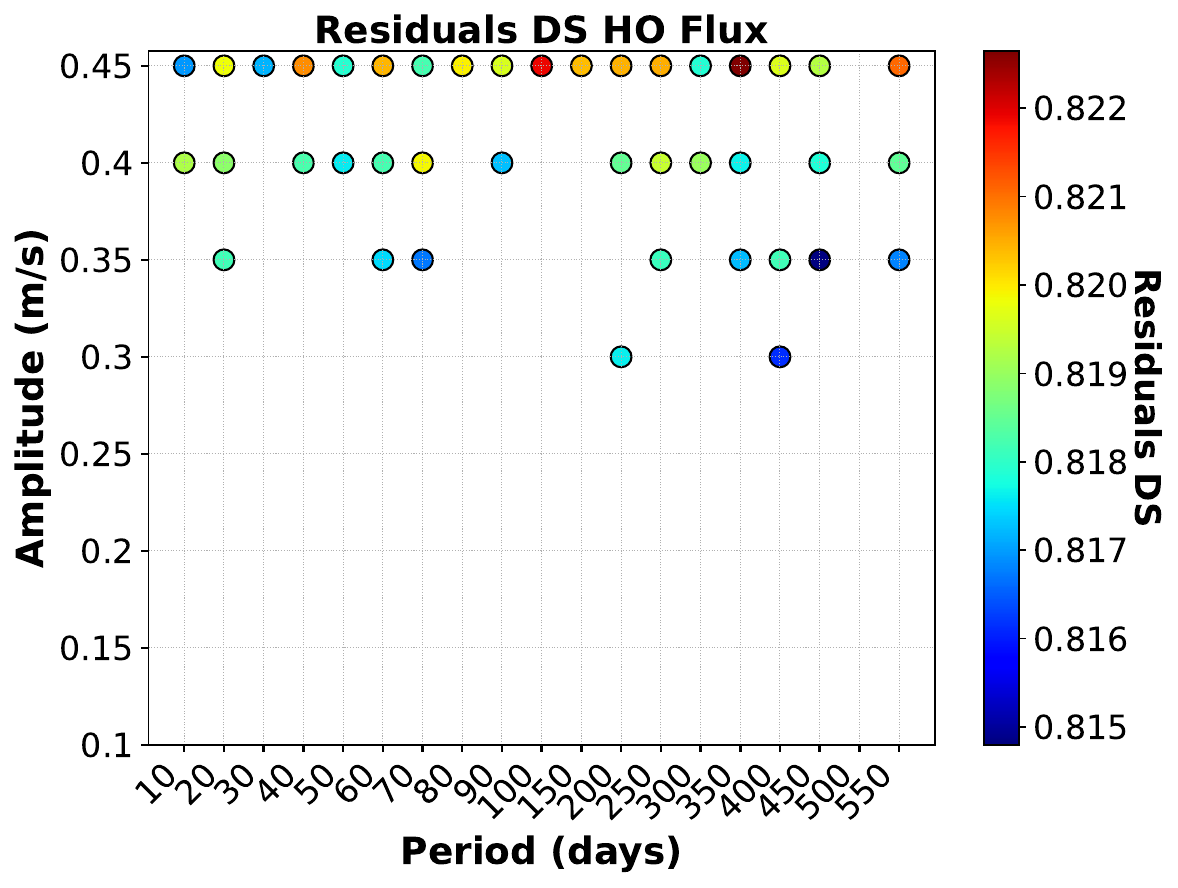}
        \includegraphics[trim=23mm 0mm 0mm 0mm, clip, width=4.9cm, height=4.8cm]{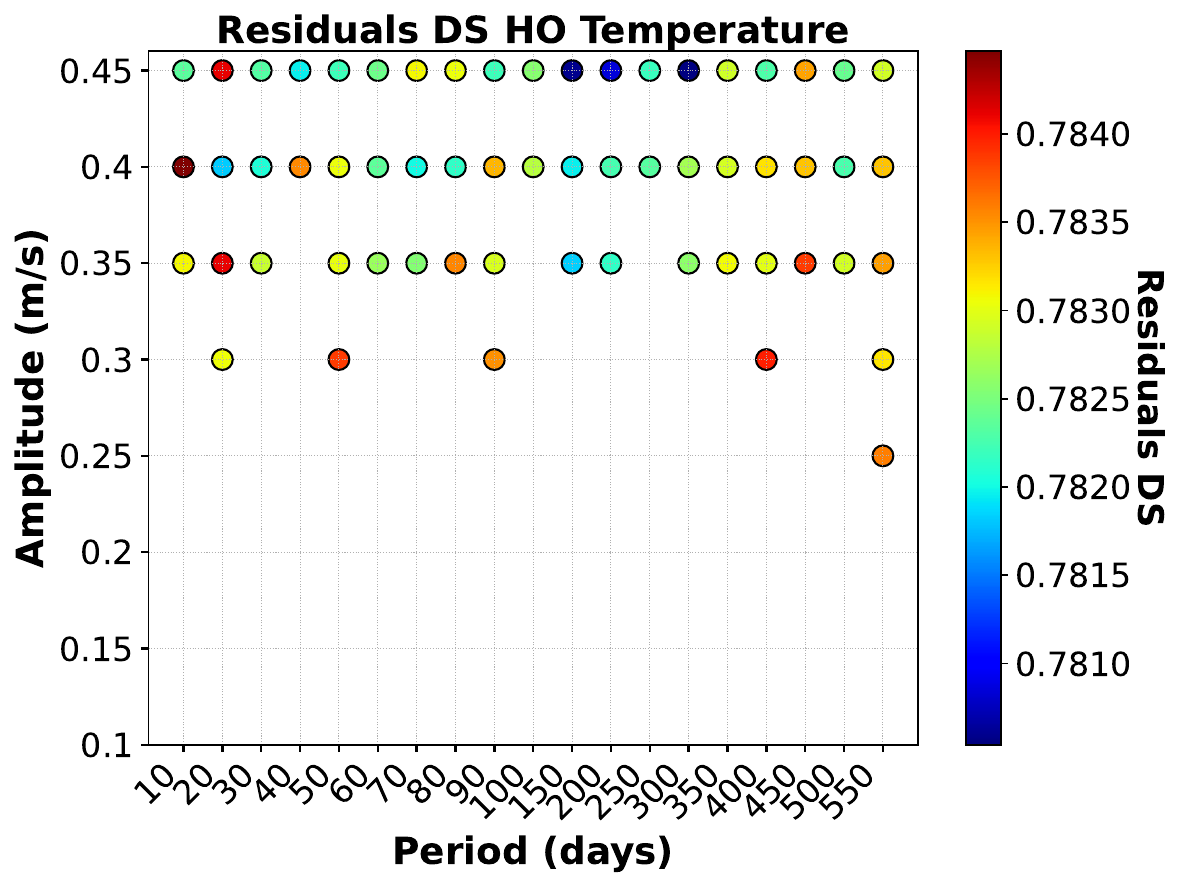}
    }

    \vspace{1em}
    \makebox[17cm][c]{
        \includegraphics[trim=0mm 0mm 8mm 0mm, clip, width=5.3cm, height=4.8cm]{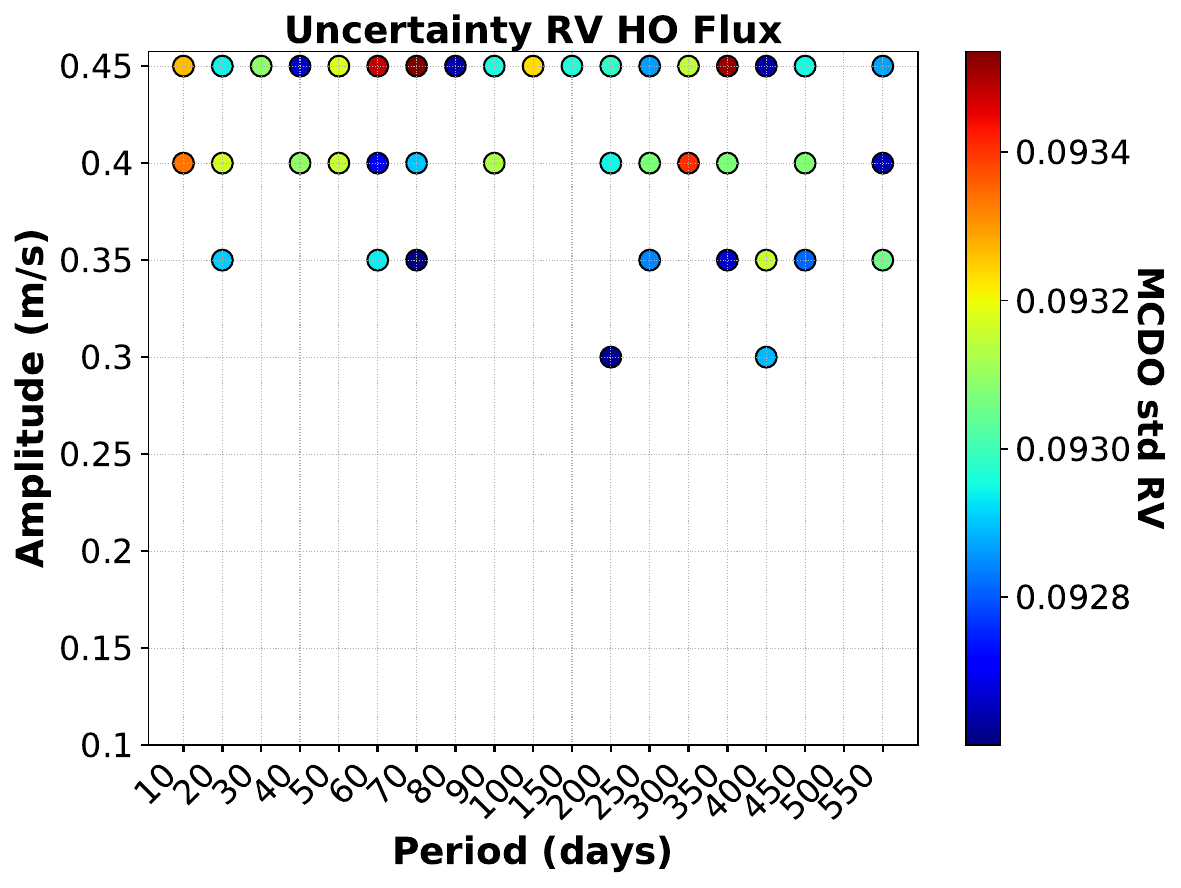}
        \includegraphics[trim=23mm 0mm 0mm 0mm, clip, width=4.9cm, height=4.8cm]{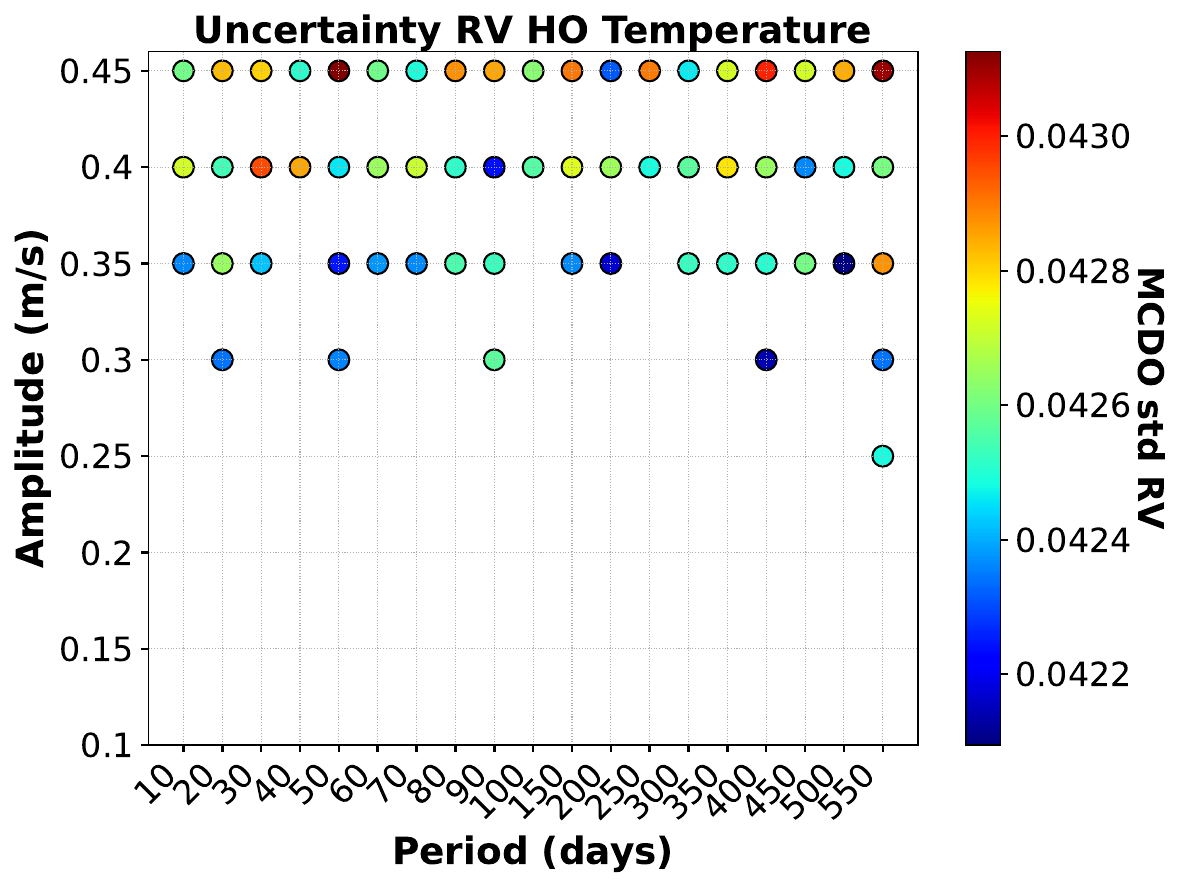}
        \includegraphics[trim=0mm 0mm 8mm 0mm, clip, width=5.3cm, height=4.8cm]{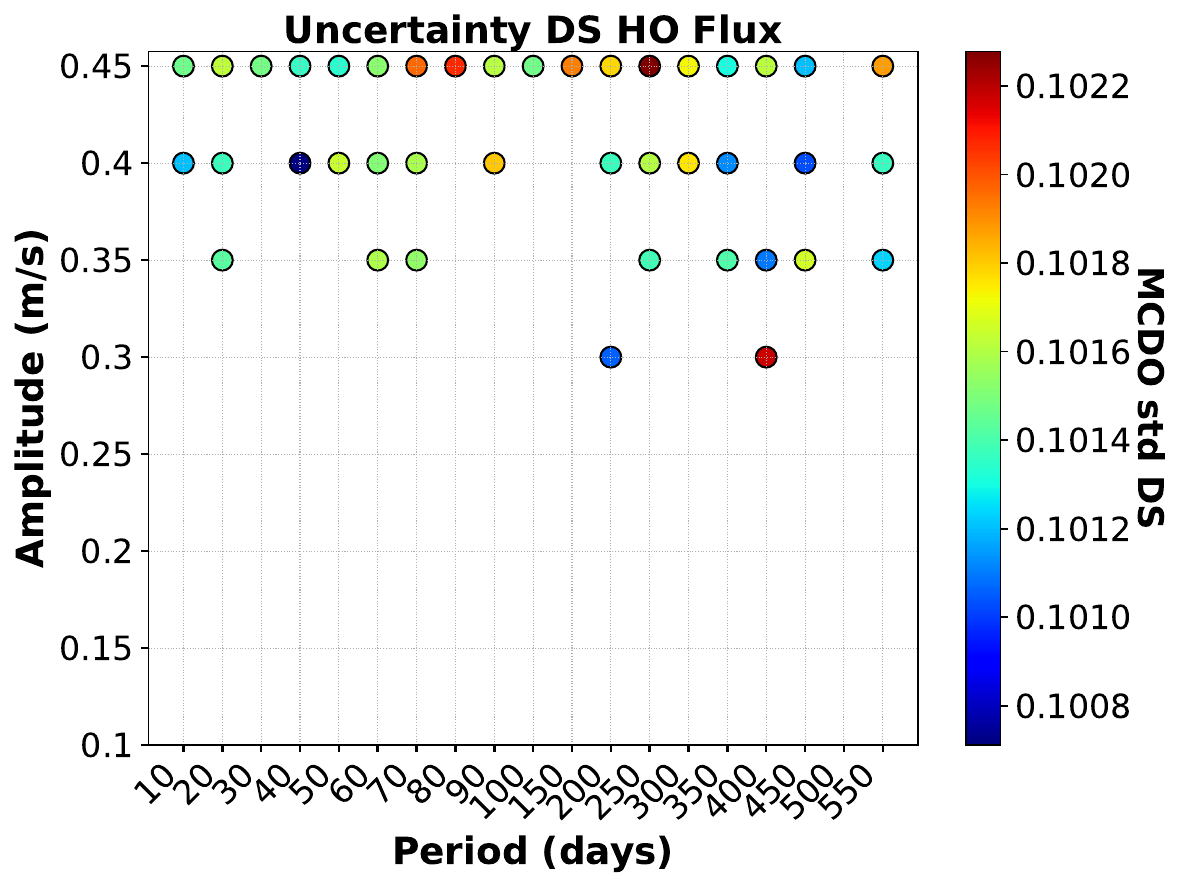}
        \includegraphics[trim=23mm 0mm 0mm 0mm, clip, width=4.9cm, height=4.8cm]{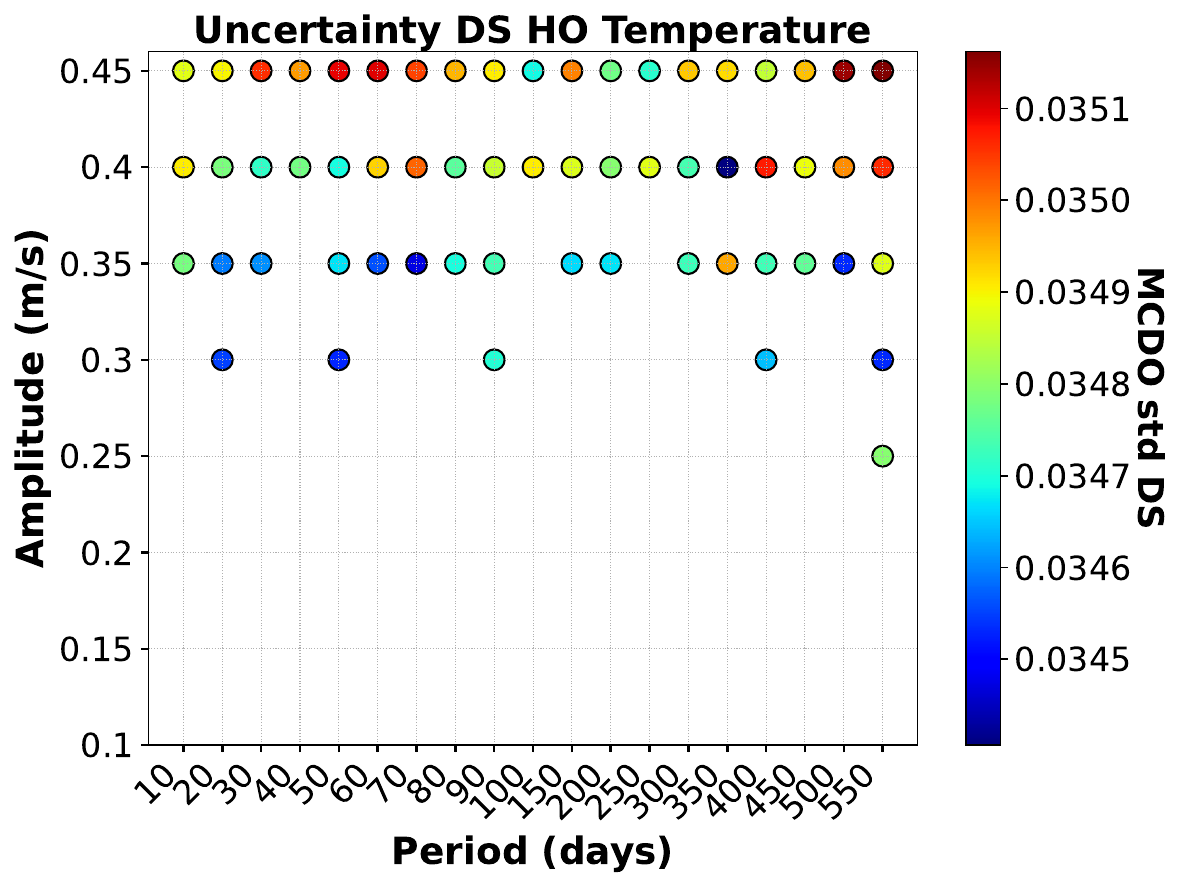}
    }

    \vspace{1em}

    \makebox[17cm][c]{
        \includegraphics[trim=0mm 0mm 8mm 0mm, clip, width=5.3cm, height=4.8cm]{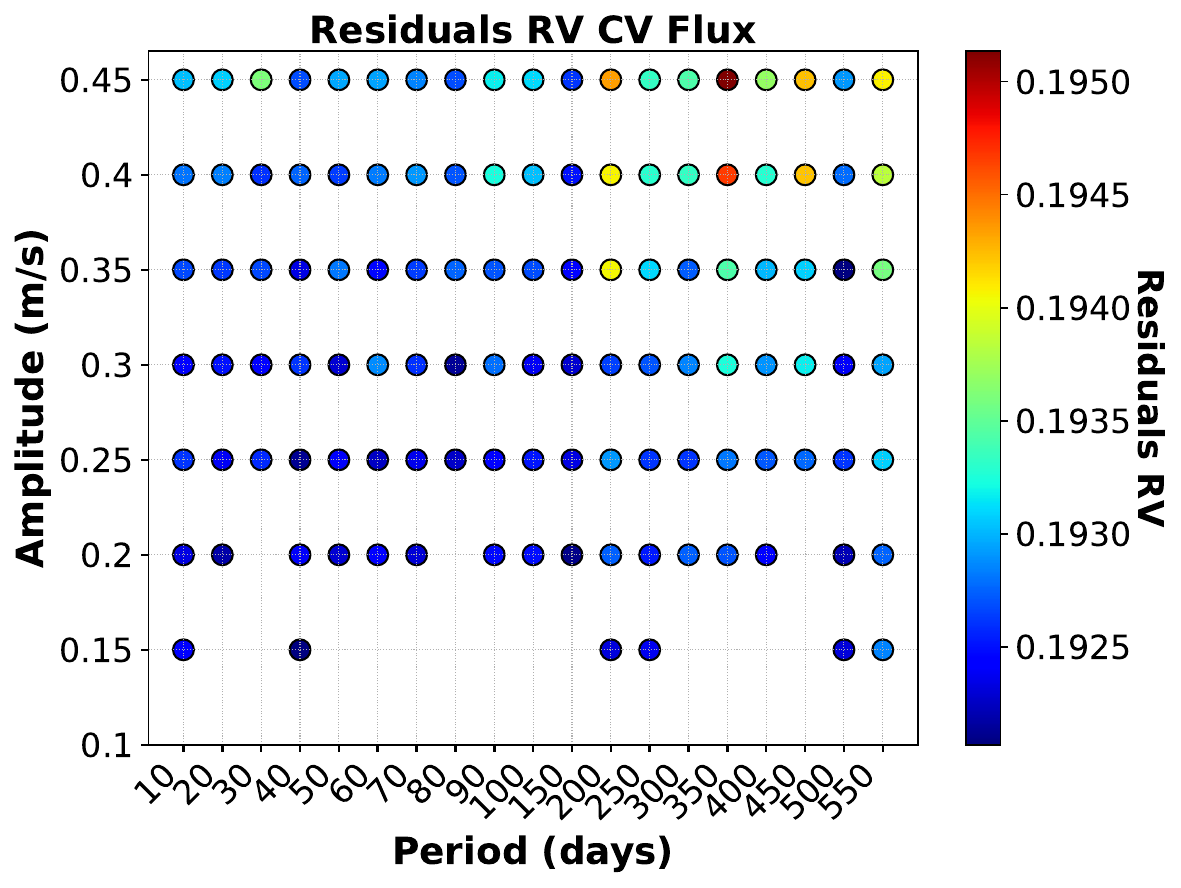}
        \includegraphics[trim=23mm 0mm 0mm 0mm, clip, width=4.9cm, height=4.8cm]{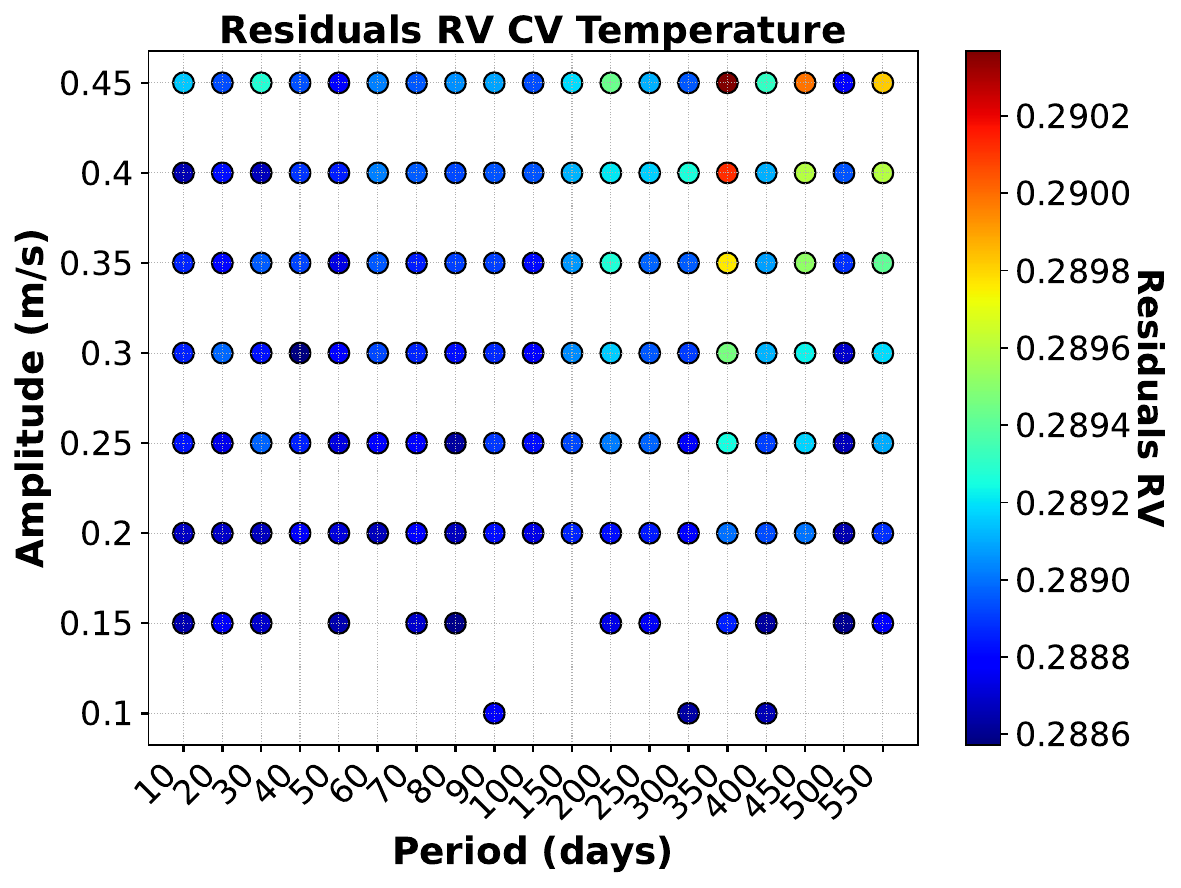}
        \includegraphics[trim=0mm 0mm 8mm 0mm, clip, width=5.3cm, height=4.8cm]{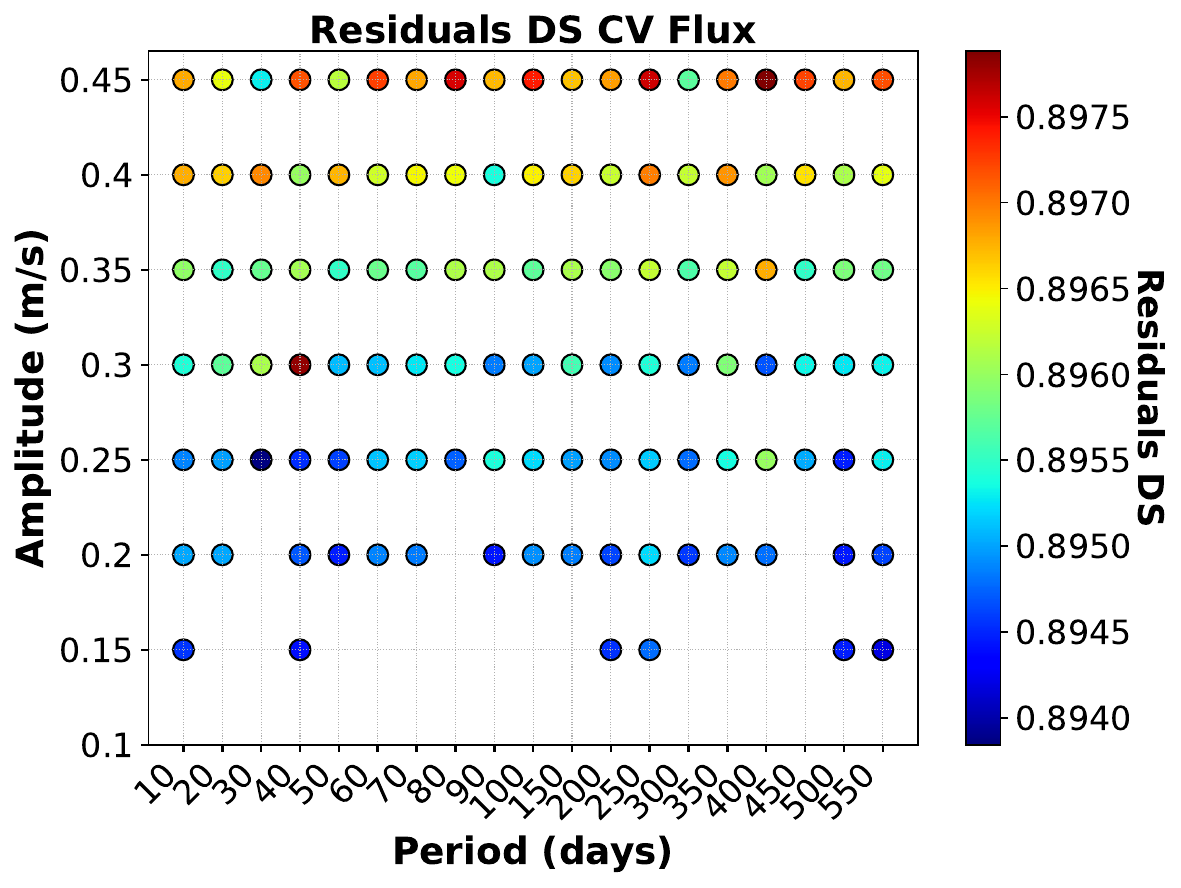}
        \includegraphics[trim=23mm 0mm 0mm 0mm, clip, width=4.9cm, height=4.8cm]{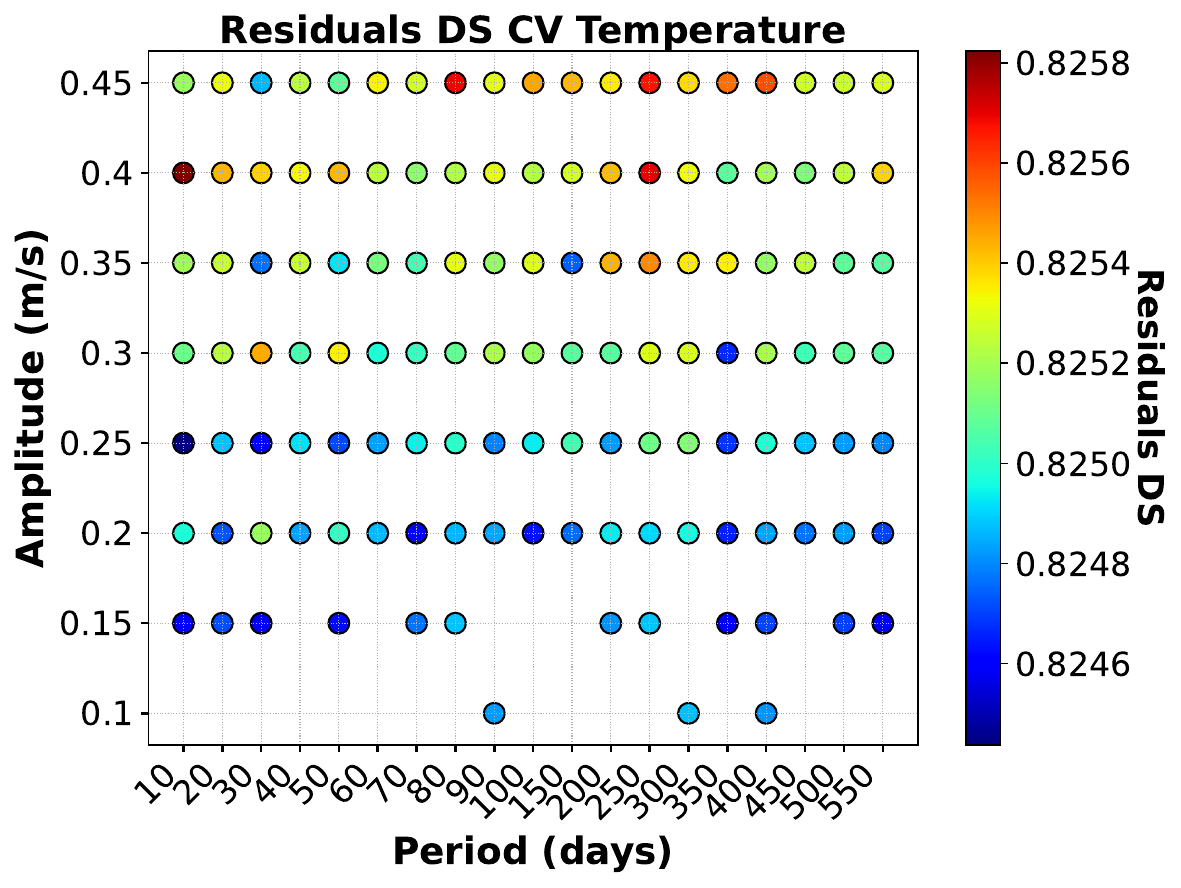}
    }

    \vspace{1em}

    \makebox[17cm][c]{
        \includegraphics[trim=0mm 0mm 8mm 0mm, clip, width=5.3cm, height=4.8cm]{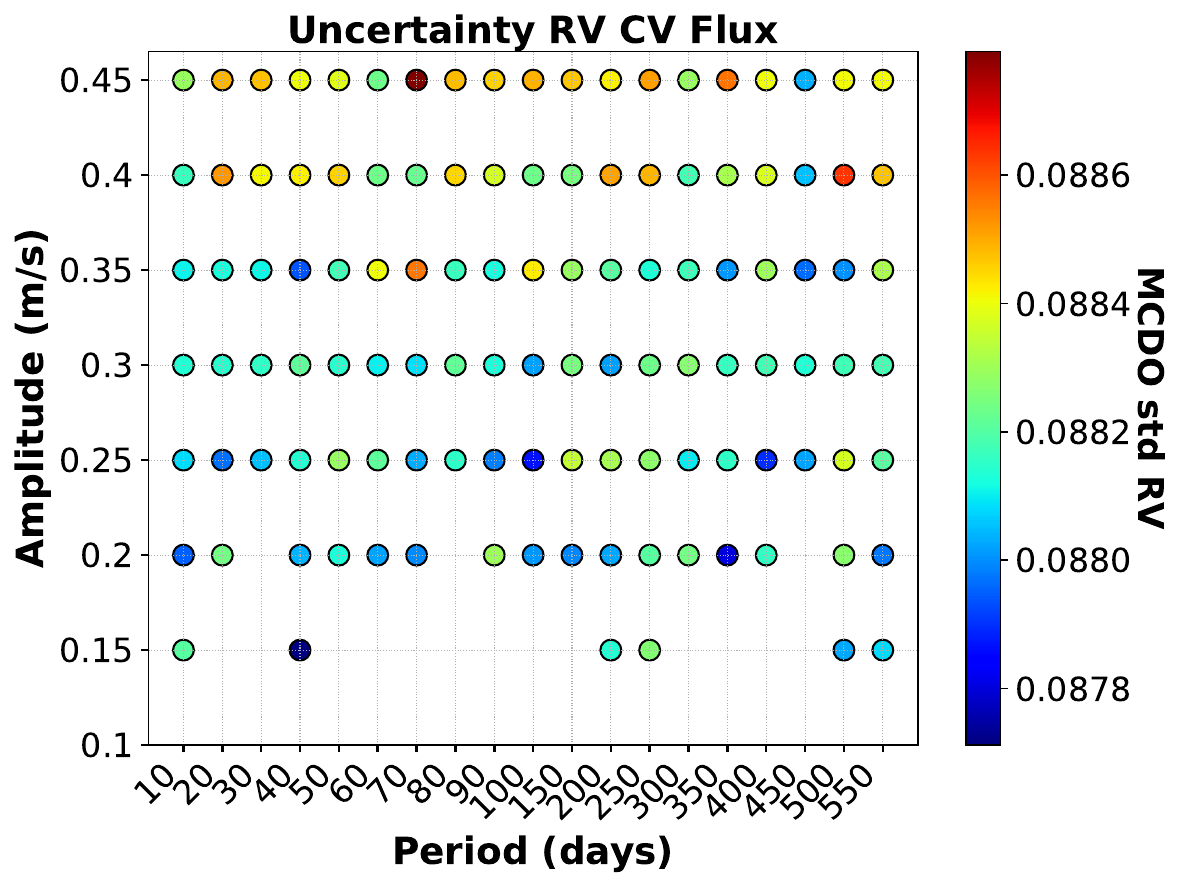}
        \includegraphics[trim=23mm 0mm 0mm 0mm, clip, width=4.9cm, height=4.8cm]{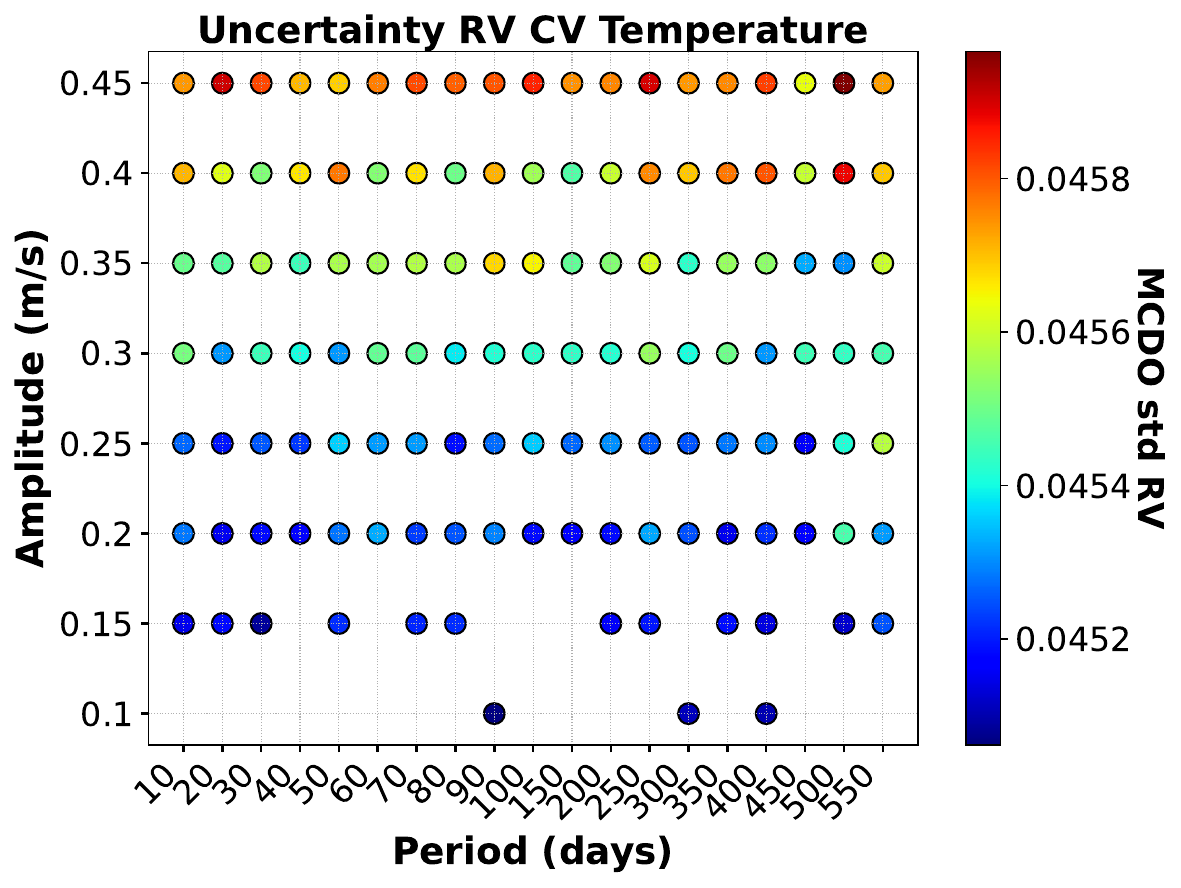}
        \includegraphics[trim=0mm 0mm 8mm 0mm, clip, width=5.3cm, height=4.8cm]{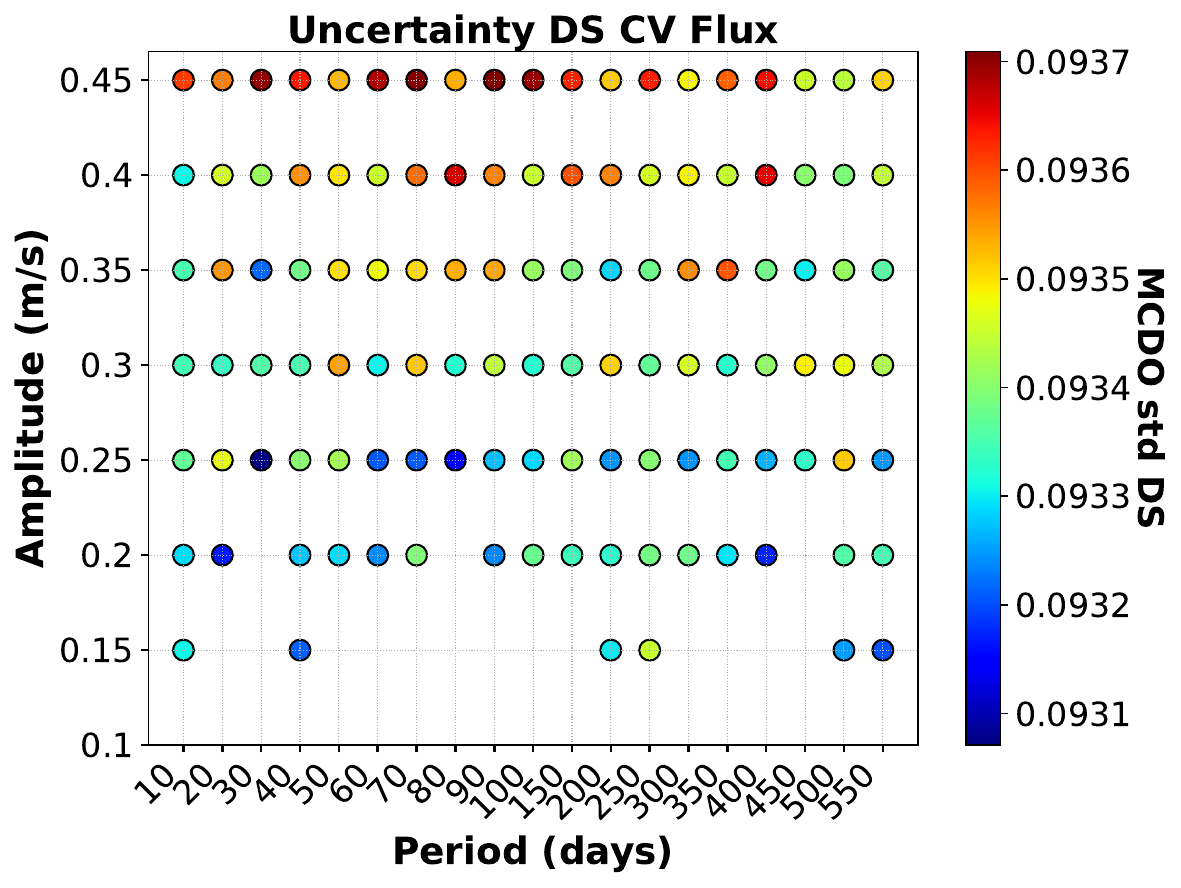}
        \includegraphics[trim=23mm 0mm 0mm 0mm, clip, width=4.9cm, height=4.8cm]{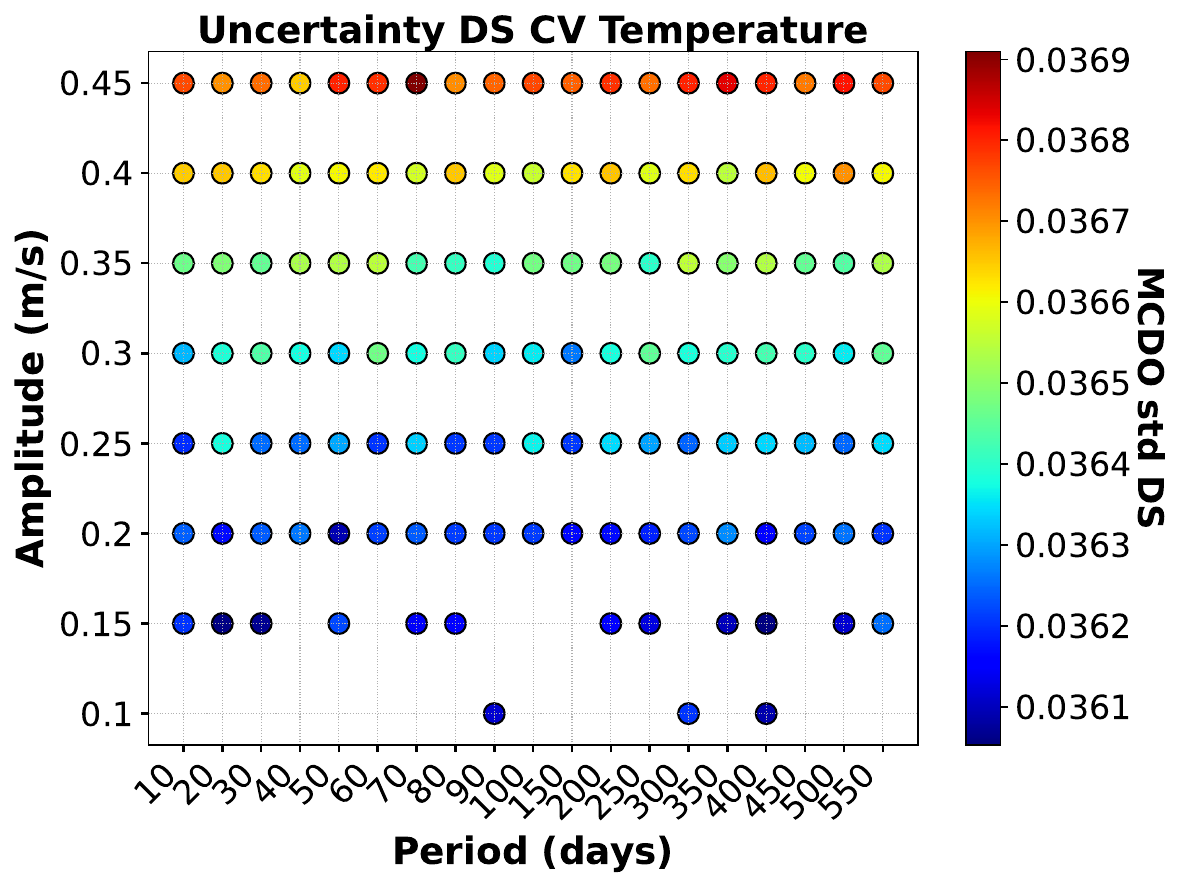}
    }

    \caption{Root mean squared errors (residuals) and predictive uncertainties for radial velocity and Doppler shift obtained using the hold-out and cross-validation approaches across the 10 different test sets. The first and second rows correspond to the HO models: the first row shows the RMSE between predicted and true values, while the second row presents the mean predictive uncertainties, defined as the standard deviation over 100 Monte Carlo dropout realizations (MCDO). The third and fourth rows correspond to the CV models: the third row shows the RMSE obtained with the 5-fold CV approach, and the fourth row presents the corresponding mean predictive uncertainties.}
    \label{fig:app_err}
\end{figure*}

For each configuration, we compute the root mean squared error (RMSE) between predicted and true values for both RV and DS, together with the mean predictive uncertainty across unseen samples. Figure~\ref{fig:app_err} summarizes these quantities for both the hold-out and cross-validation approaches. \newchanges{The RMSE values and predictive uncertainties are similar between the HO and CV strategies (Table~\ref{tab:app_metrics}), indicating consistent predictive behavior across the two validation approaches. The predictive uncertainties, estimated through MC dropout, quantify the variability of the predictions across stochastic forward passes and therefore should not be interpreted as calibrated estimates of the total prediction error.} 

In all configurations, DS predictions exhibit larger errors and predictive uncertainties than RVs. \changes{We emphasize that the neural-network architecture is explicitly designed to model the Doppler shift (DS) signal rather than the CCF-derived radial velocity. The RV predictions are included primarily as a reference quantity, directly comparable to standard spectroscopic measurements. This distinction reflects the different nature of the two targets: while the RV is directly derived from the observed spectra via the CCF, the DS corresponds to a synthetic signal injected at the spectral level, which may introduce additional sources of noise and mismatch. In this context, the improved performance of temperature-based representations in recovering the injected DS signal indicates that these representations are more sensitive to planetary Doppler shifts encoded at the spectral level, whereas flux-based representations are more closely tied to the information content of the CCF and thus to the measured RVs.}

Temperature-based shell representations consistently yield lower predictive uncertainties than flux-based shells for both HO and CV strategies, while exhibiting larger residuals in RV and lower residuals in DS (Table~\ref{tab:app_metrics}). \changes{This behavior can be understood in terms of a bias-variance trade-off, given that flux-based models more directly capture the information used to compute the CCF radial velocities, leading to lower RV residuals, whereas temperature-based representations involve an additional transformation to derive temperature that can introduce interpolation-related errors, increasing the bias in RV predictions. At the same time, temperature-based models exhibit significantly lower predictive dispersion in the MC dropout estimates, indicating reduced model variance.}

\changes{The apparent discrepancy between residuals and predictive uncertainties should therefore not be interpreted as an inconsistency or error in the uncertainty estimation. The RMSE quantifies the total prediction error with respect to the target, while MC dropout provides an estimate of epistemic uncertainty, i.e., the variability of the model predictions under stochastic forward passes. As a result, these two quantities capture different aspects of model performance and are not expected to coincide quantitatively.}

\changes{We note that the CV results correspond to an average over multiple models, which leads to slightly higher RMSE values compared to the HO case due to the aggregation of independent predictions. For this reason, direct quantitative comparison between HO and CV is not straightforward, although the relative trends between flux and temperature representations remain consistent across both strategies.}

Despite the scatter in individual DS predictions, the periodic structure of the injected signal is preserved across the time series, enabling successful recovery through periodogram-based analyses, as discussed in Section~\ref{sec:results}.

\begin{figure}[h!]
    \centering
    \includegraphics[trim=0mm 0mm 0mm 0mm, clip, width=0.42\textwidth]{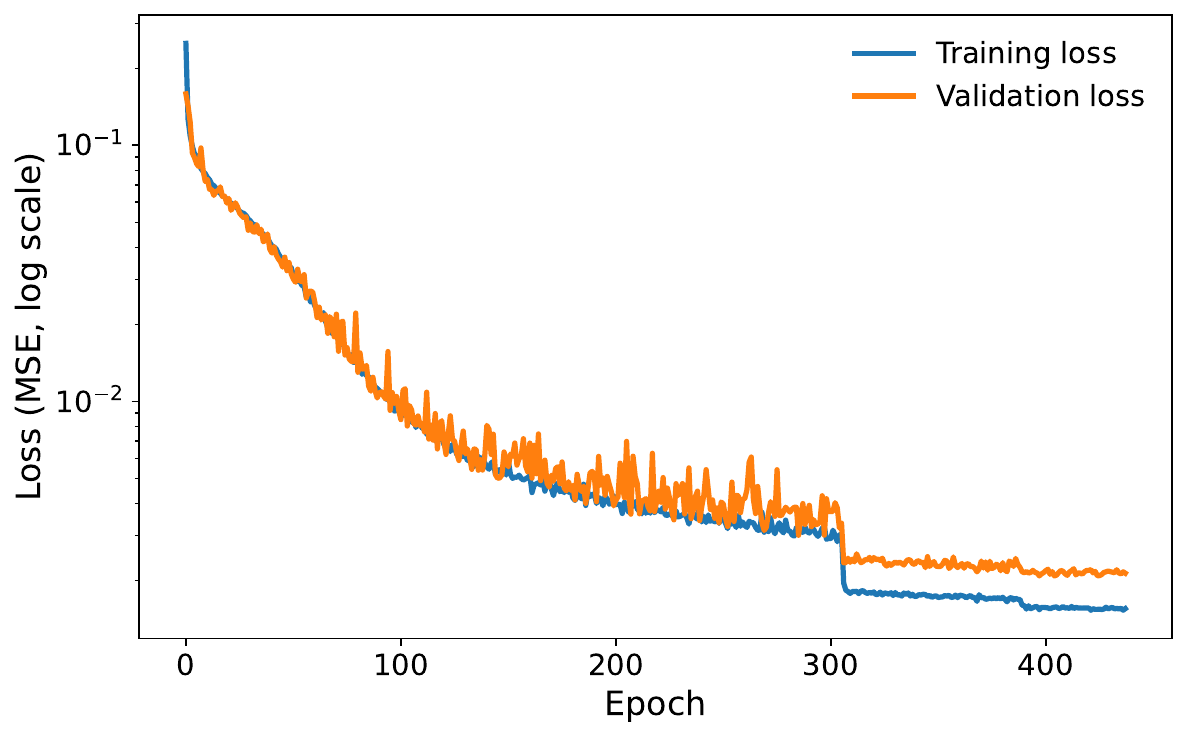}
    \caption{\changes{Training and validation loss function for an HO CNN model using temperature shells. Both curves decrease steadily and remain close throughout training, with no divergence at later epochs, indicating stable training and no sign of overfitting. }}
    \label{fig:app_loss}
\end{figure}

Finally, Fig.~\ref{fig:app_loss} shows representative training and validation loss curves \changes{for a temperature-based hold-out model corresponding to the best-performing configuration identified in Section~\ref{sec:DL_neural_design}. The loss decreases smoothly from values of order $10^{-1}$ at early epochs to a few $\times 10^{-3}$ at convergence, with both curves closely tracking each other throughout training. This behavior indicates stable convergence without signs of overfitting. Similar trends are observed for flux-based models and across the CV folds, providing additional confidence in the robustness of the training procedure.}

\begin{table}[h!]
\centering
\footnotesize
\setlength{\tabcolsep}{2.5pt}
\renewcommand{\arraystretch}{1.0}
\caption{\changes{Ranges of RMSE (residuals) and MC dropout predictive uncertainties for RV and DS across hold-out (HO) and cross-validation (CV) strategies.}}
\label{tab:app_metrics}
\begin{tabular}{lcccc}
\hline
 & \multicolumn{2}{c}{RV} & \multicolumn{2}{c}{DS} \\
\cline{2-5}
 & Flux & Temp & Flux & Temp \\
\hline
RMSE (HO)        & 0.204--0.215   & 0.294--0.302   & 0.815--0.824   & 0.7808--0.7844 \\
Unc. (HO)        & 0.0926--0.0936 & 0.042--0.0432  & 0.1006--0.1023 & 0.0344--0.0352 \\
RMSE (CV)        & 0.192--0.1952  & 0.2886--0.2904 & 0.894--0.898   & 0.8244--0.8258 \\
Unc. (CV)        & 0.0876--0.0888 & 0.045--0.046   & 0.093--0.0937  & 0.036--0.0369 \\
\hline
\end{tabular}
\end{table}

\end{document}